\documentclass[rmp,aps,twocolumn,showpacs,fleqn,epsf]{revtex4}
\usepackage{graphics,bm}
\usepackage{epsfig}
\usepackage{graphicx}
\newcommand{\beq}{\begin{equation}}
\newcommand{\eeq}{\end{equation}}
\newcommand{\bqa}{\begin{eqnarray}}
\newcommand{\eqa}{\end{eqnarray}}

\protect
\def\sumint{\hbox{$\sum$}\!\!\!\!\!\!\int}
\def\square{\vcenter{\vbox{\hrule height.4pt
          \hbox{\vrule width.4pt height8pt
          \kern8pt\vrule width.4pt}\hrule height.4pt}}}

\begin{document}

\def\bfgamma{\mbox{\boldmath$\gamma$}}
\def\bfalpha{\mbox{\boldmath$\alpha$}}
\def\bftau{\mbox{\boldmath$\tau$}}
\def\bfnabla{\mbox{\boldmath$\nabla$}}
\def\bfsigma{\mbox{\boldmath$\sigma$}}
\def\bfxi{\mbox{\boldmath$\xi$}}
\def\BN{\hbox{Bloch-Nordsiek}}
\def\vp{\mbox{$\bf v\cdot p$}}
\def\vq{\mbox{$\bf v\cdot q$}}
\def\vpq{\mbox{$\bf v\cdot(p+ q)$}}
\def\tilA{\mbox{$v\cdot A$}}
\def\tilQ{\mbox{v\cdot q}}
\def\tilQ1{\mbox{$v\cdot q_1$}}
\def\tilQ2{\mbox{$v\cdot q_2$}}
\def\bfp{\mbox{\boldmath$p$}}
%

\newenvironment{petitchar}{\begin{list}{}
{\leftmargin1.5em\rightmargin0.0cm}%
\item\small}{\end{list}}

\baselineskip=13pt
\title{Theory of the Weakly Interacting Bose Gas}
\vskip0.5cm
\author{Jens O. Andersen}
\email{jensoa@nordita.dk}
\affiliation{Nordita,\\
Blegdamsvej 17, 2100 Copenhagen, Denmark}

\date{\today}

\begin{abstract} 
We review recent advances in the theory of the three-dimensional
dilute
homogeneous Bose gas at zero and finite temperature. Effective field 
theory methods are used to formulate a systematic perturbative framework
that can be used to calculate the properties of the system at $T=0$.
The perturbative expansion of these properties is essentially
an expansion in the
gas parameter $\sqrt{na^3}$, where $a$ is the $s$-wave scattering
length and $n$ is the number density.
In particular, the leading quantum corrections to the ground state energy
density, the condensate depletion, and
long-wavelength collective excitations are rederived in an 
efficient and economical 
manner. We also discuss nonuniversal effects.
These effects are higher-order corrections that depend on properties
of the interatomic potential other 
than the scattering length, such as the effective range.
We critically examine various approaches to the dilute Bose gas 
in equilibrium at finite 
temperature. 
These include the Bogoliubov approximation, the
Popov approximation, 
the Hartree-Fock-Bogoliubov approximation,
the $\Phi$-derivable approach, optimized perturbation theory, and
renormalization
group techniques. Finally, we review recent calculations of the 
critical temperature of the dilute Bose gas, which include
$1/N$-techniques, lattice simulations, self-consistent calculations,
and variational perturbation theory.

 \end{abstract}
\vskip .5cm

\pacs{03.75.Fi, 67.40.-w, 32.80.Pj}
\maketitle

\normalsize

\tableofcontents
\section{Introduction}
The remarkable realization of Bose-Einstein condensation 
(BEC) of trapped alkali atoms~\cite{bec1,bec2,bec3} 
has created an enormous interest in the properties
of the weakly interacting Bose gas.
Although the experiments are carried out in 
magnetic and optical
harmonic traps, the homogeneous
Bose gas has also received renewed interest~\cite{grif}.
The homogeneous Bose gas is interesting in its own right and it may prove
useful to go back to this somewhat simpler system to gain insight
that carries over to the trapped case.

Bose-Einstein condensation has a very long history dating back
to the early days of quantum mechanics and the
papers by Bose and Einstein~\cite{b,e}. The canonical example of
a system that exhibits BEC is liquid $^4$He. 
At very low athmospheric pressure,
$^4$He becomes superfluid below a temperature
of 2.17K, which is called the $\lambda$-point.
The fluid then consists of two components, namely a normal
and a superfluid component. The superfluid component has zero viscosity
and it has the remarkable property that it can flow through
narrow tubes without friction.
The phase transition from a normal fluid to a superfluid
is in modern terminology
described as spontaneous breaking of the global $U(1)$-symmetry
and the occurence of a condensate of atoms residing in the
zero-momentum state. 
This description also explains the experimental fact that the 
spectrum is gapless and linear in the long-wavelength limit;
the existence of a gapless and linear dispersion relation
follows from the 
Goldstone theorem~\cite{gold}.
It turns out that $^4$He is 
strongly interacting and 
this reduces the density of the condensate
to a rather small fraction of the total number density.
At very low temperature and pressure, the condensate density is  
approximately 10\% of the total number density. 
Thus the condensate density and 
superfluid density are very different at low temperature.
Liquid $^4$He consists of extended objects with very complicated
nonlocal interactions. Moreover, the fact that it is 
strongly interacting
makes it impossible
to apply perturbative methods. This has led to the search for weakly
interacting Bose gases. The trapped alkali gases are such systems
and the advantage of them over liquid $^4$He is that
they behave like systems of point particles with 
simple local interactions.
These gases 
have therefore become a very active field of research in the past decade.
Interested readers may consult the books by Pethick and Smith~\cite{pet}, 
and by Pitaevskii and Stringari~\cite{pitab}, as well as
the review papers by Dalfovo {\it et al}.~\cite{string} 
and by Leggett~\cite{leggett}.

The concept of Bose-Einstein condensation has been applied in many 
areas of physics other than $^4$He and thus a thorough 
understanding of it is important.
For instance, many properties of a superconductor
can be understood in terms of a condensate of pairs of electrons
with opposite spin and momenta.
Similarly, many properties of the QCD vacuum can be understood on the
basis of a condensate of quark-antiquark pairs with zero total 
momentum. This condensate is called the chiral condensate 
and in massless QCD, it breaks the chiral symmetry 
of the QCD Lagrangian.
The pions are interpreted as the corresponding Goldstone particles.

The homogeneous Bose gas at zero temperature has been studied extensively
for over 50 years starting with the classic paper by Bogoliubov~\cite{bogo}.
At zero temperature, the quantum loop expansion essentially is an expansion
in the gas parameter
$\sqrt{na^3}$, where $n$ is the number density and $a$ is the 
(positive)
$s$-wave
scattering length. The leading quantum corrections to the
chemical potential, energy density and speed of sound, were calculated
many years ago by Lee and Yang
using the pseudo-potential method~\cite{leeyang}.
Part of the second quantum correction to the energy density
was obtained by Wu~\cite{wu}, by 
Hugenholz and Pines~\cite{hug2}, and by Sawada~\cite{saw}.
Only recently a complete two-loop result has been 
obtained by Braaten and Nieto~\cite{eric}.
The result depends, in addition to the scattering length $a$, also
on an energy-independent term in
the scattering amplitude for $3\rightarrow 3$ scattering.
The fact that physical quantities depend on properties other than
the $s$-wave scattering length was already pointed out by
Hugenholz and Pines~\cite{hug2}.
These effects are called nonuniversal effects.
In the past decades effective field theory 
has been established as an ideal tool for systematically calculating
low-energy observables of a physical system~\cite{eft,lep2,kap}.
Effective field theory methods have also proven very useful
for calculating
higher order corrections in powers of $\sqrt{na^3}$ as well as 
nonuniversal 
effects in the weakly interacting Bose gas~\cite{e2}.

Finite temperature corrections to the pressure of a dilute Bose gas
were first calculated by Lee and Yang~\cite{leeyang2}.
Performing an expansion of the pressure
about zero temperature, they showed that
the leading term goes as $T^4$. This shows that the thermodynamics at low
temperature is completely determined by the linear part of the spectrum.
Similarly, the low-temperature expansion of the number density was calculated
by Glassgold~\cite{glass}, and shows a $T^2$ behavior.
These calculations were all based on the Bogoliubov approximation and are 
therefore valid only at low temperature, where the depletion of the
condensate is small. The first approach that takes into account the excited
states and thus can be used in the whole temperature range up to $T_c$
is the Popov theory~\cite{grif,popov,popov2}. Very recently, 
an improved many-body
$T$-matrix approximation
was developed~\cite{jhenk,all} which does not suffer from the
infrared divergences in lower dimensions that plague
the Bogoliubov and Popov
approximations.
While the approach can be applied in any dimension, its main application
is in one and two dimensions since the predictions in three dimensions
are very similar to those of established approaches. For instance, it predicts
the same zero-temperature depletion of the condensate as the Bogoliubov
approximation and the same critical temperature as the Popov approximation.

The latest extensive review on the homogeneous Bose gas was written by
Shi and Griffin~\cite{grif} 
five years ago. In the mean time, significant 
progress has been made, and a new review paper
summarizing recent advances is appropriate. 
There will of course 
be an overlap between the review by Shi and Griffin and the present paper, 
but some new material is covered. This includes nonuniversal 
effects~\cite{e2}, renormalization group calculations~\cite{henk,jmike,alber},
some of the variational approaches~\cite{lundh}
and calculations of the critical temperature 
$T_c$~\cite{at,arnold1,baym2,baym3,svis,grut,kraut,arnlat,boris,boris2,tceric,ram1,ram2,kleinert,tcdavis,ledow}.
Traditionally, the theory of dilute Bose gases is presented using
creation and annihilation operators as well as normal and anomalous 
self-energies. The present review is different in this respect since we 
are using modern functional methods and only occasionally we will 
present the material in terms of normal and anomalous averages.
It also implies that we are working with Lagrangians and actions, rather
than with Hamiltonians. We are largely going to base
our discussion on effective field theory methods~\cite{eft,lep2,kap}.
The virtues of effective field theory are that it allows one to
systematically calculate physical quantities efficiently by taking advantage of
the separation of length scales in a particular problem.
We shall return to these issues in some detail later.

Much of the research on the theoretical side
on trapped Bose gases in recent years has been
on dynamical issues, e.g. condensate formation, damping of
collective excitations, and collapse of the condensate.
There is large body
of literature on the nonequilibrium dynamics of the dilute Bose 
gas, 
but these topics deserve a separate review. 
Similarly, following their experimental realization,
there has also been significant 
progress in the understanding of low-dimensional
Bose gases~\cite{jhenk,all,2d}.
In order to limit the material,
we will focus entirely on the three-dimensional Bose gas 
in this review. For the same
reason, we also restrict ourselves to the spinless Bose gas.

The paper is organized as follows. In Sec.~II, we briefly
review the ideal Bose
gas at finite temperature. In Sec.~III, we discuss the weakly interacting
Bose gas at zero temperature. A perturbative framework using effective
field theory methods is set up. This framework is used to rederive
the leading quantum corrections to various quantities and discuss
nonuniversal effects.
In Sec.~IV, the weakly interacting Bose gas at finite temperature
is reviewed. We discuss 
the Hartree-Fock-Bogoliubov approximation,
the Bogoliubov approximation, and the
Popov approximation 
Then we consider the $\Phi$-derivable approach and optimized 
perturbation theory, which are
variational approaches.
Finally, renormalization group techniques are reviewed.
In Sec.~V, we discuss
the calculation of the critical temperature 
$T_c$ using a variety of different techniques.
In Sec.~VI, we summarize and conclude.
Calculational details as well as notation and conventions
are included in an appendix.

\section{The Ideal Bose Gas}
In this section, we review the ideal Bose gas at finite temperature.
Although this is standard textbook material, we include a discussion
to make the paper self-contained. 
In the remainder of the paper, we set $\hbar=2m=k_B=1$.
Factors of $\hbar$, $2m$, and $k_B$
can be reinserted using dimensional analysis.

Consider $N$ bosons at temperature $T$
in a box of volume $V$, so that the number density is
$n=N/V$. We impose periodic boundary conditions. We are always 
working in the thermodynamic limit, meaning $N,V\rightarrow\infty$
in such a manner that $n$ is fixed.

At high temperature, where the thermal wavelength 
$\lambda_T=2\sqrt{\pi/T}$ is much shorter than the interparticle
spacing, the atoms behave classically and their statistics is not important.
As the temperature is lowered, the atoms can be viewed as little
wavepackets with extent of the order $\lambda_T$.
Bose-Einstein condensation takes place when the thermal wavelength
of a particle is on the order of the interparticle spacing $n^{-1/3}$
and the wavefunctions of the bosons start to overlap.
The particles then accumulate in the 
zero-momentum state. At $T=0$, all the particles reside in this state.
Since fermions behave very differently at low temperatures due to the
Pauli exclusion principle, BEC is truly a quantum phenomenon.

We can estimate $T_c^0$ by equating the thermal wavelength $\lambda_T$ with 
the average distance between the bosons $n^{-1/3}$:
\bqa
2\sqrt{\pi/T_c^0}
&\sim&n^{-1/3}\;.
\eqa
The estimate for the critical temperature $T_c^0$ then becomes
\bqa
T_c^0&\sim
4\pi n^{2/3}
\label{tc0es}
\;.
\eqa
Having estimated $T_c$, we next discuss the thermodynamics
in more detail.
In the functional approach to the imaginary-time formalism, the grand-canonical
partition function ${\cal Z}$ is given by a path integral~\cite{nege}:
\bqa
{\cal Z}&=&\int{\cal D}\psi^*{\cal D}\psi\;e^{-S[\psi^*,\psi]}\;,
\label{path0}
\eqa
where the action $S$ is given by
\bqa\nonumber
S[\psi^*,\psi]&=&\int_0^{\beta}d\tau\int d^3x\;\psi^*({\bf x},\tau)
\bigg[{\partial\over\partial\tau}
-\nabla^2
\\ &&
\hspace{4.2cm}-\mu\bigg]
\psi({\bf x},\tau)\;.
\label{fac}
\eqa
Here, $\beta=1/T$ and $\mu$ is the chemical potential.
The complex fields $\psi^*$ and $\psi$ satisfy the standard bosonic
periodicity condition that $\psi({\bf x,\tau})$ and $\psi^*({\bf x,\tau})$ 
are periodic in $\tau$ with period $\beta$. 
We next write the complex field in terms of two real 
fields\footnote{In traditional approaches one uses the single complex
field $\psi$ instead of two real fields. This is merely a matter of taste.
See also the comments at the end of page 10.} 
\bqa
\psi={1\over\sqrt{2}}\left(\psi_1+i\psi_2\right)\;.
\label{12}
\eqa
After inserting~(\ref{12}) into the action~(\ref{fac}), 
the corresponding Green's function or propagator is found to be
\bqa
\label{propf}
D_0(\omega_n,p)&=&\frac{1}{\omega^2_n
+(p^2-\mu)^2}
\left(\begin{array}{cc}
p^2-\mu&\omega_n \\
-\omega_n&p^2-\mu
\end{array}\right)\;.
\eqa
where $\omega_n=2\pi nT$ is the $n$th Matsubara frequency.
The path integral~(\ref{path0}) is Gaussian and can therefore be evaluated
exactly. The result is
\bqa
{\cal Z}&=&e^{-\int_0^{\beta}d\tau\int d^3x\det D^{-1}_0(\omega_n,p)}\;,
\eqa
where $D^{-1}_0(\omega_n,p)$ is the inverse of the propagator~(\ref{propf}) 
and $\det$ denotes a determinant in the functional sense.
The free energy density is given by
\bqa
{\cal F}&=&-{1\over V\beta}\log{\cal Z}\\
&=&
{\rm Tr}\log D_0^{-1}(\omega_n,p)\;,
\eqa
where have used that $\log\det A={\rm Tr}\log A$ for any matrix $A$
where ${\rm Tr}$
denotes the trace. 
Inverting, Eq.~(\ref{propf})
and taking the trace,
we obtain
\bqa
{\cal F}&=&{1\over2}\sumint_P\log\left[\omega_n^2
+(p^2-\mu)^2\right]\;,
\eqa
where the sum-integral means
\bqa
  \hbox{$\sum$}\!\!\!\!\!\!\int_{P}& \;\equiv\; &
M^{2\epsilon}
  T\sum_{\omega_n=2n\pi T}\:\int {d^dp \over (2 \pi)^{d}}\;.
\eqa
The sum-integral involves a summation over Matsubara frequencies and
a regularized integral over $d=3-2\epsilon$ dimensions. 
$M$ is a renormalization scale that ensures that the integral has
the canonical dimensions also for $d\neq3$. 
In the following, we absorb the factor $M^{2\epsilon}$ 
in the measure and so it will not appear explicitly.
We discuss
the details further in the appendix.
After summing over Matsubara frequencies, we obtain
\bqa\nonumber
{\cal F}&=&
\int{d^dp\over(2\pi)^d}\left\{
{1\over2}\left(p^2-\mu\right)+T\log\left[1-e^{-\beta(p^2-\mu)}\right]
\right\}\;.
\\
&&
\eqa
The first term inside the brackets is an infinite constant that is independent
of temperature. It represents the zero-point fluctuations and can be removed
by a vacuum energy counterterm 
$\Delta{\cal E}$. 
The second term is the standard
finite temperature free energy of an ideal gas of nonrelativistic bosons.

In the operator approach, the starting point would be the grand canonical
Hamiltonian which
corresponds to the action~(\ref{fac}):
\bqa
{\cal H}&=&{1\over2}\int{d^3p\over(2\pi)^3}\left(p^2-\mu\right)
\;a_{\bf p}^{\dagger}a_{\bf p}
\;.
\label{fham}
\eqa
Using the Hamiltonian~(\ref{fham}) to calculate physical quantities, one does
not encounter zero-point fluctuation terms since it has been normal ordered;
The operator $a_{\bf p}$ annihilates the vacuum.
If it had not been normal ordered, one would find the same divergent terms
as with the path-integral approach. This is the usual ambiguity of the
quantization procedure in going from classical field theory to quantum field
theory.

We have replaced momentum sums by integrals over $p$. 
Due to the measure $d^dp$, the 
integrand always vanishes at $p=0$ and the 
contribution from the ground state
is not accounted 
for. 
If we denote the condensate
density of particles in the lowest energy state by $n_0$
and the number density of particles in the excited states by $n_{\rm ex}$, 
we have $n=n_0+n_{\rm ex}$. 
The density of excited particles
is then given by minus the derivative of ${\cal F}$
with respect to the chemical potential:
\bqa
n_{\rm ex}&=&\int{d^dp\over(2\pi)^d}n(p^2-\mu)\;,
\label{nf}
\eqa
where 
\bqa
n(\omega)&=&{1\over e^{\beta\omega}-1}\;,
\eqa
is the Bose-Einstein distribution function. 
The expression~(\ref{nf}) makes sense only for 
$\mu\leq0$. If $\mu>p^2$ for some ${p}^*$, 
the occupation number of the states with ${p}<{p}^*$ would become
negative. Clearly, this is unphysical. 
Below the transition temperature, the number of particles in the
excited states are given by
the integral~(\ref{nf}) with $\mu=0$. One finds
\bqa
n_{\rm ex}&=&
{\zeta\mbox{$\left({3\over2}\right)$}\over(4\pi)^{3/2}}
T^{3/2}\;,
\label{nex}
\eqa
where $\zeta(x)$ is the Riemann zeta function with argument $x$.
The critical temperature is the temperature at which 
all the particles can be accomodated in the excited states, that is
$n=n_{\rm ex}$.
This yields
\bqa
T_c^0&=&4\pi{\left[{n\over\zeta\left({3\over2}\right)}\right]^{2/3}}\;.
\label{tc0}
\eqa
The quantity $n\lambda^3_T$ is called the {\it degeneracy parameter}.
For an ideal Bose gas, the critical number
density $n_c$ satisfies
$n_c\lambda_T^3=\zeta\left({3\over2}\right)$. 
We see that the estimate~(\ref{tc0es}) is correct within a factor of 
$\left[\zeta\left({3\over2}\right)\right]^{-2/3}
\approx0.527$ of the exact result~(\ref{tc0}).
Using Eqs.~(\ref{nex}) and~(\ref{tc0}), the
condensate density $n_0$ as a function of the temperature $T$ 
can be written as 
\bqa
n_0&=&n\left[1-\left({T\over T_c}\right)^{3/2}\right]\;.
\label{3/2}
\eqa
The exponent $3/2$ in Eq.~(\ref{3/2}) is determined solely
by the density of states. For an ideal Bose gas in a three-dimensional
isotropic harmonic trap, the exponent is 3~\cite{string}.

\section{Weakly Interacting Bose Gas at Zero Temperature}
\label{t0}
In this section, we discuss in some detail the weakly interacting Bose gas
at $T=0$. We begin with a description using effective field theory 
and formulate a perturbative framework that can be used
for practical calculations.
We then calculate the leading corrections in the 
low-density expansion to the energy density, depletion, and
long-wavelength excitations. Finally, we 
discuss nonuniversal effects.

\subsection{Effective Field Theory}
\label{eftt}
Effective field theory (EFT) is a general approach that can be used to analyze
the low-energy behavior of a physical system in a systematic 
way~\cite{eft,lep2,kap}. 
EFT takes advantage of the separation of scales in a system to make
model-independent predictions at low energy. The effective Lagrangian
${\cal L}_{\rm eff}$ that describes the low-energy physics is written
in terms of only the long-wavelength degrees of freedom and the
operators that appear are determined by these degrees of freedom and the
symmetries present at low energy.  
The effective Lagrangian generally
includes an infinite tower of nonrenormalizable interactions, but they
can be organized according to their importance at low energy; to a certain
order in a low-energy expansion, only a finite number of operators contribute
to a physical quantity and one can carry out renormalization in the standard
way order by order in this expansion.~\footnote{\label{elg}
The counterterms that
are used to cancel the divergences in the calculations of
one physical quantity are the same as those required in the
calculations of another.
Thus, having determined the counterterms once and for all, the
effective theory can be used to make predictions about other physical
quantities.}
Since the coefficients of these operators
encode the short-wavelength physics, we do not need to make any detailed
assumptions about the high-energy dynamics to make 
predictions at low energy.

In some cases, one can determine the coefficients
of the low-energy theory as functions of the coupling constants in the
underlying theory 
by a perturbative matching procedure. One calculates physical
observables at low energies perturbatively
and demand they be the same in the 
full and in the effective theory. The coefficients of the effective
theory then encode the short-distance physics. If one cannot determine
these short-distance coefficients by matching, 
they can be
taken as phenomenological parameters that are determined by experiment.

A classic example of an effective field theory is chiral perturbation theory
which is a low-energy field theory for pions~\cite{chiral}. Pions are
interacting particles whose fundamental description is provided
by QCD. However, QCD is strongly interacting and confining at low energies
and so perturbative calculations using the QCD Lagrangian are hopeless.
So instead of using the quark and gluon degrees of freedom in QCD, one
writes down the most general Lagrangian for the pions, which are the relevant
low-energy degrees of freedom. The terms that appear in the chiral Lagrangian
are determined by the global symmetries of QCD. The coefficients of the
chiral Lagrangian cannot be determined as functions of the couplings
and masses in QCD using perturbative methods.
They can in principle be determined using a nonperturbative method
such as lattice gauge theory, but in practice they are normally
determined by experiment.

Nonrelativistic 
QED~\cite{nrqed} (NRQED) 
is an example of an effective field theory whose coefficients
are tuned so they reproduce a set of low-energy scattering amplitudes
of full QED. NRQED is
tailored to perform low-energy (bound state) calculations, where
one takes advantage of the nonrelativistic nature of the
bound states by 
isolating the contributions from the relativistic
momentum scales. These are encoded in the coefficients of the
various local operators in the effective Lagrangian.
Traditional approaches involving the Bethe-Salpeter equation do not
take advantage of the separation of scales in bound state problems and are
therefore much more difficult to solve.

Landau's quasiparticle model for $^4$He can also be viewed as an effective
theory. In order to explain that the specific heat varies as $T^3$
for temperatures much smaller than $T_c$, he suggested that the
low-lying excitations are phonons with a linear dispersion relation.
More generally, he proposed a spectrum that is linear
for small wavevectors and has a local minimum around $p=p_0$.
This part of the spectrum behaves like
\bqa
\epsilon(p)&=&\Delta +{(p-p_0)^2\over2m_0}\;,
\eqa 
where $\Delta$, $p_0$, and $m_0$ are phenomenological parameters.
Excitations near $p_0$ are referred to as rotons.
Assuming that the elementary excitations are noninteracting, one can
use spectrum to calculate the specific heat.
Landau determined the parameters by 
fitting the calculated specific heat to experimental data~\cite{landau22}.

The weakly-interacting Bose gas is a system where effective field theory
methods can be applied successfully. 
The starting point is the action:
\bqa\nonumber
S[\psi^*,\psi]&=&\int dt\;
\Bigg\{\int d^dx\;
\psi^*({\bf x},t)
\left[
i{\partial\over\partial t}
+
\nabla^2+\mu
\right]
\\ &&\nonumber
\times\psi({\bf x},t)
-{1\over2}\int d^dx\int d^dx^{\prime}\;
\psi^*({\bf x},t)\psi^*({\bf x}^{\prime},t) 
\\ &&
\times
V_0({\bf x}-{\bf x}^{\prime})
\psi({\bf x},t)\psi({\bf x}^{\prime},t)+...
\Bigg\}\;.
\label{ac}
\eqa
Here, $\psi^*({\bf x},t)$ is a complex field operator that creates a 
boson at the
position ${\bf x}$ at time $t$, 
$\mu$ is the chemical potential, and $V_0({\bf x})$ 
is the two-body potential. 
The ellipses indicate terms that describe possible
interactions between
three or more bosons. 
The chemical potential
$\mu$ must be adjusted to get the correct number density $n$.

The interatomic potential can be divided into a central
part $V_0^c(x)$ and a remainder.
The central of the potential only depends on the separation $x$ of the atoms 
and their electronic spins. It conserves separately the total orbital
angular momentum and the total electronic spin of the atoms.
The noncentral part of the interaction conserves the total angular momentum,
but not separately the orbital 
angular momentum and the total electronic spin of the atoms.
An example of a term in the noncentral part of the interaction is the
magnetic dipole-dipole interaction.

\begin{figure}[htb]
\epsfysize=6.6cm
\epsffile{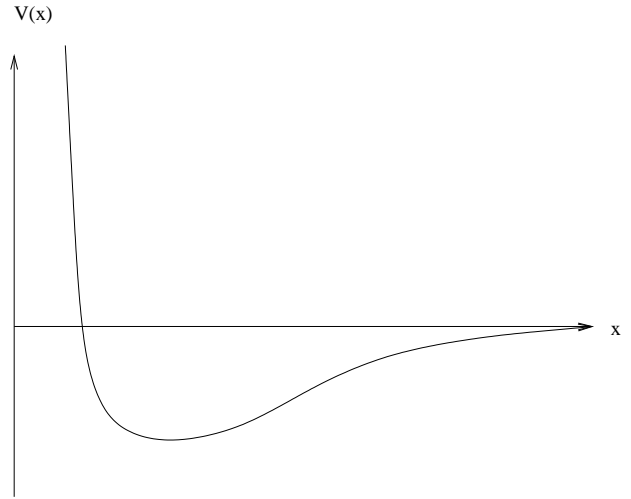}
\caption[a]{Typical behavior of a two-body potential $V_0^c(x)$.}
\label{pot1}
\end{figure}

The central part of the potential consists of a short-range
part with range $x_0$ and a long-range van der Waals tail. The latter goes as 
$1/x^6$ as $x\rightarrow\infty$. 
A typical two-body potential is
shown in Fig.~\ref{pot1}.
A model potential of this kind is
the sum of a hard-core potential with range $x_0$
and a van der Waals potential:
\bqa
V_0^c(x)&=&
\Bigg\{\begin{array}{cc}
+\infty\;,&x<x_0 \\
-{C_6\over x^6}\;,&x>x_0
\end{array}
\eqa
where $C_6$ and $x_0$ are constants. 
Another example is the hard-core square-well
potential:
\bqa
V_0^c(x)&=&
\Bigg\{\begin{array}{cc}
+\infty\;,&x<x_c \\
-V_0\;,&x_0<x<x_c\\
0\;,&x<x_c
\end{array}
\eqa
where $x_0$ and $x_c$ are constants. This potential sustains a number
of two-body bound states depending on the values of $x_0$ and $x_c$.
Many real potentials used in experiments sustain bound states and the
ground state of the system is no longer a homogeneoues gas, but rather
a state of clusters of atoms. However, if the scattering length is positive,
the homogeneous Bose gas can exist as a long-lived metastable state.

\begin{figure}[htb]
\epsfysize=6.6cm
\epsffile{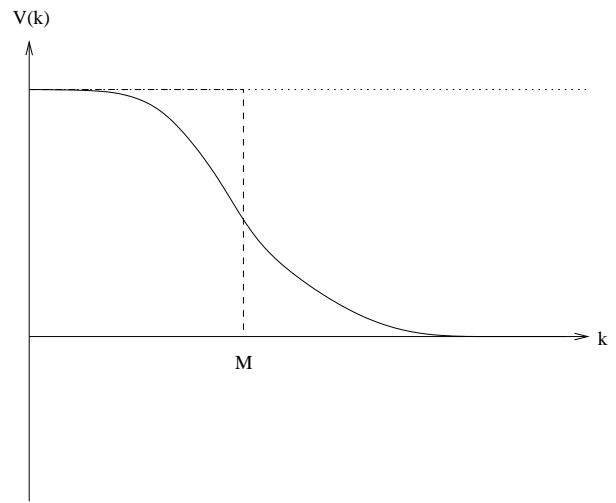}
\caption[a]{Solid line:
Typical behavior of the potential in momentum space $V({\bf k})$.
Dotted line: Fourier transform of a contact potential.}
\label{pot2}
\end{figure}

In Fig.~\ref{pot2}, we have shown the Fourier transform $V({\bf k})$ 
of a typical
short-range two-body potential with range $x_0$. Since a true interatomic
potential vanishes for large momenta $k$, one will never face 
ultraviolet divergences using it in actual (perturbative) calculations.

In this paper we restrict our calculations to the spinless Bose gas.
In experiments with trapped alkali atoms
(e.g. $^7$Li, $^{23}$Na $^{85}$Rb, $^{87}$Rb, and $^{133}$Cs), 
the situation is more complicated. In all these atoms, there is a single
$s$-electron outside closed shells. Consequently, these atoms have 
electron spin $S=1/2$. The total spin of a colliding pair of alkali atoms
is therefore either $S=0$ or $S=1$. The central part of the potential 
$V_0^c(x)$ depends on the total spin and one refers to these potentials
as the {\it singlet} and {\it triplet} potentials, respectively.
The singlet potential is generally much deeper and sustains many more 
bound states than does the triplet potential.
For example, the singlet potential of $^{87}$Rb is deeper than the
triplet potential by more than one order of magnitude. 
The scattering lengths are denoted by $a_s$ and $a_t$.
In the case of atomic hydrogen, they 
have been calculated from first
principles; $a_s=0.3a_0$ and $a_t=1.3a_0$, where $a_0$
is the Bohr radius. Similarly, the coefficient of the van der Waals
tail has also been determined; $C_6=6.499a_0$~\cite{atom2}. 

One can associate a natural length scale $l$ with any atomic potential 
$V_0^c(x)$. For a short-range potential, this length is the range $x_0$ itself.
For a long-range potential, it is a little more complicated. 
The length scale associated with the van der Waals tail is
$l_{\rm vdW}\sim C_6^{1/4}$. In this case, the length scale $l$ is
either the range or $l_{\rm vdW}$, whichever is larger.
For a generic potential, the low-energy observables such as the scattering
length and the effective range are of the order $l$.
There is nothing that forbids the these quantities to be much larger
than $l$, but it is {\it unnatural} and typically it requires fine-tuning
of one or more parameters in the potential.
In the case of the alkali atoms, $l_{\rm vdW}$ is much larger than the
range of the short-range part of the potential and is thus the
natural length scale for these atoms. The spin-singlet scattering length
for $^{85}$Rb is $a_s=+2800a_0$ and is more than one order of magnitude
larger than $l=l_{\rm vdW}=164a_0$. This can be viewed as a fine-tuning of 
the mass of the atom. This can be seen from the fact that the mass of
$^{87}$Rb is only 2.3$\%$ larger and the 
spin-singlet scattering length has the more natural value $a_s=+90.4a_0$.
One can also obtain unnaturally large scattering lengths by tuning an
external parameter in experiments. One way of doing this, is to 
tune an external magnetic field to a Feshbach resonance~\cite{fesh}.
This is currently receiving a lot of attention both 
theoretically\cite{stw,ties} and experimentally~\cite{ino}.

In the remainder of this work, we do not specify the
magntiude of the scattering length. We only demand the {\it diluteness
condition} be satisfied, namely that
\bqa
na^3\ll1\;.
\eqa
We now return to the action~(\ref{ac}), which
 is invariant under a global phase transformation
\bqa
\psi({\bf x},t)\rightarrow e^{i\alpha}\psi({\bf x},t)\;.
\label{gpha}
\eqa
The global $U(1)$ symmetry reflects the conservation of atoms. It also
ensures that the number density $n$ and current 
density ${\bf j}$ satisfy the continuity equation
\bqa
\dot{n}+\nabla\cdot{\bf j}=0\;,
\eqa
where the dot denotes differentiation with respect to time.

The nonlocal evolution equation that follows from the action~(\ref{ac}),
cannot always be used for practical calculations, but can be replaced
by a local one.
Suppose we are interested in the properties of the system at  
momenta $k$ 
such that the de Broglie wavelength $1/k$ is 
much longer than the 
range of the interatomic potential $V_0({\bf x})$.
The interactions therefore appear pointlike
on the scale of the de Broglie wavelength,
and they can 
be mimicked by local interactions
The parameters of these local
interactions must be tuned so that they reproduce low-energy observables
to sufficient accuracy.
However, if the potential is long range,
the scattering amplitude depends in a nonanalytic way on the wave vector
${\bf k}$ characterizing the incoming atoms in the CM frame. 
Such behavior cannot be reproduced by local operators.
If $V_0^c(x)$ falls of like $1/x^6$, this nonanalytic behavior 
enters first at order $k^4$~\cite{e+h}. 

The effective Lagrangian for the bosons can be constructed using the 
methods of effective field theory. Once the symmetries have been 
identified, one writes down the most general local
effective Lagrangian consistent
with these symmetries. 
At zero temperature, the symmetries are Galilean invariance, time-reversal
symmetry, and the global phase symmetry~(\ref{gpha}).
These symmetries severely restrict the possible terms in the
effective action. One finds~\cite{eric}:
\bqa\nonumber
&&S[\psi^*,\psi]=\int dt\;
\int d^dx\;\Bigg\{
\psi^*
\left[
i
{\partial\over\partial t}
+\nabla^2+\mu
\right]\psi
\\ && \nonumber
-{1\over2}
g\left(\psi^*\psi\right)^2
-{1\over2}h\left[\nabla\left(\psi^*\psi\right)\right]^2
-{g_3\over36}\left(\psi^*\psi\right)^3+...\Bigg\}
\;,
\\ &&
\label{act}
\eqa
where $g$, $h$, and $g_3$ are coupling constants that can be determined
by a matching procedure. 
The dots denote operators that are higher order in the field $\psi$ or its
derivative $\nabla\psi$ and respect the symmetries.
One demands that the effective field theory 
represented by the action~(\ref{act}) reproduces a set of low-energy
observables to some desired accuracy. 
An example is the coefficients of the expansions in $\sqrt{na^3}$ of 
the ground state energy density. Another example coefficients in the 
low-energy expansions for the scattering amplitudes in the $n$-body sector.
When the coefficients in the action~(\ref{act}) have been determined to some
accuracy in a low-energy expansion, effective field theory guarantees that
all other observables can be determined with the same accuracy.

The quantum field theory defined by the action~(\ref{act}) has ultraviolet
divergences that must be removed by renormalization of the
parameters $\mu, g, h, g_3...$ .
They arise because we are treating the interactions
between the atoms as pointlike down to arbitrarily short distances.
For instance the operator $g(\psi^{\dagger}\psi)^2$ can be thought of as 
a contact potential with strength $g$; 
$V_0({{\bf x}-{\bf x}^{\prime}})=g\delta({{\bf x}-{\bf x}^{\prime}})$.
The Fourier transform is then 
a constant in momentum space; $V_0({\bf k})=g$. 
This is illustrated is Fig.~\ref{pot2},
where the dotted line shows $V({\bf k})$. Thus $V({\bf k})$
does not vanish for large momenta and this is the reason why one encounters
ultraviolet divergences in the calculation of Feynman diagrams.
In order to make the theory well defined, we must introduce an ultraviolet
cutoff.
This is indicated in Fig.~\ref{pot2}, where we exclude wave numbers
$k>M$ in momentum integrals.

If we use a simple momentum cutoff $M$ to cut off the ultraviolet
divergences, there will be terms that are proportional to 
$M^n$, where $n$ is a positive integer.
There are also terms that are proportional to $\log(M)$.
The coefficients of the power divergences depend on the method
we use to regulate the integrals, while the coefficients
of $\log(M)$ do not. Thus the power divergence are artifacts of the
regulator, while the logarithmic divergences represent real physics.
In this paper, we will be using dimensional regularization~\cite{jerry} 
to regulate
infrared as well as ultraviolet divergences in the loop integrals. 
In dimensional regularization, one
calculates the loop integrals in $d=3-2\epsilon$ dimensions for values of
$\epsilon$ for which the integrals converge. One finally analytically continues
back to $d=3$ dimensions. In dimensional regularization, an arbitrary
momentum scale $M$ is introduced to ensure that loop integrals have their
canonical dimensions also away from three dimensions. This scale can be 
identified
with the simple momentum cutoff mentioned above. Advantages of dimensional 
regularization
are that it respects symmetries such as rotational symmetry and gauge 
invariance.
Dimensional regularization sets power divergences to zero 
and logarithmic divergences show up as poles in $\epsilon$.
With dimensional regularization, the only ultraviolet divergences
that require explicit renormalization are therefore logarithmic
divergences. This simplifies calculations significantly as
we shall see. In fact all the divergences encountered
in the one-loop calculations that we present here, can be
removed by the renormalization of $\mu$, $g$, and the vacuum
energy ${\cal E}$
and in three dimensions, these are power divergences.
Thus no explicit renormalization is required.

We now return to the determination of the parameters $g, h, g_3...\;.$ 
We follow~\cite{e2} and determine them
by demanding that the
effective field theory~(\ref{act}) 
reproduces the low-momentum expansions for the
scattering amplitudes in the vacuum 
for $2\rightarrow2$ scattering, $3\rightarrow3$ scattering, 
etc. to some desired accuracy. 

\begin{figure}[htb]
\epsfysize=1.4cm
\vspace{0.1cm}
\epsffile{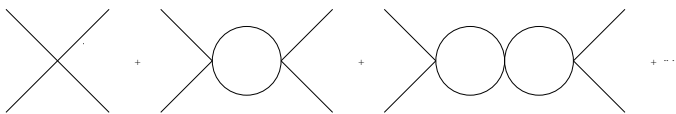}
\caption[a]{Diagrams contributing to the $2\rightarrow2$
scattering amplitude.}
\label{loopfig}
\end{figure}

To calculate the coupling constant $g$,
consider the scattering of two atoms in the vacuum
with initial wave numbers ${\bf k}_1$
and ${\bf k}_2$, and final wave numbers ${\bf k}_1^{\prime}$
and ${\bf k}_2^{\prime}$. The probability amplitude for the
$2\rightarrow2$ scattering is given by the T-matrix element 
${\cal T}$. The tree-level contribution to ${\cal T}$ comes from
the leftmost diagram in Fig.~\ref{loopfig} and is given by
\bqa
{\cal T}_0&=&-2g\;,
\eqa
where the subscript indicates the number of loops. The quantum corrections
to the tree-level result come from the loop diagrams in Fig.~\ref{loopfig}.
The leading quantum correction comes from the one-loop diagram and reads
\bqa\nonumber
{\cal T}_1(q)&=&-2ig^2\int{d\omega\over2\pi}\int{d^dk\over(2\pi)^d}
{1\over\omega-k^2+i\varepsilon} 
\\ &&
\times
{1\over(\omega-k_1^2-k^2_2)+({\bf k}-{\bf k}_1-{\bf k}_2)^2+i\varepsilon} 
\;,
\eqa
where $q={1\over2}|{\bf k}_1-{\bf k}_2|$, and we have used that the
free propagator $\Delta_0(\omega,{\bf k})$ 
in vacuum corresponding to the action~(\ref{act}) is
\bqa
\Delta_0(\omega,{\bf k})&=&{i\over \omega-k^2+i\epsilon}\;.
\eqa
The integral over $\omega$ is performed using contour integration.
After changing variables 
${\bf k}\rightarrow {\bf k}+|{\bf k}_1+{\bf k}_2|/2$, we obtain
\bqa
{\cal T}_1(q)&=&g^2\int{d^dk\over(2\pi)^d}
{1\over k^2-q^2-i\varepsilon}\;.
\label{cvar}
\eqa
Note that the integral over $k$ is linearly divergent in the ultraviolet. 
This divergence is set to zero in dimensional regularization.
Using dimensional regularization, 
the result of the integration of $k$ can be written as
\bqa
{\cal T}_1(q)&=&g^2M^{2\epsilon}{\Gamma(1-{d\over2})\over(4\pi)^{d/2}}\left[
-q^2-i\varepsilon
\right]^{d-2\over2}
\label{m11}
\;,
\eqa
where $\Gamma(x)$ is the gamma function.
The limit $d\rightarrow3$ is regular and  Eq.~(\ref{m11}) reduces to
\bqa
{\cal T}_1(q)&=&i{g^2q\over16\pi}\;.
\eqa
This expression is simply Fermi's golden rule.
The quantum corrections from higher orders are given by the
diagrams like the two-loop graph in Fig.~\ref{loopfig}.
They form a geometric series that can be summed up exactly and the
exact $2\rightarrow2$ amplitude becomes
\bqa
{\cal T}(q)&=&-{2g^2\over g+{\cal T}_1(q)}\;.
\label{match}
\eqa
The scattering amplitude for $2\rightarrow2$ scattering in the
underlying theory described by the action~(\ref{ac}) 
can be calculated by solving the two-body scattering problem in the
potential $V_0({\bf x}-{\bf x}^{\prime})$.
The contribution from $s$-wave scattering is~\cite{landau}
\bqa
{\cal T}&=&{16\pi\over q}e^{i\delta_0(q)}\sin[\delta_0(q)]\;,
\label{m1}
\eqa
where $\delta_0(q)$ is the $s$-wave phase shift.
We next write the low-momentum expansion as follows
\bqa
q\cot[\delta_0(q)]&=&
\left[-{1\over a}+{1\over2}r_sq^2+..\right]\;.
\label{lmexp}
\eqa
This expansion defines
the scattering length $a$ and the effective range $r_s$.
Using the identity $e^{ix}\sin(x)=1/(\cot(x)-i)$, 
we can expand the $T$-matrix in powers of momentum $q$.
Matching the expressions~(\ref{match}) and~(\ref{m1}) 
through first order in the external momentum $q$ using~(\ref{lmexp}),
we obtain 
\bqa
g=8\pi a\;.
\label{ga}
\eqa
The parameter $h$ can be determined by going to the next order in the 
low-momentum expansion. We will do this in subsec.~\ref{noneff}.
Similarly, the parameter $g_3$ can be determined by solving the 
three-body scattering problem in the potential 
$V_0({\bf x}-{\bf x}^{\prime})$ and demand that the $3\rightarrow3$
scattering amplitudes in the full and in the effective theory
be the same at low momentum.

Traditionally, the starting point has been the interatomic potential
$V_{0}({\bf x}-{\bf x}^{\prime})$. At low densities, it can be shown
that the ladder diagrams are of equal importance and must be summed. 
The summation of these diagrams can be expressed in terms of an effective
interaction $\Gamma$ 
that  satisfies an integral equation which also involves
$V_{0}({\bf x}-{\bf x}^{\prime})$~\cite{beli,fetter}. 
The interatomic potential appearing in $\Gamma$ can be eliminated
in favor of the scattering amplitude for two-particle scattering.
In the low-momentum limit, the effective interaction reduces to the
$8\pi a$. The mean-field self-energies and therefore the first order
propagator and spectrum
can be expressed in terms of the effective interaction.
Thus the spectrum reduces in the low-momentum limit to the Bogoliubov
spectrum.

\subsection{Perturbative Framework}
\label{pf}
We next discuss the perturbative framework set up by Braaten and 
Nieto~\cite{eric}
that can be used to systematically
calculate the low-energy properties of a weakly interacting Bose gas.

We first parameterize the quantum field $\psi$
in terms of a time-independent
condensate $v$ and a quantum fluctuating field $\tilde{\psi}$:
\bqa
\psi=v+\tilde{\psi}\;.
\label{shift}
\eqa
The fluctuating field $\tilde{\psi}$ can 
be conveniently written in terms of two real fields:
\bqa
\tilde{\psi}={1\over\sqrt{2}}\left(\psi_1+i\psi_2\right)\;.
\label{split}
\eqa
Substituting Eqs.~(\ref{shift}) and~(\ref{split}) 
into Eq.~(\ref{act}), the action can 
be decomposed into three terms
\bqa
\label{terms}
S[v,\psi_1,\psi_2]=S[v]+S_{\rm free}[v,\psi_1,\psi_2]
+S_{\rm int}[v,\psi_1,\psi_2]\;,
\eqa
where we have indicated that the action depends on $v$ as well, and
switched to the variables $\psi_1$ and $\psi_2$ instead of
$\psi^*$ and $\psi$.
$S[v]$ is the classical action
\bqa
S[v]=\int dt\int d^dx
\left[\mu v^2-{1\over2}gv^4
\right]\;,
\label{cact}
\eqa
while the free part of the action is
\bqa\nonumber
S_{\rm free}[v,\psi_1,\psi_2]&=&\int dt\int d^dx
\left[
{1\over2}\left(\dot{\psi}_1\psi_2-\psi_1\dot{\psi_2}\right)
\right.\\ 
&&\left.
\hspace{-2cm}+{1\over2}
\psi_1\left(\nabla^2+X\right)\psi_1
+{1\over2}
\psi_2\left(
\nabla^2+Y\right)\psi_2
\right]\;.
\label{free}
\eqa
where
\bqa
X&=&\mu-3gv^2\;,
\label{x}
\\
Y&=&\mu-gv^2\;.
\label{y}
\eqa
The terms $3gv^2$ and $gv^2$ in $X$ and $Y$ are often referred to as 
mean-field self-energies.
The interaction part of the action is
\bqa\nonumber
S_{\rm int}[v,\psi_1,\psi_2]&=&
\int dt\int d^dx\left[
\sqrt{2}J\psi_1
\right. \\ &&\left.
\hspace{-1cm}
+{1\over\sqrt{2}}Z\psi_1\left(\psi_1^2+\psi_2^2\right)
-{1\over8}g\left(\psi_1^2+\psi_2^2\right)^2
\right]\;,
\label{inter}
\eqa
The sources in Eq.~(\ref{inter}) are
\bqa
J&=&\left[\mu-gv^2\right]v
\label{t}
\;,\\
Z&=&-gv\;.
\eqa
The propagator that corresponds to the free action 
$S_{\rm free}[v,\psi_1,\psi_2]$
in Eq.~(\ref{free}) is
\bqa
\label{prop}
D(\omega,p)&=&\frac{i}{\omega^2
-\epsilon^2(p)+i\varepsilon}
\left(\begin{array}{cc}
p^2-Y&-i\omega \\
i\omega&p^2-X
\end{array}\right)\;.
\eqa
Here, $p=|{\bf p}|$, where ${\bf p}$ is the wavevector, 
$\omega$ 
is the frequency, and
$\epsilon(p)$ is the dispersion relation:
\bqa
\label{disp}
\epsilon(p)=\sqrt{(p^2-X)(p^2-Y)}\;.
\eqa
Note that one can diagonalize the matrix~(\ref{prop})
by a field redefinition, which is equivalent to a Bogoliubov transformation
in the operator approach. Such a field redefinition makes, however,
the interaction terms much more complicated and 
increases the number of diagrams that one needs to evaluate.
For practical purposes, we therefore stick to the above propagator.

The partition function ${\cal Z}$ can be expressed as a path integral
over the quantum fields $\psi_1$ and $\psi_2$~\cite{nege}:
\bqa
{\cal Z}=\int{\cal D}\psi_1{\cal D}\psi_2\,e^{iS[v,\psi_1,\psi_2]}\;.
\eqa
All the thermodynamic observables can be derived from 
the partition function
${\cal Z}$.
For instance, the free energy density ${\cal F}$ is given by
\bqa
{\cal F}(\mu)=i{\log{\cal Z}\over{\cal V}}\;,
\eqa
where ${\cal V}$ 
is the spacetime volume of the system. The pressure ${\cal P}(\mu)$
is
\bqa
{\cal P}(\mu)&=&-{\cal F}(\mu)\;.
\eqa
The number density $n$ is given by the expectation value 
$\langle\psi^*\psi\rangle$ in the ground state:
\bqa
n(\mu)&=&\int{\cal D}\psi_1{\cal D}\psi_2\,(\psi^*\psi)\;e^{iS[v,\psi_1,\psi_2]}\;.
\eqa
It can therefore be expressed
as 
\bqa
\label{rhodef}
n(\mu)=-{\partial {\cal F}(\mu)\over \partial \mu}\;.
\eqa
The energy density ${\cal E}$ is given by the Legendre transform of
the free energy density ${\cal F}$:
\bqa
{\cal E}(n)={\cal F}(\mu)+n\mu\;.
\label{rela}
\eqa

The free energy ${\cal F}$ is given by all 
{\it connected vacuum diagrams} which are
Feynman diagrams with no external legs.
The sum of the vacuum graphs is independent of the condensate $v$.
At this point it is convenient to
introduce the thermodynamic potential $\Omega(\mu,v)$.
The thermodynamic potential is given by all 
{\it one-particle irreducible vacuum diagrams} 
and can be expanded in number of loops:
\bqa
\Omega(\mu,v)&=&
\Omega_0(\mu,v)+\Omega_1(\mu,v)+\Omega_2(\mu,v)+...\;,
\label{loopp}
\eqa
where the subscript $n$ denotes the contribution from the $n$th order in the
loop expansion.  
If $\Omega$ is evaluated at a value of the condensate that satisfies the
condition
\bqa
\bar{v}&=&\langle\psi\rangle\;,
\label{condit}
\eqa
all {\it one-particle reducible diagrams} (those that are disconnected 
by cutting a single propagator line) vanish. 
Thus evaluating the thermodynamic potential at the value
of the condensate that satisfies~(\ref{condit}), one obtains the free
energy:
\bqa
{\cal F}(\mu)&=&
\Omega_0(\mu,\bar{v})+\Omega_1(\mu,\bar{v})+\Omega_2(\mu,\bar{v})+...\;.
\label{loop}
\eqa
Using Eq.~(\ref{split}), the 
condition~(\ref{condit}) reduces to 
$\langle\psi_2\rangle=0$ and $\langle\psi_1\rangle=0$.
The first condition can be automatically satisfied by a suitable
choice of the phase of $\psi$.  
The second condition is then equivalent to
\bqa
{\partial\Omega(\mu,v)\over\partial v}&=&0\;.
\label{tad1}
\eqa
The value of the condensate that satisfies~(\ref{tad1}) is denoted 
by $\bar{v}$.
The free energy can also be expanded in
powers of quantum corrections around the 
mean-field result ${\cal F}_0(\mu)$:
\bqa
{\cal F}(\mu)&=&{\cal F}_0(\mu)+{\cal F}_1(\mu)+{\cal F}_2(\mu)+...
\label{fh}
\eqa
The loop expansion~(\ref{loop}) of ${\cal F}(\mu)$
does not coincide with the expansion~(\ref{fh})
of ${\cal F}(\mu)$ in powers of quantum corrections because of its dependence 
on $\bar{v}$. 
To obtain the expansion of ${\cal F}$ in powers of
quantum corrections, we must
expand the condensate $\bar{v}$ in powers of
quantum corrections:
\bqa
\bar{v}=v_0+v_1+v_2+...\;,
\label{vexp}
\eqa
where $v_0$ is the classical minimum, which satisfies
\bqa
{\partial\Omega_0(\mu,v)\over\partial v}&=&0\;.
\eqa
By expanding Eq.~(\ref{tad1})
about $v_0$, one obtains the quantum corrections $v_1, v_2,...$
to the condensate. For instance,
the first quantum correction $v_1$ to the classical minimum $v_0$ is
\bqa
v_1=-{\partial\Omega_1(\mu,v)\over\partial v}\Bigg|_{v=v_0}\Bigg/
{\partial^2\Omega_0(\mu,v)\over\partial v^2}\Bigg|_{v=v_0}
\;.
\label{v11}
\eqa
The mean-field free energy density is
\bqa
{\cal F}_0(\mu)&=&\Omega_0(\mu,v_0)
\eqa
Inserting~(\ref{vexp}) into~(\ref{loop}) and expanding in powers
of $v_1, v_2...$, we obtain the quantum expansion of the free
energy density.
The first quantum correction to the free energy density is 
\bqa
{\cal F}_1(\mu)=\Omega_1(\mu,v_0)\;,
\eqa
and the second quantum correction to the free energy density
is
\bqa\nonumber
{\cal F}_2(\mu)&=&\Omega_2(\mu,v_0)+
v_1{\partial \Omega_1(\mu,v)\over\partial v}\Bigg|_{v=v_0}\\
&&+
{1\over2}v^2_1{\partial^2\Omega_0(\mu,v)\over\partial v^2}\Bigg|_{v=v_0}\;.
\eqa
Expressions for higher order corrections to the free energy can be 
derived in the same way.

The value of the condensate $v$ that minimizes the classical 
action~(\ref{cact}) is given by $v_0=\sqrt{\mu/g}$. 
The linear term
in Eq.~(\ref{inter}) then vanishes since $J=0$. 
At the minimum of the classical action,
both the propagator~(\ref{prop})
and the dispersion
relation~(\ref{disp}) simplify significantly since we also have $Y=0$.
At the minimum,
these equations reduce to
\bqa
D(\omega,p)&=&\frac{i}{\omega^2
-\epsilon^2(p)+i\varepsilon}
\left(\begin{array}{cc}
p^2&-i\omega \\
i\omega&\epsilon^2(p)/p^2
\end{array}\right)\;,
\label{prop2}
\\
\epsilon(p)&=&p\sqrt{p^2+2\mu}\;.
\label{disps}
\eqa
The spectrum~(\ref{disps}) was first derived by Bogoliubov in 1947~\cite{bogo}.
The dispersion relation is gapless and linear for small wave vectors.
This reflects the spontaneous breakdown of the $U(1)$-symmetry.
The dispersion relation changes from being linear to being quadratic
in the vicinity of $p=\sqrt{2\mu}$, which is called the
inverse coherence length. 
For very large wave vectors, the
dispersion relation is approximately $\epsilon(p)=p^2+\mu$, where
the second term represents the mean-field energy due to the
interaction with the condensed particles.

We next comment on how traditional approaches fit into the more
general perturbative framework  discussed in this section.
The starting point is the second quantized grand canonical Hamiltonian
that includes a two-body potential $V_0({\bf x})$:
\bqa\nonumber
{\cal H}&=&
\int d^dx\;
\psi^*({\bf x})
\left[
-\nabla^2-\mu
\right]\psi({\bf x})
\\ &&\hspace{-0.5cm}\nonumber
+{1\over2}\int d^dx\int d^dx^{\prime}\;
\psi^*({\bf x})\psi^*({\bf x}^{\prime}) 
V_0({\bf x}-{\bf x}^{\prime})
\psi({\bf x})\psi({\bf x}^{\prime})\;.
\\ &&
\label{ham1}
\eqa
This Hamiltonian is nonlocal and is often replaced by a local one. This is
done by approximating the true two-body potential by a local 
two-body interaction, whose strength $g$ is tuned to reproduce the 
scattering length of $V_{0}({\bf x}-{\bf x}^{\prime})$.
This yields
\bqa
{\cal H}&=&
\int d^dx\;\left\{
\psi^*
\left[
-\nabla^2-\mu
\right]\psi
+{1\over2}g\left(\psi^*\psi\right)^2
\right\}\;.
\label{ham2}
\eqa
The grand canonical Hamiltonian
is expressed in terms of creation and annihilation operators that satisfy
the standard equal-time commutation relations. Bogoliubov's idea was 
to treat the ${\bf p}=0$ momentum separately~\cite{bogo}.
Since this state is macroscopically occupied, the 
creation and annihilation operators for bosons with
${\bf p}=0$ commute to very good approximation.
Thus they can be treated classically and be replaced by a constant which
is the condensate density. This step is equivalent to splitting the
quantum field $\psi$ into a condensate $v$ and a fluctuating field 
$\tilde{\psi}$, Eq.~(\ref{shift}).

In the Bogoliubov approximation~\cite{bogo}, 
one makes a quadratic approximation to the Hamiltonian by neglecting
terms with three and four operators.
Since the resulting Hamiltonian contains products of two annihilation
and products of two creation operators, it must be
diagonalized by a canonical transformation.
The resulting quasi-particle spectrum is then given by Eq.~(\ref{disps}).

In the Beliaev approximation~\cite{beli}, one goes one step
further by calculating the leading quantum corrections to the
quasi-particle spectrum~(\ref{disps}).
This is done by including all one-loop diagrams, that is
all diagrams up to second order in the interaction, 
in the self-energies and calculating the poles of the propagator.
The 
correction to the Bogoliubov spectrum~(\ref{disps}) was
first calculated by
Beliaev~\cite{beli} and coincides with a leading-order calculation in the 
approach outlined here. We shall return to that calculation in 
Sec.~\ref{belsec}.

The perturbative framework has been formulated in terms of two real fields
$\psi_1$ and $\psi_2$. 
As noted in the introduction, one  has traditionally
presented the theory in terms of normal and anomalous 
Green's functions $G_{11}(\omega,p)$ and $G_{12}(\omega,p)$
and the self-energies
$\Sigma_{11}(\omega,p)$ and $\Sigma_{12}(\omega,p)$. 
This formulation corresponds to using the fields $\psi^{*}$
and $\psi$ instead of $\psi_1$ and $\psi_2$. 
In either formulation, one ends up with a $2\times2$ matrix for the
propagator and self-energies and the amount of work to calculate most
quantities is comparable.
For readers who want to translate intermediate results to
a more familiar language, we note that
\bqa
\Sigma_{11}(\omega,p)&=&
{1\over2}\left[\Pi_{11}(\omega,p)+\Pi_{22}(\omega,p)
\right]\;, 
\label{r1}
\\
\Sigma_{12}(\omega,p)&=&
{1\over2}\left[\Pi_{11}(\omega,p)-\Pi_{22}(\omega,p)\right]
\label{r2}
\;,
\eqa
where $\Pi_{ij}(\omega,p)$ 
are the components of the $2\times2$ self-energy matrix.

We next comment on the Hugenholz-Pines theorem~\cite{hug}.
The Hugenholz-Pines theorem ensures that the spectrum does not exhibit a
gap and it has been proven to all orders in perturbation theory.
It is simply the Goldstone theorem for a dilute Bose gas with
a spontaneously broken continuous symmetry.
The Hugenholz-Pines theorem is normally given in terms of
normal and anomalous self-energies:
\bqa\nonumber
\mu&=&
\Sigma_{11}(0,0)-\Sigma_{12}(0,0)\;.
\eqa
In terms of the self-energies $\Pi_{ij}(\omega,p)$, the theorem
takes a particular
simple form:
\bqa
\mu&=&\Pi_{22}(0,0)\;.
\eqa

\subsection{Ground State Energy Density and Condensate Depletion}
\label{subcon}
In this subsection, we calculate the leading quantum correction to the
ground state energy and the depletion of the condensate. While these 
results are standard textbook material~\cite{fetter}, it is instructive
to see how they are derived within the present framework.

The mean-field thermodynamic potential $\Omega_0(\mu,v)$ is given 
by the terms in the classical action~(\ref{cact}):
\bqa
\Omega_0(\mu,v)=-\mu v^2+{1\over2}gv^4\;.
\label{o0}
\eqa
The minimum of the mean-field thermodynamic potential is given
by $v_0=\sqrt{\mu/g}$ and 
the mean-field free energy  ${\cal F}_0$ is obtained by 
evaluating Eq.~(\ref{o0}) at the minimum: 
\bqa
{\cal F}_0(\mu)&=&-{\mu^2\over 2g}\;.
\label{f0}
\eqa
The mean-field number density follows from differentiating~(\ref{f0})
with respect to $\mu$. The chemical potential in the mean-field
approximation is then obtained by
inversion:
\bqa
\mu_0&=&gn\;.
\eqa
The mean-field energy ${\cal E}_0$ is easily from~(\ref{rela}) found to be
\bqa\nonumber
{\cal E}_0(n)&=&{1\over2}gn^2\\
&=&4\pi an^2\;.
\label{e0}
\eqa
The mean-field result for the energy density was first obtained by 
Bogoliubov~\cite{bogo}.

The one-loop contribution to the thermodynamic potential $\Omega(\mu,v)$ is
given by \bqa
\Omega_1(\mu,v)&=&i{\log{\cal Z}_0\over{\cal V}}\;,
\eqa
where ${\cal Z}_0$ is the path integral involving
the quadratic quantum fluctuations around the mean field.
\bqa\nonumber
{\cal Z}_0&=&\int{\cal D}\psi_1{\cal D}\psi_2\;
e^{iS_{\rm free}[v,\psi_1,\psi_2]}\\
&=&e^{i\int dt\int d^3x\det D^{-1}(\omega,p)}
\eqa
where $D^{-1}(\omega,p)$ 
is the inverse of the propagator~(\ref{prop}) 
and $\det$ denotes a determinant in the functional sense.
Using the fact that ${\rm Tr}\log A=\log\det A$ for any matrix $A$,
we obtain
\bqa\nonumber
\Omega_1(\mu,v)&=&-{1\over2}i\int{d\omega\over2\pi}
\int{d^dk\over(2\pi)^d}\log\det D^{-1}(\omega,k)
\\ &&
+\Delta_1\Omega\;.
\label{1om}
\eqa
Here we have added 
$\Delta_1\Omega$, which is the one-loop counterterm.
The counterterm is added to cancel the ultraviolet divergences
that one encounters when evaluating the integral in~(\ref{1om}).
After integrating over $\omega$ using~(\ref{wi1}), we obtain
\bqa\nonumber
\Omega_1(\mu,v)&=&{1\over2}\int{d^dk\over(2\pi)^d}\sqrt{(k^2-X)(k^2-Y)}
\\&&
+\Delta_1\Omega(\mu,v)\;,
\label{2om}
\eqa
The one-loop contribution ${\cal F}_1$ to the free energy is obtained by
evaluating Eq.~(\ref{2om}) at the classical minimum,
where $Y=0$ and $X=-2\mu$.
The one-loop free energy then reduces to 
\bqa\nonumber
{\cal F}_{0+1}(\mu)&=&
-{\mu^2\over 2g}+{1\over2}\int{d^dk\over(2\pi)^d}k\sqrt{k^2+2\mu}
+\Delta_1{\cal F}(\mu)\\
&=&-{\mu^2\over 2g}+{1\over2}I_{0,-1}(2\mu)+\Delta_1{\cal F}(\mu)\;,
\label{f01}
\eqa
where the integral $I_{m,n}(\Lambda)$ is defined in the appendix and 
$\Delta_1{\cal F}$ is the one-loop counterterm.
The integral $I_{0,-1}(\Lambda)$ has quintic, cubic, and linear ultraviolet 
divergences that are
set to zero in dimensional 
regularization.
The counterterm $\Delta_1{\cal F}$ is therefore zero and
the limit $d\rightarrow3$ is 
regular. We obtain
\bqa
{\cal F}_{0+1}(\mu)&=&-{\mu^2\over2g}
\left[1-{4\sqrt{2\mu g^2}\over15\pi^2}
\right]\;.
\label{fr0+1}
\eqa
It might
be useful to see how the renormalization procedure works with a simple 
ultraviolet cutoff. In that case, the one-loop contribution to the
free energy ${\cal F}_{1}$
can be written as
\bqa
{\cal F}_{1}(\mu)&=&{1\over2}\int^M{d^3k\over(2\pi)^3}
k\sqrt{k^2+2\mu}+\Delta_1{\cal F}
\;,
\eqa
where the integral is calculated in $d=3$ dimensions and the superscript
$M$ indicates that $|{\bf k}|<M$ has been imposed.
We can now rewrite this as
\bqa\nonumber
{\cal F}_{1}(\mu)&=&{1\over2}\int^M{d^3k\over(2\pi)^3}
\left[k\sqrt{k^2+2\mu}
-k^2-\mu+{\mu^2\over2k^2}
\right]\\ &&
+\int^M{d^3k\over(2\pi)^3}
\left[
k^2+\mu-{\mu^2\over2k^2}
\right]
+\Delta_1{\cal F}
\eqa
The first integral is now convergent in the limit $M\rightarrow\infty$
and the divergences have been isolated in the second integral.
This first term goes like $M^5$ 
and is independent of $\mu$ and $g$. It 
can therefore be removed by a vacuum energy counterterm $\Delta_1{\cal E}$.
The form of $\Delta_1{\cal F}$ can be found by substituting 
$g\rightarrow g+\Delta_1g$ and $\mu\rightarrow \mu+\Delta_1{\mu}$
in ${\cal F}_{0}(\mu)$ 
and expanding to first order in $\Delta_1g$ and $\Delta_1\mu$.
Including the vacuum counterterm, we obtain
\bqa
\Delta{\cal F}_1(\mu)&=&-{\mu\over g}\Delta_1\mu
+{\mu^2\over2g^2}\Delta_1g+\Delta_1{\cal E}\;.
\eqa
The counterterms needed to cancel the quintic, cubic and linear
divergences can then be found by inspection. One obtains
\bqa
\Delta_1{\cal E}&=&
{1\over2}\int^M{d^3k\over(2\pi)^3}{k^2}\;,
\\
\label{dgcut}
\Delta_1g&=&{1\over2}g^2\int^M{d^3k\over(2\pi)^3}{1\over k^2}\;,\\
\Delta_1\mu&=&{1\over2}g\int^M{d^3k\over(2\pi)^3}\;.
\eqa
Note that the counterterm $\Delta_1g$ in Eq.~(\ref{dgcut}) is
precisely what is needed to cancel the divergence appearing
in the one-loop correction to the scattering length, Eq.~(\ref{cvar}).
This is in accord with our comment on counterterms in footnote~\ref{elg}.
The renormalized one-loop 
free energy can then be written as
\bqa\nonumber
{\cal F}_{0+1}(\mu)&=&-{\mu^2\over2g}+
{1\over2}\int^M{d^3k\over(2\pi)^3}\left[\epsilon(k)-
k^2
\right.\\ &&
\left.
-\mu+{\mu^2\over2k^2}
\right]\;.
\eqa
We can now take the limit $M\rightarrow\infty$.
Evaluating the integral,
we recover Eq.~(\ref{fr0+1}).

Finally, we mention that the {\it pseudo-potential method}~\cite{huangyang} 
is an alternative way of treating the
ultraviolet divergences. One replaces the contact potential
$g\delta^3({\bf x})/2$ by the pseudo potential
$g\delta^3({\bf x})\left(\partial/\partial x\right)x/2$ and one can 
avoid the ultraviolet divergences by evaluating the partial derivative
at the right stage of the calculation. We shall not discuss this method
any further.

Using Eqs.~(\ref{rhodef}),~(\ref{f01}) and~(\ref{alge1}), 
we obtain the number 
density in the 
one-loop approximation
\bqa\nonumber
n_{0+1}(\mu)&=&{\mu\over g}-{1\over2}I_{1,1}(2\mu)\\
&=&{\mu\over g}\left[1-{\sqrt{2\mu g^2}\over3\pi^2}
\right]
\;.
\label{1lden}
\eqa
Inverting Eq.~(\ref{1lden}) to obtain the chemical potential as a function
of the number density, one finds
\bqa\nonumber
\mu_{0+1}(n)&=&gn+{1\over2}gI_{1,1}(2gn)\\
&=&8\pi an\left[
1+{32\over3}\sqrt{{na^3\over\pi}}
\right]\;.
\label{muuu}
\eqa 
Note that here and in the following, we are replacing the argument
$\mu$ by $gn$ in the loop integrals $I_{m,n}$, since corrections are
of higher order in $\sqrt{na^3}$.
Using Eqs.~(\ref{rela}),~(\ref{f01}) and~(\ref{1lden}), 
we obtain the energy density 
in the one-loop approximation:
\bqa\nonumber
{\cal E}_{0+1}(n)&=&{1\over2}gn^2+{1\over2}I_{0,-1}(2gn)\\
&=&4\pi an^2\left[
1+{128\over15}\sqrt{{na^3\over\pi}}
\right]
\;.
\label{ed}
\eqa
The leading quantum correction to the mean-field result~(\ref{e0}) 
was first derived 
by Lee, Huang and Yang~\cite{leeyang,leehu} for a hard-sphere potential. 
Later it has been shown that it is {\it universal} in the sense that
it applies to all short-range potential with scattering length 
$a$~\cite{bs,beli,lieb}.
Note that the result~(\ref{ed}) is 
nonanalytic
in the scattering length $a$. This shows that the result is nonperturbative
from the point of view of ``naive'' perturbation theory, where one uses
free particle propagators.
It corresponds to the summation of an infinite set of one-loop diagrams,
with repeated insertions of the operator ${3\over2}gv^2\psi_1^2$.
The structure of these diagrams is the same as the ring 
diagrams first discussed by Gell-Mann and Brueckner
for the nonrelativistic electron gas~\cite{gm}, 
and are summed in the same manner.
It is interesting to note that each diagram in the series
is increasingly divergent in the infrared, but the sum is infrared
convergent.

\begin{figure}[htb]
\epsfysize=1.6cm
\epsffile{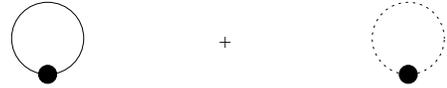}
\caption[a]{One-loop diagrams contributing to
expectation values $\langle\psi_1^2\rangle$
and $\langle\psi_2^2\rangle$.
}
\label{depl}
\end{figure}

Using~(\ref{shift}) and the parametrization~(\ref{split}), the 
expression for the number density becomes
\bqa
n&=&
v^2+{1\over2}\langle\psi^2_1+\psi_2^2\rangle\;.
\label{ndef}
\eqa
The condensate density $n_0$ is given by the expectation value
$v^2=|\langle\psi\rangle|^2$ 
At the mean-field level,
one neglects the fluctuations of $\psi_1$ and $\psi_2$ and 
replaces $\langle\psi^*\psi\rangle$ by $|\langle\psi\rangle|^2$. 
The total number density
is then 
equal to the condensate density. Taking quantum fluctuations into account,
this is no longer the case. Due to interactions, some of the particles
are kicked out of the condensate and are not in the ${\bf k}=0$
momentum state. The difference $n-n_0$
is called the depletion of the condensate and is proportional to the 
diluteness parameter $\sqrt{na^3}$. 
The one-loop diagrams that contribute to the 
expectation values $\langle\psi_1^2\rangle$ and $\langle\psi_2^2\rangle$
are shown in Fig.~\ref{depl}.
The solid line denotes the diagonal propagator
for $\psi_1$ and the dashed line the diagonal propagator for $\psi_2$.
The blobs denote an insertion of the operator 
$\psi_1^2$ or $\psi_2^2$.
Taking these one-loop effects into account,
Eq.~(\ref{ndef}) reduces to
\bqa\nonumber
n_{0+1}(n_0)&=&n_0+{1\over2}i
\int{d\omega\over2\pi}\int{d^dk\over(2\pi)^d} 
\left[{k^2+\epsilon^2(k)/k^2\over\omega^2-\epsilon^2(k)+i\varepsilon}\right]\\
&=&n_0+{1\over4}\left[I_{1,1}(2gn_0)+I_{-1,-1}(2gn_0)\right]
\;.
\label{t0n}
\eqa
Using the expressions for the integrals in the appendix, we obtain
\bqa
n_{0+1}(n_0)&=&n_0\left[1+{8\over3}\sqrt{n_0a^3\over\pi}\right]
\label{deplbog}
\;.
\eqa
The result~(\ref{deplbog}) for the depletion
was first obtained by Bogoliubov in 1947~\cite{bogo}.
In recent experiments, Cornish {\it et al.}~\cite{corny} were able to vary
the $s$-wave scattering length
$a$ for $^{85}$Rb atoms
over a large range by applying a strong external magnetic field
and exploiting the existence of a Feshbach resonance at $B\sim 155\;G$.
Values for $\sqrt{na^3}$ up to approximately
0.1 were obtained and should be sufficiently large to see deviations
from the mean field in experiments. In order to observe this quantum
phenomenon, it is essential that experiments are carried out at sufficiently
low temperature so that the thermal depletion of the
condensate is negligible.

\subsection{Collective Excitations}
\label{belsec}
The Bogoliubov spectrum is given by Eq.~(\ref{disps})
and was derived from the microscopic theory represented by the
action~(\ref{act}). 
It is linear for 
small momentum $p$ with slope $\sqrt{2\mu}$. 
The spectrum $\omega(p)$ of collective excitations is given by
the poles of the propagator. The poles are the solutions to the
equation
\bqa
\det\left[D^{-1}_0(\omega,p)-\Pi(\omega,p)\right]&=&0\;,
\label{colll}
\eqa
where $D_0(\omega,p)$ is the real-time version of the free
propagator~(\ref{propf})
and $\Pi(\omega,p)$ is the $2\times2$ self-energy matrix.
The dispersion relation $\omega(p)$ is generally complex and can 
therefore be written as
\bqa
\omega(p)&=&{\rm Re}\,\omega(p)-i\gamma(p)\;.
\eqa
The real part ${\rm Re}\,\omega(p)$
gives the energies of the excitations, while
the imaginary part $\gamma(p)$ 
represents the damping of the excitations.
The Bogoliubov spectrum is purely real and in this approximation,
the excitations therefore have an infinite lifetime.

\begin{figure}[htb]
\epsfysize=3.0cm
\epsffile{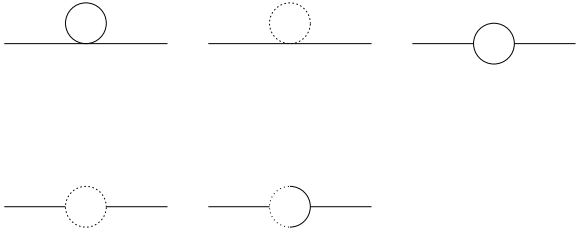}
\caption[a]{One-loop diagrams contributing to the self-energy
$\Pi_{11}(\omega,p)$.}
\label{selfff}
\end{figure}

\begin{figure}[htb]
\epsfysize=3.0cm
\epsffile{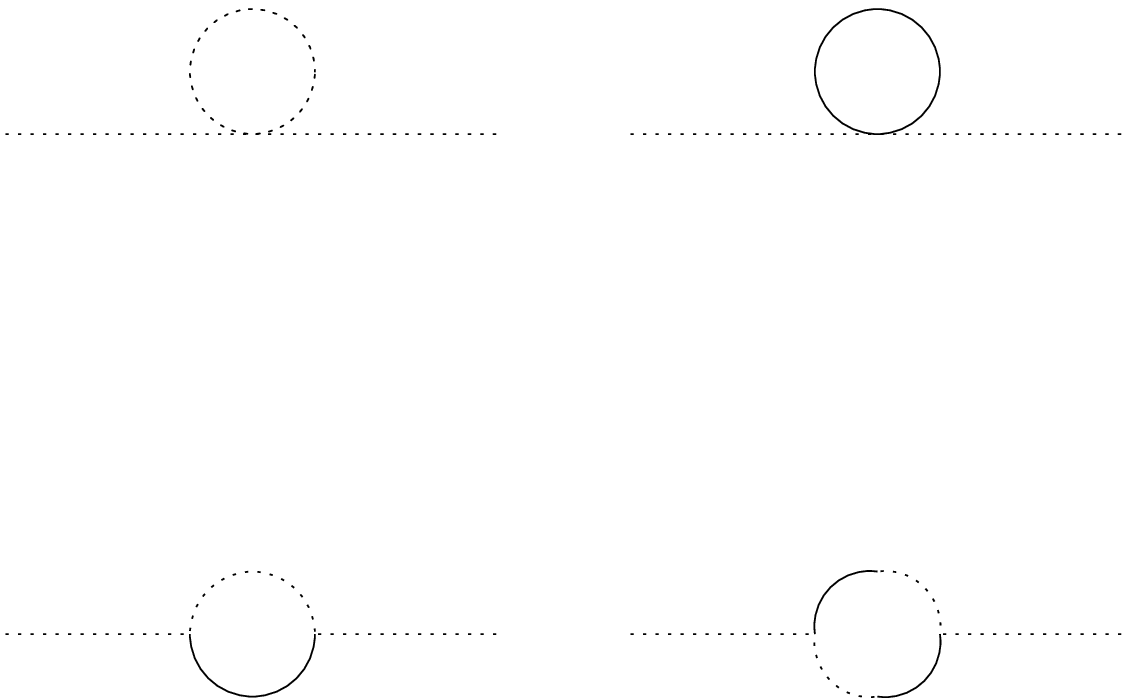}
\caption[a]{One-loop diagrams contributing to the self-energy
$\Pi_{22}(\omega,p)$.}
\label{s222}
\end{figure}

\begin{figure}[htb]
\epsfysize=0.6cm
\epsffile{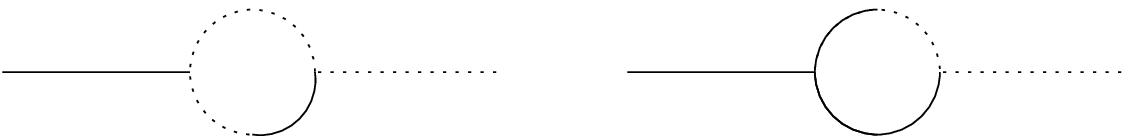}
\caption[a]{One-loop diagrams contributing to the self-energy
$\Pi_{12}(\omega,p)$.}
\label{s12}
\end{figure}

In the following, we calculate the leading quantum correction to the
real part of the spectrum in the long-wavelength limit
and thus reproduce Beliaev's classic result~\cite{beli}.
The one-loop Feynman diagrams that contribute to the self-energies
are shown in 
Figs.~\ref{selfff}--\ref{s12}\footnote{Note that the sum of all 
{\it one-particle reducible} diagrams that contribute to the self-energies
vanishes. This sum is proportional to the derivative of the effective
potential and is zero when evaluated at the minimum of the effective
potential.}. 
The solid line denotes the diagonal propagator
for $\psi_1$ and the dashed line the diagonal propagator for $\psi_2$.
The off-diagonal propagators for $\psi_1$ and $\psi_2$ are represented
by lines that are half solid and half dashed and vice versa.
One-loop contributions to self-energies are down by a factor of $\sqrt{na^3}$
compared to the mean-field terms in the inverse propagator.
It is therefore consistent to evaluate the self-energies
using the mean-field dispersion relation $\epsilon(p)$ since 
corrections would be suppressed by at least a factor of $na^3$.\\

\vspace{1cm}

\vspace{1cm}

In the following, we calculate the off-diagonal self-energy 
$\Pi_{12}(\omega,p)$ explicitly. The expression is
\bqa\nonumber
\Pi_{12}^{}(\epsilon(p),p)&=&g^2v^2
\int{d\omega\over2\pi}
\int{d^dk\over(2\pi)^d}
\\ &&\nonumber
\hspace{-2.4cm}
\times\Bigg\{
{3\left[\omega+\epsilon(p)\right]k^2
\over[(\omega+\epsilon(p))^2-\epsilon^2(|{\bf p}+{\bf k}|)+i\varepsilon]
[\omega^2-\epsilon^2(k)+i\varepsilon]}
\\ &&
\hspace{-2.5cm}
-
{\left[\omega+\epsilon(p)\right]\epsilon^2(k)/k^2
\over\left[(\omega+\epsilon(p))^2-\epsilon^2(|{\bf p}+{\bf k}|)
+i\varepsilon\right]
\left[\omega^2-\epsilon^2(k)+i\varepsilon\right]}
\Bigg\}\;,
\eqa
After integrating over $\omega$, the imaginary 
part of $\Pi_{12}(\epsilon(p),p)$
becomes
\bqa\nonumber
{\rm Im}\,\Pi_{12}^{}(\epsilon(p),p)&=&-g^2v^2
\int{d^dk\over(2\pi)^d}
\\ && \nonumber
\times\Bigg\{
{[\epsilon(p)+\epsilon(k)][3k^2/\epsilon(k)-\epsilon(k)/k^2]
\over\left[\epsilon^2(|{\bf p}+{\bf k}|)-\epsilon^2(p)-\epsilon^2(k)\right]}
\\ &&\nonumber
+{3k^2-\epsilon^2(k)/k^2
\over\left[(\epsilon(k)+\epsilon(p))^2
-\epsilon^2(|{\bf p}+{\bf k}|)\right]}
\Bigg\}\;.
\\ &&
\label{pi12t}
\eqa
Finally, we expand Eq.~(\ref{pi12t}) 
in powers of the external momentum $p$. After
simplifying using Eqs.~(\ref{alge1})--(\ref{alge3}), the self-energy
reduces to
\bqa\nonumber
{\rm Im}\,\Pi_{12}^{}(\epsilon(p),p)&=&{1\over8}g(2gv^2)^{3/2}
\left[3I_{1,3}-I_{-1,1}
\right]p
\\ &&
+{\cal O}\left(p^3/g^{3/2}v^3\right)\;.
\label{p1}
\eqa
Notice in particular that $\Pi_{12}^{}(0,0)$ vanishes. This 
property holds to all orders in perturbation theory and follows from 
time-reversal invariance. Note also that 
$\Pi_{21}^{}(\omega,p)=-\Pi_{12}^{}(\omega,p)$, which also holds
to all orders in loop expansion.
The real part of the self-energies $\Pi_{11}(\epsilon(p),p)$ and
$\Pi_{22}(\epsilon(p),p)$ are expanded 
about zero external momentum in the same way.
Including the mean-field self-energies, one finds:
\bqa
\nonumber
{\rm Re}\,\Pi_{11}(\epsilon(p),p)&=&3gv^2+
{1\over4}g\Big[
3I_{1,1}+I_{-1,-1}-gv^2\left(9I_{2,3}
\right.\\ && \left.
-6I_{0,1}+I_{-2,-1}
\right)
\Big]
+{\cal O}\left(p^2/gv^2\right)
\label{p2}
\;,\\ \nonumber
{\rm Re}\,\Pi_{22}(\epsilon(p),p)&=&gv^2+
{1\over12}g\Big[9I_{1,1}+3I_{-1,-1}
\\ &&
\nonumber
-p^2gv^2\left(6I_{3,5}-11I_{1,3}+3I_{-1,1}
\right)\Big]
\\ &&
+{\cal O}\left(p^4/(g^2v^4\right)
\;.
\label{p3}
\eqa
The expressions for the 
self-energies $\Pi_{11}(\epsilon(p),p)$, $\Pi_{22}(\epsilon(p),p)$
and $\Pi_{12}(\epsilon(p),p)$ 
are infrared
divergent. The integrals $I_{-2,-1}(2gn_0)$  and $I_{-1,1}(2gn_0)$ 
both have 
a logarithmic divergence as the loop momentum $k$ goes to zero. 
These divergences show up as poles in $\epsilon$ in 
Eqs.~(\ref{ir1})--(\ref{ir2}).
However, it is important to point out that they
cancel in the final results for physical quantities, as we shall see
below.

The real part of equation~(\ref{colll}) can now be written as
\bqa\nonumber
[\omega-{\rm Im}\,\Pi_{12}(\epsilon(p),p)]^2&=&
\left[p^2-\mu+{\rm Re}\,\Pi_{11}(\epsilon(p),p)\right]
\\ &&\hspace{-1cm}
\times
\left[p^2-\mu+{\rm Re}\,\Pi_{22}(\epsilon(p),p)\right]\;.
\label{cnew}
\eqa
The next step is to eliminate the chemical potential from Eq.~(\ref{cnew})
by minimizing the one-loop thermodynamic potential 
$\Omega_{0+1}(\mu,v)$, This is 
found by differentiating the sum of Eqs.~(\ref{o0}) and~(\ref{2om}) 
with respect to the condensate $v$
and setting it to zero. This yields
\bqa\nonumber
0&=&-\mu+gv^2
+{1\over4}g
\int{d^dp\over(2\pi)^d}{3(p^2-Y)+(p^2-X)
\over\sqrt{(p^2-X)(p^2-Y)}}
\;.
\\ &&
\label{gol}
\eqa
It is consistent to the order in quantum corrections at which we are 
calculating, to evaluate the one-loop contribution to~(\ref{gol})
at the classical minimum $v=v_0$.
Eq.~(\ref{gol}) then reduces to
\bqa\nonumber
0&=&-\mu+gv^2+{1\over4}g\left[3I_{1,1}+I_{-1,-1}\right]\\
&=&-\mu+\Pi_{22}(0,0)\;.
\label{eqs}
\eqa
Eq.~(\ref{eqs}) shows that the Goldstone theorem is satisfied 
at the stationary point of the thermodynamic potential.
Using (\ref{t0n}), we see that the chemical potential obtained 
from~(\ref{muuu}) agrees with~(\ref{eqs}).
Solving for the chemical potential and substituting the result as well
as the self-energies given by Eqs.~(\ref{p1})--(\ref{p3}) 
into Eq.~(\ref{cnew}), we obtain
\bqa\nonumber
&&
\left[\omega+{1\over8}g(2gv^2)^{3/2}
\left(3I_{1,3}-I_{-1,1}
\right)p\right]^2=
\\ && \nonumber
p^2\left[p^2+2gv^2\left(1-{1\over8}g
(9I_{2,3}+I_{-2,-1}-6I_{0,1})\right)\right] \\
&&
\times
\left[1-{1\over12}g^2v^2\left(6I_{3,5}-11I_{1,3}+3I_{-1,1}
\right)\right]\;.
\label{om}
\eqa
We can now solve for $\omega$ in Eq.~(\ref{om}).
In the long-wavelength limit, $p\ll\sqrt{2gv^2}$, we obtain
\bqa\nonumber
{\rm Re}\,\omega(p)&=&p\sqrt{2gv^2}\left[1-{g^2v^2\over24}
\left(6I_{3,5}+7I_{1,3}-3I_{-1,1}
\right)
\right]\\ && \nonumber
\times
\left[1-{1\over16}g\left(
9I_{2,3}-6I_{0,1}+I_{-2,-1}
\right)\right]\\ 
&=&p\sqrt{2gn_0}\left[
1+{28\over3}\sqrt{n_0a^3\over\pi}
\right]\;.
\label{pre}
\eqa
The infrared divergent terms that appear in the 
expressions for the self-energies
cancel algebraically after having used
the relations~(\ref{alge1})--(\ref{alge3}).
The ultraviolet divergences associated with the integrals in
Eq.~(\ref{pre}) are again power divergences. Thus one immediately
obtains a finite result and this is another example of the 
convenience of employing dimensional regularization.
The result~(\ref{pre}) was first obtained by Beliaev~\cite{beli}.
One can easily check that the slope of $\omega$ in Eq.~(\ref{pre}) is the same
as the macroscopic speed of sound, $c$, that one obtains from differentiating
the pressure with respect to the number density~\cite{fetter}. 
This equality has been proven
to all orders in perturbation theory by Gavoret and Nozieres~\cite{gav}, and by
Hohenberg and Martin~\cite{hohen}.

The effective field theory approach presented in this section 
is based on a cartesian parametrization of the quantum field 
$\tilde{\psi}$ in Eq.~(\ref{split}).
A similar effective field theory approach 
was developed by Popov~\cite{popov2} and later extended by Liu~\cite{w1,w2}.
Instead of using a cartesian parametrization of the field $\psi$,
it was parametrized using the density $n$
and the phase $\chi$:
\bqa
\psi({\bf x},t)&=&\sqrt{n({\bf x},t)}\;e^{i\chi({\bf x},t)}\;.
\label{np}
\eqa
The density field is shifted in analogy with the shift in Eq.~(\ref{shift}):
\bqa
n({\bf x},t)&=&v^2+\sigma({\bf x},t)\;.
\label{shift2}
\eqa
We recall that the condition~(\ref{condit}) simplifies calculations,
since it makes all one-particle reducible diagrams vanish.
This condition is replaced by
\bqa
\langle\sigma\rangle&=&0\;.
\eqa
Using the parametrization~(\ref{np}) together
with~(\ref{shift2}), the action~(\ref{act}) takes the form
\bqa\nonumber
S&=&S[v]+\int dt\int d^dx\left[{1\over v}J\sigma
+{1\over2}\left(\chi\dot{\sigma}-\dot{\chi}\sigma
\right)
-{1\over2}g\sigma^2
\right.\\ \nonumber
&&\left. \hspace{2.3cm}
-v^2(\nabla\chi)^2-{1\over4v^2}(\nabla\sigma)^2
-\sigma(\nabla\chi)^2
\right.\\ &&\left.\hspace{2.3cm}
+{1\over4v^4}\sigma\left(\nabla\sigma\right)^2+...
\right]\;,
\label{s22}
\eqa
where the dots indicate an infinite series of higher order
operators.
The classical action $S[v]$ and the source $J$ are given by Eqs.~(\ref{cact}) 
and~(\ref{t}), respectively. 
After a rescaling of the fields $\sigma$ and $\chi$, the free propagator
corresponding to the action~(\ref{s22}) is identical to Eq.~(\ref{prop2}).
However, the interaction vertices
are different. In particular, the vertex 
corresponding to the operator
$\sigma(\nabla\chi)^2$ is momentum
dependent. One feature of the perturbative expansion that follows from 
the action~(\ref{s22}) is the absence of infrared divergences in 
individual 
diagrams~\cite{popov,ericag}. The momentum dependence of the trilinear 
interaction $\sigma(\nabla\chi)^2$ compensates the singular behavior
of the propagator at low momenta. 
This difference should not be viewed as
being fundamental since the infrared divergences always cancel in physical 
quantities. We have already seen one example of this, when we 
considered the quantum correction to the Bogoliubov dispersion relation.
On the other hand, individual diagrams are more severely ultraviolet 
divergent, but these cancel when the diagrams are added. 
The above features merely represent a different 
way of organizing the perturbative calculations. The equality of
calculations of physical quantities order by order in perturbation theory
simply reflects the reparametrization invariance
of the functional integral.

As mentioned before, the imaginary part of the dispersion relation
represents the damping of the collective excitations.
In the case of a dilute Bose gas, 
this term was first calculated by Beliaev~\cite{beli}.
In the long-wavelength limit, his 
calculations showed that the damping rate is proportional to $p^5$:
\bqa
\gamma(p)&=&{3p^5\over320\pi n_0}\;.
\label{imw}
\eqa
The imaginary part $\gamma(p)$ 
is connected with one phonon decaying into two phonons
with lower energy. It is often referred to as Beliaev damping.
The action~(\ref{s22}) has later been used by several 
authors~\cite{popov,w1,w2} to rederive
the result~(\ref{imw}).
The calculations represent a significant
simplification compared to the original derivation.

\subsection{Nonuniversal Effects}
\label{noneff}
In the previous subsection, it was shown that the dominant effects of the
interaction between the atoms could be subsumed in a single 
coupling constant called the $s$-wave scattering length.
Thus all interatomic potentials with the same $s$-wave scattering length
will have the same properties to leading order in the 
low-density expansion, and this property is called universality.
However, at higher orders in the low-density expansion, physical
quantities will depend on the 
details of the interatomic potential such as the effective range $r_s$.
These are called nonuniversal effects. A detailed analysis of nonuniversal
effects can be found in the paper by Braaten, Hammer and Hermans~\cite{e2}.
We discuss these next.

Including the operator 
$\left[\nabla(\psi^*\psi)\right]^2$ in Eq.~(\ref{act}), we can 
again calculate exactly the scattering amplitude for $s$-wave scattering.
Summing the contributions from the diagrams in Fig.~\ref{loopfig},
we obtain~\cite{e2}
\bqa
{\cal T}(q)&=&-\left[{1\over2g+4hq^2}+i{q\over16\pi}\right]^{-1}\;.
\label{m2}
\eqa
The coupling constant $h$ is then related to the effective range 
$r_s$ of the true potential~\cite{e2} and it is determined by 
matching Eqs.~(\ref{m1}) and~(\ref{m2}) through third order in $q$.
This yields
\bqa
h&=&
2\pi a^2r_s\;.
\label{hdef}
\eqa
After performing the shift~(\ref{shift}), the operator 
$\left[\nabla(\psi^*\psi)\right]^2$ also
contributes to the free part of the action.
Including the effects of this operator, one obtains a modified
propagator and a modified dispersion relation: 
\bqa
D(\omega,p)&=&\frac{i}{\omega^2
-\epsilon^2(p)+i\varepsilon}
\left(\begin{array}{cc}
p^2&-i\omega \\
i\omega&\epsilon^2(p)/p^2
\end{array}\right)\;,
\label{propexp}
\eqa
where the new dispersion relation is
\bqa
\epsilon(p)&=&p\sqrt{(1+2hv^2)p^2+2gv^2}
\label{newspec}
\;.
\eqa
The spectrum~(\ref{newspec}) has the Bogoliubov form with modified 
coefficients. It is therefore straightforward to recalculate the 
ground state energy density and the depletion of the condensate
by rescaling the momentum ${\bf p}\rightarrow {\bf p}\sqrt{1+2hv^2}$. 
For instance, the expression for the number density is in analogy with
equation~(\ref{t0n}) given by
\bqa\nonumber
n_{0+1}(n_0)&=&n_0
+{1\over4}
\left[\int{d^dp\over(2\pi)^d}
{p^2\over\epsilon(p)}+{\epsilon^2(p)\over p^2}
\right]\\ \nonumber
&=&n_0+{1\over4}\left[1+2hv^2\right]^{-{d+1\over2}}I_{1,1}
\\ &&
+{1\over4}\left[1+2hv^2\right]^{-{d-1\over2}}I_{-1,-1}\;.
\eqa
The limit $d\rightarrow3$ is regular and we obtain
\bqa
n_{0+1}(n_0)&=&n_0\left[1+
{1\over24\pi^2}{\sqrt{8n_0g^3}\over(1+2hv^2)^2}\left(1-2hv^2\right)
\right]\;.
\eqa
Using Eqs.~(\ref{ga}) and~(\ref{hdef}), and
expanding to first order in the effective range
$r_s$, we find
\bqa\nonumber
n_{0+1}(n_0)&=&n_0\left[1+
{8\over3}\sqrt{n_0a^3\over\pi}
\right.\\ &&\left.
-{32\pi^2r_s\over a}\left(
{n_0a^3\over\pi}\right)^{3/2}\right]
\;.
\eqa
Similarly, we can calculate the ground state energy density.
Expanding to first order in the effective range
$r_s$, one finds~\cite{e2}:
\bqa\nonumber
{\cal E}_{0+1}(n)&=& 
4\pi an^2\left[
1+{128\over15}\sqrt{{na^3\over\pi}}
\right.
\\ &&
-{1024\pi^2r_s\over15a}\left({na^3\over\pi}\right)^{3/2}
\Bigg]
\label{expl}
\;.
\eqa
Effective field theory can be used to determine at which order in the 
low-density expansion a given operator starts to contribute to 
various physical quantities. As an example, we consider the 
energy density ${\cal E}$. Each power of $\psi^*\psi$ contributes
a factor of $n$. Each power of $\nabla$ contributes a factor of $\sqrt{na}$.
Each loop order in the quantum loop expansion contribitutes a factor of
$\sqrt{na^3}$. The derivative interaction $[\nabla(\psi^*\psi)]^2$
does not contribute to the energy density at the mean-field level for
a homogeneous Bose gas since $\nabla v$ obviously vanishes in this case.
It first contributes at the one-loop level and this gives one factor
of $\sqrt{na^3}$
There are two powers of $\psi^*\psi$ and two powers of $\nabla$, which give
two factors of $n$ and $\sqrt{na}$, respectively. This yields a contribution
proportional to $r_sn^2({na^3})^{3/2}$ in accordance with the explicit
calculation~(\ref{expl}).

We next comment on the nonuniversal effects that arise in 
higher-order calculations
of the energy density. For simplicity, we ignore 
all other coupling constants than $g$.
The leading term is the
mean-field contribution given in~(\ref{e0}).
A one-loop calculation gives rise to a
universal correction proportional to $\sqrt{na^3}$ shown in~(\ref{ed}).
At the two-loop level, one 
encounters two terms. The first is a universal logarithmic
corrections proportional to $na^3\log na^3$. This 
term was first calculated by
Wu~\cite{wu}. The second term is a nonuniversal term proportional
to $na^3$.
This term was first calculated
by Braaten and Nieto~\cite{eric}.
It comes about as follows. At the two-loop level, there is a
logarithmic ultraviolet divergence 
that cannot be cancelled by a local
two-body counterterm of the form $\delta g(\psi^*\psi)^2$.
The only way to cancel it, is to add to the Lagrangian
a local counterterm of the 
form $\delta g_3(\psi^*\psi)^3$ and absorb the divergence in the
coefficient of this operator. One can see the necessity of such
an operator by considering 
$3\rightarrow3$ scattering. At the two-loop level, or fourth order in $a$, 
there are Feynman diagrams that depend logarithmically on the
ultraviolet cutoff. They give an additional momentum-independent
contribution to the $3\rightarrow3$
scattering amplitude.
In order to reproduce the low-energy scattering
of three atoms, the local momentum-independent
operator $\delta g_3(\psi^*\psi)^3$
must be included in the effective Lagrangian. The counterterm of this
operator that removes the logarithimic divergences from the
$3\rightarrow3$ scattering amplitude is then exactly the same
counterterm needed to remove the logarithmic divergence
in the energy density.

The coefficient $g_3$ generally depends
on the properties of the two-body and
three-body potentials. 
The operator $(\psi^*\psi)^3$ 
in Eq.~(\ref{act}) takes into account not only the contribution from
$3\rightarrow3$ scattering 
from a possible three-body potential, but also the contribution from 
the successive $2\rightarrow2$ scattering via the potential $V_0({\bf x})$.
One way to determine the coefficient $g_3$ would be to solve the
$3\rightarrow3$ scattering problem for the potential $V_0({\bf x})$.
Alternatively, one can determine it from 
calculating the ground state energy density of 
bosons interacting through $V_0({\bf x})$.
Such a strategy was recently used by Braaten, Hammer and 
Hermans~\cite{e2} using the Monte Carlo calculations of the 
condensate fraction and energy density for four different model potentials
by Giorgini, Boronat, and Casulleras~\cite{georg}.
The four potential were a hard-sphere potential with radius $a$, 
two soft-sphere potentials with height $V_0$ and radii $R=5a$ and $R=10a$, and
a hard-sphere square-well potential with depth $V_0$ with inner and outer radii
of $R=a/50$ and $R=a/10$, respectively.
These four potentials 
all have the same $s$-wave scattering length $a$, 
but different effective range $r_s$. 
By calculating the energy density for the homogeneous Bose in the low-density
expansion and matching it onto the Monte Carlo results, 
Braaten, Hammer and Hermans were able
to estimate the coefficient $g_3$. Due to the large statistical errors,
they could not find any deviation from universality in the three-body
contact parameter. In order to determine
the coefficient more accurately, one needs
data with higher statistics at various densities.

\section{Weakly Interacting Bose Gas at Finite Temperature}
In this section, we discuss the weakly interacting Bose gas at finite 
temperature.
We first review the 
the Hartree-Fock-Bogoliubov (HFB) approximation, the
Bogoliubov approximation, and the Popov approximation.
We then discuss Wilson's renormalization
group approach applied to this problem. Finally, we discuss improved
variational approaches to the finite temperature Bose gas. 

\subsection{Hartree-Fock Bogoliubov Approximation}
The self-consistent Hartree-Fock-Bogoliubov approximation
and its relation to the Bogoliubov approximation and the Popov approximation
have been discussed in detail~\cite{hbf1,hottot,grif}.
The starting point is the action~(\ref{act}) in imaginary time:
\bqa\nonumber
S[\psi^*,\psi]&=&\int_0^{\beta}d\tau\int d^dx\Bigg\{
\psi^*\left[{\partial\over\partial\tau}
-\mu-\nabla^2\right]\psi
\\ &&
+{1\over2}g(\psi^*\psi)^2
\Bigg\}\;.
\label{s1}
\eqa
The next step is to treat the interaction term using a self-consistent
quadratic approximation. After performing the shift~(\ref{shift}),
there are terms that are cubic and quartic in the quantum field
$\tilde{\psi}$.
These terms are approximated as follows
\bqa
\tilde{\psi}^*\tilde{\psi}\tilde{\psi}
&\approx&
2\langle\tilde{\psi}^*\tilde{\psi}\rangle\tilde{\psi}+
\langle\tilde{\psi}\tilde{\psi}\rangle\tilde{\psi}^*\;,
\label{ap1}
\\ \nonumber
\tilde{\psi}^*\tilde{\psi}\tilde{\psi}^*\tilde{\psi}
&\approx&4\langle\tilde{\psi}^*\tilde{\psi}\rangle\tilde{\psi}^*\tilde{\psi}
+\langle\tilde{\psi}^*\tilde{\psi}^*\rangle\tilde{\psi}\tilde{\psi}
+\langle\tilde{\psi}\tilde{\psi}\rangle\tilde{\psi}^*\tilde{\psi}^*\;.
\\ &&
\label{ap2}
\eqa
Terms involving the expectation value of a single field
have been omitted
since $\langle\tilde{\psi}\rangle=\langle\tilde{\psi}^*\rangle=0$.
Moreover, terms involving the expectation values of three or four
fields are omitted.
The quantities $\langle\tilde{\psi}^*\tilde{\psi}\rangle$
and $\langle\tilde{\psi}\tilde{\psi}\rangle$ are often referred
to as the normal and anomalous average, respectively.

We next write the quantum field $\tilde{\psi}=(\psi_1+i\psi_2)/\sqrt{2}$ 
and insert Eqs.~(\ref{ap1})--(\ref{ap2}) into the action~(\ref{s1}).
One then obtains an approximate action which is quadratic in the 
fluctuating fields:
\bqa\nonumber
S[v,\psi_1,\psi_2]&=&
S[v]+S_{\rm free}[v,\psi_1,\psi_2]+S_{\rm int}[v,\psi_1,\psi_2]\;,
\\&&
\eqa
where in analogy with the zero-temperature case, we have defined
the classical, free, and interacting parts of the action by
\bqa
S[v]&=&\int_0^{\beta}d\tau\int d^dx
\left[-\mu v^2+{1\over2}gv^4\right]\;,\\ \nonumber
S_{\rm free}[v,\psi_1\psi_2]&=&\int_0^{\beta}d\tau\int d^dx 
\left[
{1\over2}i\left(\psi_1\dot{\psi_2}-\dot{\psi}_1\psi_2\right)
\right. 
\\ &&\left.\nonumber
+{1\over2}iW\psi_1\psi_2
+{1\over2}
\psi_1\left(-\nabla^2+X\right)\psi_1
\right.\\ &&
\left.
+{1\over2}
\psi_2\left(-
\nabla^2+Y\right)\psi_2
\right] \;,
\label{action22}
\\ \nonumber
S_{\rm int}[v,\psi_1,\psi_2]&=&
\int_0^{\beta}d\tau\int d^dx\Bigg\{\bigg[-\mu+g\left(v^2+
2\langle\tilde{\psi}^*\tilde{\psi}\rangle
\right.\\ &&\left.\nonumber
+{1\over2}\langle\tilde{\psi}\tilde{\psi}\rangle
+{1\over2}\langle\tilde{\psi}^*\tilde{\psi}^*\rangle\right)
\bigg]\sqrt{2}v\psi_1
\\ &&
+{1\over2}\left[\langle\tilde{\psi}^*\tilde{\psi}^*\rangle
-\langle\tilde{\psi}\tilde{\psi}\rangle\right]\sqrt{2}v\psi_2
\Bigg\}\;,
\label{demand}
\eqa
where 
\bqa
W&=&g\left[
\langle\tilde{\psi}\tilde{\psi}\rangle
-\langle\tilde{\psi}^*\tilde{\psi}^*\rangle\right]
\;,\\
X&=&-\mu+g\left[3v^2+
2\langle\tilde{\psi}^*\tilde{\psi}\rangle
+{1\over2}\langle\tilde{\psi}\tilde{\psi}\rangle
+{1\over2}\langle\tilde{\psi}^*\tilde{\psi}^*\rangle
\right]\;, 
\label{a3}\\ 
Y&=&
-\mu+g\left[v^2+
2\langle\tilde{\psi}^*\tilde{\psi}\rangle
-{1\over2}\langle\tilde{\psi}\tilde{\psi}\rangle
-{1\over2}\langle\tilde{\psi}^*\tilde{\psi}^*\rangle
\right]\;.
\label{a4}
\eqa
Demanding that the terms linear in $\psi_1$ and $\psi_2$
in Eq.~(\ref{demand}) vanish, immediately yields
\bqa
\label{w22}
0&=&\langle\tilde{\psi}^*\tilde{\psi}^*\rangle-
\langle\tilde{\psi}\tilde{\psi}\rangle\;,\\
0&=&-\mu+g\left[v^2
+2\langle\tilde{\psi}^*\tilde{\psi}\rangle
+{1\over2}\langle\tilde{\psi}\tilde{\psi}\rangle
+{1\over2}\langle\tilde{\psi}^*\tilde{\psi}^*\rangle
\right]\;.
\label{heos3p}
\eqa
The equation of motion for the quantum field $\tilde{\psi}$ follows
from~(\ref{action22}) and reads
\bqa\nonumber
{\partial\tilde{\psi}\over\partial\tau}&=&
-\left[\nabla^2+\mu\right]\tilde{\psi}
+g\left[
2\left(v^2+\langle\tilde{\psi}^*\tilde{\psi}\rangle\right)\tilde{\psi}
\right.\\ &&\left.
+\left(v^2+\langle\tilde{\psi}\tilde{\psi}\rangle
\right)\tilde{\psi}^*
\right]\;.
\label{heos3}
\eqa
The propagator that corresponds to the free part of the action is
\bqa\nonumber
D(\omega_n,p)&=&\frac{1}{\omega^2_n
+\epsilon^2(p)}
\left(\begin{array}{cc}
p^2-2g\langle\tilde{\psi}\tilde{\psi}\rangle
&\omega_n
 \\ 
-\omega_n
&p^2+2gv^2
\end{array}\right)\;,
\\ &&
\eqa
where we have used Eq.~(\ref{w22}) to simplify.
The dispersion relation is given by
\bqa
\epsilon^2(p)&=&
\left[p^2-2g\langle\tilde{\psi}\tilde{\psi}\rangle
\right]\left[p^2+2gv^2\right]
\;.
\label{hbfd}
\eqa
It can be shown that the anomalous average 
$\langle\tilde{\psi}\tilde{\psi}\rangle$
is negative~\cite{hbf1}, so that
the dispersion relation~(\ref{hbfd}) makes sense.

The Hartree-Fock-Bogoliubov approximation is
given by Eqs.~(\ref{heos3p})--(\ref{hbfd}). One of the attractive 
features of the HFB approximation is that it is a conserving
approximation that is guaranteed to respect the conservation laws
that follow from the underlying symmetries of the field theory.
One of the problems with the HFB approximation is that 
there is a gap in the spectrum~(\ref{hbfd}),
which is a consequence of the spontaneously broken symmetry.
Thus the HFB approximation violates the Goldstone theorem.
Griffin and Shi~\cite{grif} have argued that the problem of a gap
in the HFB approximation is due to overcounting diagrams that contribute
to the anomalous self-energy and are second order in the interaction.
Thus the HFB approximation is inconsistent to that order in the 
interaction.

Another problem with the HFB approximation is that is it computationally
difficult to apply. One starts with some initial guess for the
condensate density, the chemical potential 
as well as the normal and anomalous averages satisfying Eq.~(\ref{heos3p}).
One then solves for the infinitely many normal modes in Eq.~(\ref{heos3}).
These normal modes are then used to calculate the normal and anomalous
averages. The procedure is iterated to 
self-consistency~\footnote{In practice one faces the problem of 
ultraviolet divergences in the anomalous average. Various renormalization
procedures have been discussed by Hutchinson {\it et al.}~\cite{hottot}.}.
In this way, one obtains information about the different normal modes.
However, since the HFB approximation has a gap, some of this information
must be qualitatively incorrect.

\begin{figure}[htb]
\epsfysize=5.cm
\epsffile{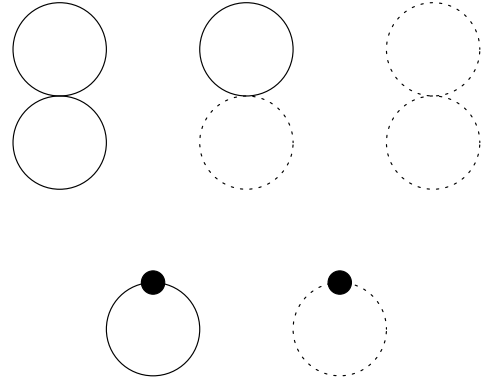}
\caption[a]{Vacuum diagrams included in the Gaussian approximation.}
\label{gauss}
\end{figure}

The HFB approximation can also be viewed as a variational one~\cite{rip},
although the equivalence seems to have gone unnoticed in some of the
literature. 
The idea is to make a Gaussian {\it ansatz} for the 
ground state wave functional and the excitations. 
The variational parameters are 
the normal and anomalous self-energies
$\Sigma_{11}(\omega_n,p)$ and $\Sigma_{12}(\omega_n,p)$.
They are normally taken to be independent of frequency and momentum.
The corresponding terms 
$\Sigma_{11}(\omega_n,p)\psi^*\psi$ and $\Sigma_{12}(\omega_n,p)\psi\psi$
are quadratic in fields
and are simply added to and subtracted from the Lagrangian. 
Expressing the normal and anomalous self-energies in terms
of $\Pi_{11}$ and $\Pi_{22}$,
the action can be split into a free part and an interacting
part according to
\bqa\nonumber
S_{\rm free}[v,\psi_1\psi_2]&=&\int_0^{\beta}d\tau\int d^dx 
\left[
{1\over2}i\left(\psi_1\dot{\psi_2}-\dot{\psi}_1\psi_2\right)
\right. 
\\ &&\left.
\hspace{-2cm}+{1\over2}
\psi_1\left(-\nabla^2-X\right)\psi_1
+{1\over2}
\psi_2\left(-
\nabla^2-Y\right)\psi_2
\right] \;,
\label{action2222}
\\ \nonumber
S_{\rm int}[v,\psi_1,\psi_2]&=&
\int_0^{\beta}d\tau\int d^dx\Bigg[
{1\over2}\left(3gv^2-\Pi_{11}\right)\psi_1^2
\\ && \nonumber
\hspace{-1cm}+{1\over2}\left(gv^2-\Pi_{22}\right)\psi_2^2
-{1\over\sqrt{2}}Z\psi_1\left(\psi_1^2+\psi_2^2\right)
\\ &&
+{1\over8}g\left(\psi_1^2+\psi_2^2\right)^2
\Bigg]\;,
\label{demand2}
\eqa
where we have defined
\bqa
X&=&\mu-\Pi_{11}\;,\\
Y&=&\mu-\Pi_{22}\;,\\
Z&=&-gv^2\;.
\eqa
The propagator that corresponds to the free part of the action is
\bqa\nonumber
D(\omega_n,p)&=&\frac{1}{\omega^2_n
+\epsilon^2(p)}
\left(\begin{array}{cc}
p^2-Y
&\omega_n
 \\ 
-\omega_n
&p^2-X
\end{array}\right)\;,
\eqa
where the dispersion relation is given  by
\bqa
\epsilon(p)&=&\sqrt{\left(p^2-X\right)
\left(p^2-Y\right)}\;.
\label{gaussdisp}
\eqa
The next step is to calculate the thermodynamic potential
in some approximate way including the interaction term~(\ref{demand2})
which consists of three and four-point vertices
together with the subtracted self-energies.
This is done 
by calculating the thermodynamic potential approximately according to
\bqa
\Omega&=&-\mu v^2+{1\over2}gv^4+
{\rm Tr}\log D^{-1}+\langle S_{\rm int}\rangle\;,
\eqa
where $\langle A\rangle$ is the thermal average of the operator $A$.
The thermodynamic potential then
becomes
\bqa\nonumber
\Omega&=&-\mu v^2+{1\over2}gv^4+
{1\over2}{\rm Tr}\log D^{-1}
\\ && \nonumber
+{1\over4}g\left[
\sumint_{P}
{p^2-Y\over\omega_n^2+\epsilon^2(p)}\right]
\sumint_{P}
\left[{p^2-X\over\omega_n^2+\epsilon^2(p)}\right]
\\ && \nonumber
+{3\over8}g\left[
\sumint_{P}
{p^2-Y\over\omega_n^2+\epsilon^2(p)}
\right]^2
+{3\over8}g\left[
\sumint_{P}
{p^2-X\over\omega_n^2+\epsilon^2(p)}
\right]^2
\\ \nonumber
&&
+{1\over2}(3gv^2-\Pi_{11})
\sumint_{P}\left[{p^2-Y\over\omega_n^2+\epsilon^2(p)}\right]
\\ &&
+{1\over2}\left(gv^2-\Pi_{22}\right)
\sumint_{P}\left[{p^2-X\over\omega_n^2+\epsilon^2(p)}\right]\;,
\label{gint}
\eqa
The Feynman diagrams that correspond to the interaction
term in Eq.~(\ref{gint})
are shown in Fig.~\ref{gauss}. The blob denotes an insertion of
either $3gv^2-\Pi_{11}$ or $gv^2-\Pi_{22}$.

\begin{figure}[htb]
\epsfysize=3.0cm
\epsffile{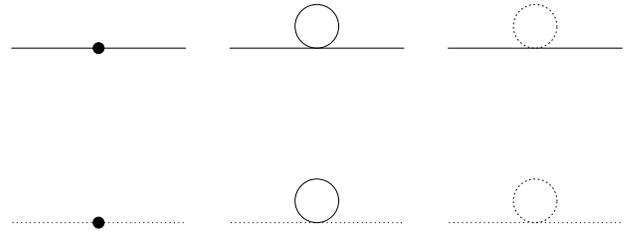}
\caption[a]{Feynman diagrams included in the self-energies $\Pi_{11}$
and $\Pi_{22}$
in the Gaussian approximation.}
\label{sgauss}
\end{figure}

The condensate density $v$ is determined by the 
stationarity condition
\bqa
{\partial\Omega\over\partial v}&=&0\;.
\label{statig}
\eqa
Using the expression~(\ref{gint}) for the thermodynamic
potential, Eq.~(\ref{statig}) becomes
\bqa\nonumber
0&=&-\mu+gv^2+
{3\over2}g\sumint_{P}\left[{p^2-Y\over\omega_n^2+\epsilon^2(p)}\right]
\\ &&
+{1\over2}g\sumint_{P}\left[{p^2-X\over\omega_n^2+\epsilon^2(p)}\right]\;.
\label{min}
\eqa
The self-energies $\Pi_{11}$ and $\Pi_{22}$ are determined 
variationally by demanding that they minimize the free energy:
\bqa
{\partial \Omega\over\partial\Pi_{11}}&=&0\;,
\label{var1}
\\
{\partial \Omega\over\partial\Pi_{22}}&=&0\;.
\label{var2}
\eqa
The equations~(\ref{var1}) and~(\ref{var2}) are often referred to as
gap equations. The solutions are
\bqa\nonumber
\Pi_{11}&=&3gv^2+
{3\over2}g\sumint_{P}\left[{p^2-Y\over\omega_n^2+\epsilon^2(p)}\right]
\\ &&
+{1\over2}g\sumint_{P}\left[{p^2-X\over\omega_n^2+\epsilon^2(p)}\right]\;,
\label{sg1}
\\ \nonumber
\Pi_{22}&=&gv^2+
{1\over2}g\sumint_{P}\left[{p^2-Y\over\omega_n^2+\epsilon^2(p)}\right]
\\ &&
+{3\over2}g\sumint_{P}\left[{p^2-X\over\omega_n^2+\epsilon^2(p)}\right]\;.
\label{sg2}
\eqa
The Feynman diagrams that correspond to the gap 
equations~(\ref{sg1}) and~(\ref{sg2}) are shown in Fig.~\ref{sgauss}.
Inserting~(\ref{sg1}) and~(\ref{sg2}) into~(\ref{gaussdisp})
and using~(\ref{min}) to eliminate the chemical potential, we obtain
the Hartree-Fock-Bogoliubov spectrum~(\ref{hbfd}).

The Gaussian approximation can be made the starting point for a systematic
expansion procedure. This expansion was first formulated by
Okopinska~\cite{oko}.
Stancu and Stevenson~\cite{steve} calculated the leading
corrections to the Gaussian approximation for 
a relativistic $\phi^4$-theory in four dimensions.
The Feynman diagrams that are included in this next-to-leading order
calculations are the three-loop diagrams, the two-loop setting sun diagrams,
the two-loop double bubbles with a single self-energy insertion,
as well as the one-loop diagrams with two insertions of a self-energy.

We next discuss the Bogoliubov and Popov approximations, which are
obtained by making certain approximations in the full HFB approximation.

\subsubsection{Bogoliubov approximation}
The Bogoliubov appoximation amounts to neglecting both the normal and
anomalous averages in Eqs.~(\ref{heos3p})--(\ref{hbfd}).
The chemical potential~(\ref{heos3p}) then reduces to
\bqa
\mu&=&gv^2\;.
\label{bmuu}
\eqa
The spectrum~(\ref{hbfd}) is given by
\bqa
\epsilon(p)=p\sqrt{p^2+2\mu}\;,
\eqa
which is gapless. 
The Bogoliubov approximation has a natural interpretation in terms of
Feynman diagrams. It is obtained by neglecting the two-loop diagrams
contributing to the thermodynamic potential~(\ref{gint}).
The self-energies that follow from the gap equations
are then given by their mean-field values, 
so we have $\Pi_{11}(\omega_n,p)=3gv^2$ 
and $\Pi_{22}(\omega_n,p)=gv^2$. 
The stationarity condition~(\ref{statig}) simply reduces to~(\ref{bmuu}) 
The free energy is obtained by substituting the mean-field
values for the self-energies into the thermodynamic potential
evaluated at the minimum. This yields
\bqa\nonumber
{\cal F}&=&-{\mu^2\over2g}+{1\over2}
\sumint_{P}\log\left[\omega_n^2+\epsilon^2(p)\right]\\ \nonumber
&=&
-{\mu^2\over2g}+{1\over2}I_{0,-1}+
{T\over2\pi^2}\int_0^{\infty} dp\;p^2\log\left[1-e^{-\beta\epsilon(p)}\right]\;.\\ &&
\label{fbbb}
\eqa
In the Bogoliubov approximation, one assumes that most particles
are in the zero-momentum state.
Clearly, this approximation is only valid at very low temperatures where
one can ignore the thermal depletion of the condensate.

One can expand the free energy~(\ref{fbbb}) about zero temperature.
At sufficiently low temperatures, the thermodynamics is dominated by 
the phonon part of the spectrum. 
Using Eqs.~(\ref{freedr}) and~(\ref{ftt1}), 
the free energy reduces in the limit $T\ll2\mu$ to
\bqa
{\cal F}&=&-{\mu^2\over2g}
\left[1-{4\sqrt{2\mu g^2}\over15\pi^2}\right]
-{\pi^2T^4\over90(2\mu)^{3/2}}\;.
\label{fbog}
\eqa
The other thermodynamic functions follow from the free energy~(\ref{fbbb}).
In the low-temperature limit, the number density becomes
\bqa
n&=&{\mu\over g}\left[1-{\sqrt{2\mu g^2}\over3\pi^2}\right]
+{\pi^2T^4\over30(2\mu)^{5/2}}\;.
\label{nofmu}
\eqa
Inverting~(\ref{nofmu}), we obtain the chemical potential
in the limit $T\ll2gn$:
\bqa
\mu&=&gn\left[1+{\sqrt{2ng^3}\over3\pi^2}\right]
+{\pi^2T^4\over30(2gn)^{5/2}}
\eqa
Similarly, for temperatures $T\ll2gn$, the equilibrium energy density is
\bqa
{\cal E}&=&
4\pi an^2\left[
1+{128\over15}\sqrt{{na^3\over\pi}}
\right]
+{\pi^2T^4\over30(2gn)^{3/2}}
\;.
\label{ebog}
\eqa
Using Eq.~(\ref{ndef}), one can calculate the total number density as function
of the condensate density and temperature:
\bqa\nonumber
n&=&n_0+{1\over2}\sumint_{P}{p^2+\epsilon^2(p)\over\omega^2_n+\epsilon^2(p)}\\
\nonumber
&=&n_0+{1\over4}\left[I_{1,1}+I_{-1,-1}\right]\\
&&+{1\over4\pi^2}\int_0^{\infty}
dp\;{p^2(p^2+\epsilon^2(p))\over\epsilon(p)}n(\epsilon(p))
\;.
\eqa
In the limit $T\ll2gn_0$, one finds
\bqa
n&=&n_0\left[1+{8\over3}\sqrt{n_0a^3\over\pi}\right]
+{T^2\over24\sqrt{2gn_0}}\;.
\label{nbog}
\eqa 
Eqs.~(\ref{fbog})--(\ref{nbog}) were first obtained by Lee and 
Yang~\cite{leeyang2}. The second term inside the brackets is the
quantum depletion of the condensate that was calculated in
Sec.~\ref{subcon}. The last term is the thermal depletion.

\subsubsection{Popov approximation}
\label{secpop}
In the Popov approximation, one neglects the anomalous average 
in Eqs.~(\ref{heos3p})--(\ref{hbfd}). 
The chemical potential~(\ref{heos3p}) reduces to
\bqa
\mu&=&g\left[v^2+2\langle\tilde{\psi}^*\tilde{\psi}\rangle\right] 
\;.
\label{popmu}
\eqa
Similarly, the spectrum~(\ref{hbfd}) becomes
\bqa
\epsilon(p)&=&p\sqrt{p^2+2gv^2}\;,
\eqa
and is gapless. 
Note that the spectrum formally has the same form as the
Bogoliubov spectrum, but now the condensate density 
depends on the temperature. 
The number density $n$ satisfies
\bqa
n&=&n_0+\langle
\tilde{\psi}^*\tilde{\psi}
\rangle
\;.
\label{popden}
\eqa
Eqs.~(\ref{popmu}) and~(\ref{popden}) constitute the equation of state
for the weakly interacting Bose gas in the Popov approximation.
For a given temperature $T$ and total number density $n$, they must be solved
simultaneously for the condensate density $n_0$ and the chemical potential.
Note in particular that the condensate density has a strong
temperature dependence. For small values of the gas parameter,
the condensate density as a function of temperature
deviates typically only by a few percent from the result~(\ref{3/2})
for the ideal gas.

At $T=0$, the expectation value in Eq.~(\ref{popmu}) is 
suppressed by a factor of $\sqrt{na^3}$ compared to the mean-field term.
This implies that the Popov approximation gives the same results
as the Bogoliubov approximation 
for all thermodynamic quantities up to corrections of order 
the gas parameter $\sqrt{na^3}$. Since the Popov approximation
does not include all corrections of order $\sqrt{na^3}$, it is no more
accurate than the 
Bogoliubov approximation at zero temperature.

The Popov approximation ought to give a good 
description at higher temperatures than the Bogoliubov approximation
since it takes into account the temperature dependence of the condensate.
However, the Popov approximation 
breaks down in a narrow temperature region
around $T_c$. This can be seen from the
fact that the Popov approximation predicts a first-order phase transition
for the weakly interacting Bose gas, while universality arguments based on the
$O(2)$-symmetry tell us that the phase transition is of second 
order~\footnote{In Sec.~\ref{rgapp}, we show that the long-distance properties
of the dilute Bose gas is given by a classical field theory in three
dimensions with an $O(2)$ symmetry. In particular, the infrared
Wilson-Fisher fixed point is that of a three-dimensional $O(2)$ model.
This model is known to have a second order phase transition.}.

The Popov approximation can also be interpreted in terms of Feynman
diagrams. This is done by expressing propagators and interactions
in terms of normal and anomalous self-energies rather than
$\Pi_{11}$ and $\Pi_{22}$. The Feynman diagrams that contribute
to the thermodynamic potential in the HFB approximation are still those
shown in Fig.~\ref{gauss}, but the sum-integrals are
now in terms of normal and anomalous propagators
and the symmetry factors are different.
The Popov approximation for the
thermodynamic potential is then defined by keeping the diagrams
that involve normal propagators and neglecting those that involve
anomalous ones. The corresponding gap equation that follow
from the variational principle, gives the normal self-energy 
as the sum of the mean-field contribution $2gv^2$ and the
one-loop tadpole diagram that involves the normal propagator.
The anomalous self-energy is given by the mean-field contribution 
$gv^2$ alone. This definition has been given before~\cite{grif}.
Finally, the free energy is obtained by substituting the 
expressions for the self-energies into the thermodynamic potential
evaluated at the minimum. The free energy in the Popov approximation
has the Bogoliubov form~(\ref{fbbb}).

\subsubsection{Many-body $T$-matrix and modified Popov approximation}
In Sec.~\ref{eftt}, we calculated the exact scattering amplitude for 
$2\rightarrow2$ scattering in the vacuum. In the limit where
the external momentum goes to zero, the two-body scattering
matrix goes to a constant.
In the next section, we will show using renormalization group methods
that the effective coupling constant for a weakly interacting Bose gas
is temperature dependent, and in particular that it vanishes at
the critical temperature. This behavior is expected at a second-order phase
transition where the correlation length goes to infinity 
v(the effective chemical potential or the effective mass goes to zero)
due to the fact that $\phi^4$-theory is a trivial theory.
It is therefore important to improve on the Popov theory by using an 
effective temperature-dependent coupling constant.
This can be done by using the many-body $T$-matrix.
The many-body $T$-matrix takes medium effects into account
by summing repeated two-body scattering processes of 
quasi-particles in the gas rather than in the vacuum. 
This is done by calculating the diagrams in Fig.~\ref{loopfig}
at finite temperature using the propagator~(\ref{prop2}).
The many-body $T$-matrix $T^{\rm MB}(k,k ,K;z)$
depends on the relative momenta 
of the two atoms ${\bf k}$ and ${\bf k}$ before and
after the collision, and the total center-of-mass momentum ${\bf K}$
and center-of-mass energy $z$.
In the following, we neglect this energy and momentum dependence which 
is a good approximation~\cite{grif,sb}.

At zero external momentum ${p}$, 
the one-loop diagram in Fig.~\ref{loopfig} is
\bqa\nonumber
{\cal T}_1(0)&=&2g^2
\sumint_P{1\over\omega_n^2+\epsilon^2(p)} \\
&=&
gI_{0,1}+
g{1\over\pi^2}\int_0^{\infty}dp\;{p^2\over\epsilon(p)}n(\epsilon(p))\;.
\eqa
By summing the geometric series corrresponding to the
diagrams shown in Fig.~\ref{loopfig},
one finds~\cite{grif,sb}
\bqa
T^{\rm MB}(0)&=&{2g^2\over2g-{\cal T}_1(0)}\;.
\label{mt}
\eqa
We next consider the many-body $T$-matrix at low temperature.
For $T\ll2gn_0$, we obtain
\bqa
T^{\rm MB}(0)&=&g\left\{
1+{\sqrt{2n_0g^3}\over4\pi^2}
\left[1
-{\pi^2\over6}\left({T\over n_0g}\right)^2
\right]
\right\}\;.
\eqa
We note in particular that at $T=0$, the many-body $T$-matrix reduces to $g$
up to corrections of order $\sqrt{na^3}$. 

The many-body $T$-matrix has been used to obtain an improved approximation
from the Gaussian approximation~\cite{sb}.
The normal and anomalous self-energies can be easily calculated from 
Eqs.~(\ref{sg1}) and~(\ref{sg2}). We obtain 
\bqa\nonumber
\Sigma_{11}&=&2gv^2+g\sumint_P
\left[{p^2-Y\over\omega_n^2+\epsilon^2(p)}\right] 
+g\sumint_P\left[{p^2-X\over\omega_n^2+\epsilon^2(p)}\right] \\
&=&2g\left[v^2+\langle
\tilde{\psi}^*\tilde{\psi}
\rangle\right]\;,\\ \nonumber
\Sigma_{12}&=&gv^2+
g\sumint_P\left[{p^2-Y\over\omega_n^2+\epsilon^2(p)}\right] 
-g\sumint_P\left[{p^2-X\over\omega_n^2+\epsilon^2(p)}\right] \\ \nonumber
&=&gv^2+{1\over2g}\Sigma_{12}{\cal T}_1(0)\;.
\eqa
The last equation can be easily solved for $\Sigma_{12}$. Using
(\ref{mt}), we obtain
\bqa
\Sigma_{12}
&=&T^{\rm MB}(0)v^2
\label{solving}
\;.
\eqa
Thus the normal self-energy $\Sigma_{11}(0,0)$
is given in the Hartree-Fock or in the one-loop approximation.
while the anomalous self-energy $\Sigma_{12}(0,0)$
is given in the many-body $T$-matrix 
approximation.
This observation motivated Bijlsma and Stoof~\cite{henk} to redo the
calculation using the many-body $T$-matrix as an effective interaction.
This gives both the normal and anomalous self-energies in the 
many-body $T$-matrix approximation:
\bqa
\Sigma_{11}&=&2T^{\rm MB}(0)
\left[v^2+\langle\tilde{\psi}^*\tilde{\psi}\rangle\right]\;,
\label{sg11}
\\ 
\Sigma_{12}&=&T^{\rm MB}(0)v^2\;.
\label{sg12}
\eqa
These modified self-energies define a modified Gaussian approximation,
which was used to investigate the thermodynamic
properties of the homogeneous Bose gas in two and three dimensions.
In three dimensions, this approach yields a second order phase transition,
but the critical temperature is the same as that of an ideal gas.
At the same time, the effective coupling constant, which is precisely 
the many-body $T$-matrix, vanishes at the temperature given by 
Eq.~(\ref{tc0})~\cite{grif}. 
However, it turns out that the Hugenholz-Pines
theorem is not always satisfied. At very low temperature, it can be 
shown~\cite{henk} that the value of $v^2$ that minimizes the thermodynamic
potential does not exactly correspond to the condition
$\mu=\Sigma_{11}(0,0)-\Sigma_{12}(0,0)$.

The self-energies~(\ref{sg11}) and~(\ref{sg12}) can also be obtained from the
HFB equations~(\ref{a3}) and~(\ref{a4}) by
neglecting the anomalous average and replacing the 
coupling constant $g$ by the many-body $T$-matrix.
By making the substituting $g\rightarrow T^{\rm MB}$ and neglecting the
anomalous average in the remaining 
equations that define the HFB approximation,
a gapless approximation was 
obtained~\cite{ext1,ext2}~\footnote{It is {\it not} gapless
in the sense that the value of the condensate $v$ that minimizes the effective
potential coincides with $\mu=\Sigma_{11}(0,0)-\Sigma_{12}(0,0)$, as
explained above.}
Eqs.~(\ref{heos3p}), ~(\ref{heos3}) and~(\ref{hbfd}) now become
\bqa
\label{mod1}
0&=&-\mu+T^{\rm MB}\left[v^2+2\langle\tilde{\psi}^*\tilde{\psi}\rangle
\right]\;,\\ \nonumber
{\partial\tilde{\psi}\over\partial\tau}&=&
-\left[\nabla^2+\mu\right]\tilde{\psi}+T^{\rm MB}
\left[
2\left(v^2+\langle\tilde{\psi}^*\tilde{\psi}\rangle\right)\tilde{\psi}
\right.\\ &&\left.
+v^2\tilde{\psi}^*\right]\;,
\label{mod2}\\ 
\epsilon^2(p)&=&p^2\left[p^2+2T^{\rm MB}v^2\right]\;.
\label{mod3}
\eqa
The modified mean-field approximation does not reproduce known perturbative
results at $T=0$. For instance, the prediction for the correction to the
Bogoliubov spectrum at long wavelengths differs from Eq.~(\ref{pre}).
Thus at $T=0$, this approximation is no more accurate than the 
Bogoliubov or Popov approximations.

 \subsection{Other Variational Approaches}
In this subsection, we discuss other 
variational approaches to the weakly interacting
Bose gas. The idea is to define a thermodynamic potential $\Omega$ that 
depends on a set of variational parameters $a_i$. The free energy and other
thermodynamic variables are then given by the thermodynamic potential
and its derivative evaluated at the variational minimum 
$\partial\Omega/\partial a_i=0$. A variational method can be successful
only if the essential physics can be captured by the variational 
parameters.

\subsubsection{$\Phi$-derivable approach}
The $\Phi$-derivable approximation is an approach
in which the full propagator serves as an infinite set of varitional 
parameters. It was first formulated by 
Luttinger and Ward~\cite{lut}, and by Baym~\cite{baym} for nonrelativistic
fermions, and later generalized to relativistic field theories by 
Cornwall, Jackiw, and Tomboulis~\cite{CJT-74}. 
A property of the $\Phi$-derivable approximation is that it is
conserving, which means that it respects the conservation laws that
follow from the global symmetries of the system.

We next apply the $\Phi$-derivable approach to the dilute Bose gas.
We will rederive some of the results obtained by Lundh and Rammer~\cite{lundh},
but the formulation is somewhat different.

The $\Phi$-derivable thermodynamic potential $\Omega[D]$ has the form
\bqa\nonumber 
\Omega[D]&=&-\mu v^2+{1\over2}gv^4+
{1\over2}{\rm Tr}\log D^{-1}
\\ &&-{1\over2}{\rm Tr}\Pi D+\Phi[D]
\label{phie}
\;,
\eqa
where $\Pi$ is the exact self-energy, 
$D$ is the exact propagator, and the interaction potential
$\Phi[D]$ is the sum of all 
{\it two-particle irreducible (2PI) vacuum diagrams}. 
These are diagrams that do not fall apart by cutting two propagator
lines.
It is understood that both
$\Pi$ and $D$ are $2\times2$ matrices.
${\rm Tr}$ denotes the trace in configuration space.
The condensate $v$ is found by minimizing the thermodynamic potential 
in the usual way
\bqa
{\partial\Omega[D]\over\partial v}&=&0\;.
\label{vardv}
\eqa
Similarly,
the variational principle requires
that the thermodynamic potential be stationary under variations of the
full propagator at fixed $D_0$. This can be expressed as
\bqa
{\partial\Omega[D]\over\partial D}&=&0\;.
\label{fulll}
\eqa
If we denote the free propagator by $D_0$,
the Schwinger-Dyson equation for the exact propagator $D$ can be written as
\bqa
D^{-1}=D_0^{-1}+\Pi\;.
\label{sd}
\eqa
Using Eq.~(\ref{sd}), we can rewrite Eq.~(\ref{fulll}) 
\bqa
{\partial\Phi[D]\over\partial D}&=&{1\over2}\Pi
\label{peq}
\;.
\eqa
Eq.~(\ref{peq}) for the self-energy cannot be solved exactly, but one 
can resort to some systematic approximation.
The $n$-loop $\Phi$-derivable approximation
is such an approximation. It is defined by 
keeping all 2PI diagrams
up to $n$ loops in the thermodynamic potential.
Differentiation with respect to the components of the
propagator is equivalent to cutting the corresponding lines in the
Feynman diagrams. Thus the gap equations for the self-energy
in the $n$-loop $\Phi$-derivable approximation
contain all 2PI diagrams up to $n$-1 loops.
It leads to 
a set of integral equations for the self-energies
that generally are extremely difficult to
solve. In certain simple cases, where the self-energies are momentum
independent, one can solve these equations. 

\begin{figure}

\hspace{-1cm}
\psfig{figure=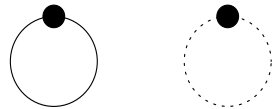,width=3.5cm,height=1.4cm}
\caption{
Vacuum diagrams contributing to the one-loop $\Phi$-derivable approximation
$\Phi_1[D]$.}
\label{111loop}
\end{figure}

Consider the one-loop $\Phi$-derivable approximation. 
The diagrams that contribute to $\Phi_1[D]$ are shown in Fig.~\ref{111loop} and
it reads
\bqa\nonumber
\Phi_1[D]&=&
{3\over2}gv^2\sumint_{P}{p^2-Y\over\omega_n^2+\epsilon^2(p)}
+{1\over2}gv^2\sumint_{P}{p^2-X\over\omega_n^2+\epsilon^2(p)}\;.
\\ &&
\eqa
The self-energies are given by
\bqa
\Pi_{11}&=&3gv^2\;,
\label{sf1}\\
\Pi_{22}&=&3gv^2\;,
\label{sf2}
\eqa
while $\Pi_{12}$ and $\Pi_{21}$ both vanish. Thus the self-energies are
those of the Bogoliubov approximation.
We next substitute the self-energies~(\ref{sf1}) and~(\ref{sf2}) into
the thermodynamic potential~(\ref{phie}). It turns out that the
two terms $-{1\over2}{\rm Tr}\Pi D$ and $\Phi_1[D]$ cancel each other
and the one-loop $\Phi$-derivable approximation for the free energy
reduces to the Bogoliubov approximation~(\ref{fbbb}).

\begin{figure}

\hspace{-1cm}
\psfig{figure=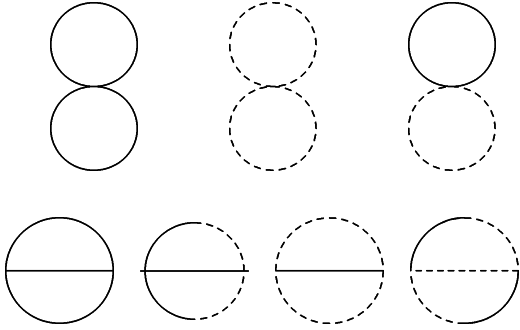,width=10.5cm,height=15cm}
\vspace{-9.5cm}
\caption{
Vacuum diagrams contributing to the two-loop $\Phi$-derivable approximation
$\Phi_2[D]$.}
\label{2loop}
\end{figure}

The two-loop $\Phi$-derivable approximation $\Phi_2$ is very complicated.
The two-loop diagrams that contribute to the thermodynamic potential are
shown in Fig.~\ref{2loop}. The corresponding equation for the 
self-energy matrix~(\ref{peq}) is obtained by cutting the lines in the
diagrams. The diagrams that are first order in the interaction are
momentum independent and can be easily calculated. The diagrams that are
second order in the interaction are momentum dependent. The difficult momentum
dependence comes from the self-energy in the propagators. This leads to 
intractable integral equations for the self-energies. 
There have been attempts to simplify the two-loop $\Phi$-derivable
approximation by making the {\it ansatz} that the self-energy is 
a momentum-independent mass term~\cite{camel}. However, such approximations
lead to a mass gap and hence violate the Hugenholz-Pines theorem.

Finally, we notice that if one neglects the setting sun diagrams in
Fig.~\ref{2loop}, 
the two-loop $\Phi$-derivable approximation reduces to the 
Gaussian approximation that was discussed in detail in the previous subsection.

\subsubsection{Optimized perturbation theory}
\label{subopt}
The complexity  of the $\Phi$-derivable approach calls for the use of
simpler variational approaches. One introduces a finite number of variational
parameters. In its simplest form, one introduces a single variational parameter
$m$ or $\mu$. In a relativistic field theory, $m$ is a variational mass 
parameter, while in a nonrelativistic field theory, $\mu$ is a
variational chemical potential. 
It was first formulated by Yukalov~\cite{yukyuk}.
The method 
can be extended to 
include variational coupling constants and a variational
kinetic energy term~\cite{chick2}.
At this point, it is important to emphasize that the parameters
introduced in optimized perturbation theory are completely arbitrary and
that one needs a prescription for them, in order to complete a calculation.
One prescription is the principle of minimal sensitivity (PMS), where one 
requires that the parameters satisfy a stationarity condition. For instance,
one can demand 
that the free energy be stationary with respect to variations of these
parameters. Another criterion is the principle of fastest apparent
convergence (FAC). This condition requires that the difference between
a physical quantity calculated at two different loop orders be as small
as possible.

The starting point is the action~(\ref{s1}).
We rewrite the corresponding Lagrangian by introducing
an effective chemical potential
$\mu_1$, an effective coupling constant $\lambda$,  
and a counting variable $\delta$ in the following manner:
\bqa\nonumber
{\cal L}&=&
\psi^*\left[{\partial\over
\partial \tau}
-\nabla^2-\mu_1-\delta(\mu-\mu_1)\right]\psi
\\ &&
+{1\over2}\lambda(\psi^*\psi)^2+{1\over2}\delta(g-\lambda)(\psi^*\psi)^2
+...\;.
\label{gene}
\eqa
If we set $\delta=1$, we immediately recover the Lagrangian
that corresponds to the action~(\ref{s1}). 
Optimized perturbation
theory is defined by the power counting rule that $\delta$ be of the order 
$\lambda\sim g$. We carry out calculations in powers of $\lambda$
and at the end of the calculations, we set $\delta=1$.
It is important to note that the power counting rules also should be applied
to the counterterms.
The ultraviolet divergences that appear in optimized perturbation theory are
removed by the counterterms determined in the standard loop expansion
for perturbation theory at $T=0$. 

The introduction of $\mu_1$ and $\lambda$ represents a reorganization of the
perturbative series, which 
is a selective resummation of higher order graphs. 
But it is important to emphasize that there is no overcounting
of diagrams. Every diagram is counted once with the correct symmetry
factor.
If calculated to all
orders, the results for physical quantities
would be independent of these parameters. However,
at any finite order in perturbation theory they do
depend on these parameters.

Performing the shift~(\ref{shift}), the
free propagator now takes the form
\bqa
D(\omega_n,p)&=&\frac{1}{\omega^2_n
+\epsilon^2(p)}\left(\begin{array}{cc}
p^2-Y&\omega_n \\
-\omega_n&p^2-X
\end{array}\right)\;,
\label{propfree}
\eqa
where 
\bqa
X&=&\mu_1-3\lambda v^2\;,\\
Y&=&\mu_1-\lambda v^2\;,\\
\epsilon(p)&=&\sqrt{(p^2-X)(p^2-Y)}
\label{epep}
\;.
\eqa
The mean-field thermodynamic potential is
\bqa
\Omega_0&=&-{\mu}_1v^2+{1\over2}\lambda v^4
\;.
\label{v0}
\eqa
The one-loop contribution to the thermodynamic potential reads
\bqa\nonumber
\Omega_1&=&-\delta\left(\mu-\mu_1\right)v^2+{1\over2}\delta(g-\lambda)v^4
\\ &&
+{1\over2}\sumint_{P}\log\left[\omega_n^2+(p^2-X)(p^2-Y)\right]\;.
\label{v1}
\eqa
The one-loop thermodynamic potential is then given by the sum of 
Eqs.~(\ref{v0}) and~(\ref{v1}). Setting $\delta=1$, we obtain
\bqa\nonumber
\Omega_{0+1}&=&-{\mu}v^2+{1\over2}gv^4
\\ &&
+{1\over2}\sumint_{P}\log\left[
\omega_n^2+(p^2-X)(p^2-Y)
\right]
\;.
\label{v01}
\eqa
At this point, we would like to discuss the Goldstone theorem in connection
with optimized perturbation theory. There is some confusion in the literature
whether OPT violates Goldstone's theorem. Depending on the choice of the
parameters $\mu_1$ and $\lambda$, the mean-field dispersion relation
$\epsilon(p)$ in Eq.~(\ref{epep})
may or may not be gapless when evaluated at the minimum of the
effective potential. Beyond the mean-field approximation, the 
dispersion relation is, however, not given by Eq.~(\ref{epep}), but rather
by
\bqa
\det\left[D_0^{-1}(\omega_n,p)-\Pi(\omega_n,p)\right]&=&0\;,
\label{colll2}
\eqa
where $D_0(\omega,p)$ is the propagator~(\ref{propf})
and
$\Pi(\omega_n,p)$ is the $2\times2$ self-energy matrix.
The solution to Eq.~(\ref{colll2}) is gapless order by order in 
optimized perturbation theory. We can easily check that at the one-loop
level. Differentiating the effective potential~(\ref{v01}) with respect
to the condensate and demanding that $v$ be a stationary point yields
\bqa\nonumber
0&=&-\mu+gv^2+{1\over2}\lambda\sumint_{P}{3(p^2-Y)+(p^2-X)
\over\omega_n^2+\epsilon^2(p)}\\
&=&-\mu+\Pi_{22}(0,0)\;,
\label{gaplesso}
\eqa
where $\Pi_{22}(0,0)$ is imaginary-time version of~(\ref{p3})
and corresponds to the
self-energy diagrams displayed in Fig.~\ref{s222}. 
Eq.~(\ref{gaplesso}) ensures that Goldstone's theorem is satisfied
at one loop.

We next discuss various choices of the parameters $\mu_1$ and $\lambda$.
Although they in principle are arbitrary, some choices are better motivated
from a physics point of view than others. One requirement one might
impose is that the results reduce in the limit
$T\rightarrow0$ to the results one obtains by applying the
perturbative framework discussed in the previous section.
One very simple choice is
\bqa
\mu_1&=&gv^2\;,
\label{fv} \\ 
\lambda&=&g
\;.
\eqa
This choice leads to the dispersion relation~(\ref{disps}).
Another choice is motivated by the principle of minimal sensitivity,
where one demands that the parameters satisfy 
\bqa
{\partial \Omega\over\partial\mu_1}=0\;,
\label{triv1}
&& \\
{\partial \Omega\over\partial\lambda}=0\;.
\label{triv2}
\eqa
At the one-loop level, the only solution to Eqs.~(\ref{triv1})--(\ref{triv2})
is the trivial solution $\mu_1=\lambda=0$~\cite{chick2}.
Only at higher orders are there nontrivial solutions.
A very promising choice is 
\bqa
\mu_1&=&\mu\;,
\label{cg1}\\
\lambda&=&T^{\rm MB}\;.
\label{cg2}
\eqa
The many-body $T$-matrix has several desirable features. Up to corrections
of order $\sqrt{na^3}$, it reduces to the $s$-wave
scattering length at $T=0$.
Thus the choice~(\ref{cg1})--(\ref{cg2}) will reproduce the
results for the weakly interacting Bose gas at $T=0$.
Furthermore, it also takes into account medium effects
by summing repeated two-particle scattering in the gas.

Relativistic $\phi^4$-theory at finite temperature has been studied 
using OPT~\cite{CK-98,chick2}.
The one-loop calculation that was carried out~\cite{CK-98} 
predicts a first-order phase 
transition, while the two-loop calculation~\cite{chick2} 
predicts a second-order phase transition.
Thus it is very likely that 
a two-loop calculation using optimized
perturbation theory is capable of describing correctly this important aspect of
the phase transition, while incorporating the Goldstone theorem.
Secondly, OPT provides an ideal framework for calculating the
energy shifts and damping rates of collective excitations of
a Bose-Einstein condensate.
Clearly, however, more work is needed. 

In the paper by Haugset {\it et al.}~\cite{haug}, 
the authors construct an improved 
one-loop thermodynamic potential
by including the self-energy $\Pi_{22}(0,0)$ in the propagator. 
The calculation of the improved one-loop effective 
potential is equivalent to
summing up all the ring diagrams, except that the two-loop diagram is 
counted twice. One must then subtract by hand
the term that was overincluded.
From the viewpoint of OPT, this overcounting comes
about because the one-loop diagram with a self-energy insertion was omitted
(in addition to the setting sun diagrams). Thus a consistent power
counting according to the rules above is necessary to avoid 
problems with overcounting of Feynman diagrams.

\subsection{Renormalization Group Approach}\label{rgapp}
One very powerful method of quantum field theory is the Wilson
renormalization group (RG)~\cite{wilson,joe}. 
The basic idea is to separate the momentum modes in the path integral
into fast modes and slow modes by a cutoff. One then integrates out the
fast modes. This yields an effective action for the
slow modes, in which the coefficients of the original operators are 
renormalized and new operators are induced.
By lowering the cutoff infinitesimally, one obtains a set of differential
equations for the parameters in the effective action. Integrating out all
modes down to $k=0$ yields the full effective action.

Renormalization group techniques have been applied
to the homogeneous Bose gas at finite temperature
by several authors~\cite{henk,jmike,alber}. The first quantitative study
was carried out by Bijlsma and Stoof.
In that paper, they considered the one-loop diagrams that contribute to the 
effective chemical potential, effective 
four-point vertex etc. By introducing a cutoff,
as explained above, a set of coupled differential equations was derived.
By solving these equations numerically, they obtained 
the condensate
density as a function of the temperature and the critical
temperature as a function of the $s$-wave scattering length.
They also derived equations for the fixed points and
calculated critical exponents.

A somewhat different renormalization group approach was used 
by Andersen and Strickland.~\cite{jmike}. 
It is based on a derivative expansion of the
effective action $\Gamma[v]$~\cite{morris22}, 
which is obtained by integrating out the
quantum and thermal fluctuations. If one imposes an infrared cutoff $k$,
one can expand the corresponding effective action $\Gamma_{k}[v]$
as
\bqa\nonumber
\Gamma_{k}[v]&=&\int_0^{\beta}d\tau\int d^dx 
\left\{V_{k}(v)+
{1\over2}iZ^{(1)}_{k}(v)\epsilon_{ij}v_i{\partial\over\partial\tau}v_j
\right.\\ &&\left.
+
{1\over2}Z^{(2)}_{k}(v)(\nabla v_i)^2
+\ldots
\right\}\;,
\label{ea}
\eqa
where $i,j=1,2$, and repeated indices are summed over.
$V_k$ is the effective potential and 
$Z_k^{(1)}$ and $Z_k^{(2)}$ are
wavefunction normalization constants.
$\epsilon_{ij}$ is the Levi-Civita symbol and $v_i$ is the $i$th component
of condensate $v$. 
The dots indicate all higher order terms in the derivative expansion.
Here and in the following, the subscript $k$ indicates a dependence on 
the infrared cutoff.
By lowering the cutoff $k$, one obtains a set of coupled integral
equations for the functions $V_k$, $Z_k^{(1)}$, $Z_k^{(2)}\cdots$.
The leading order 
in the derivative expansion is defined by setting 
the coefficients $Z_{k}^{(1)}$ and $Z_{k}^{(2)}$ to unity
and the coefficients of all higher derivative operators in Eq.~(\ref{ea})
to zero. This is called the local potential approximation.
In the following, we restrict ourselves to the
local potential approximation and 
derive a flow equation for the effective potential $V_k(v)$.

\subsubsection{One-loop effective potential}
\label{1lrg}
We are now ready to calculate quantum 
and thermal corrections to the classical potential.
We compute the one-loop effective potential which we will
``RG improve'' in the Section~\ref{rgrgrg}. 
This method of deriving RG flow equations
is conceptually and technically simpler than the direct application of 
exact or momentum-shell RG techniques~\cite{jmike}.

\begin{figure}[]
\epsfysize=5cm
\epsffile{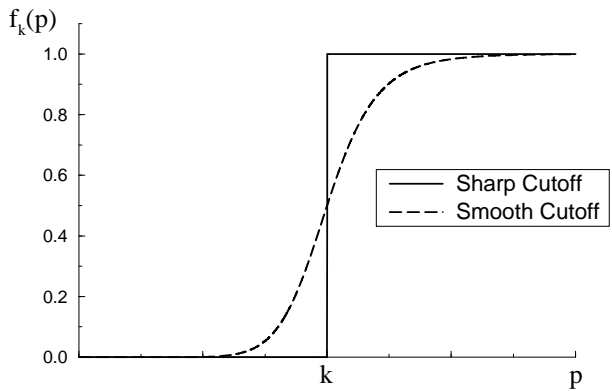}
\caption[a]{Sharp blocking function (solid line) and a typical 
smooth blocking function (dashed line).
}
\label{cut}
\end{figure}

The one-loop effective potential reads
\bqa\nonumber
V&=&V_0+{1\over2}\mbox{Tr}\log D^{-1}(\omega_n,p)\\
&=&-\mu v^2+{1\over2}gv^4
+{1\over 2}\sumint_{P}
\log\left[\omega_n^2+\epsilon^2(p)\right]\;,
\label{efpot1}
\eqa
where the mean-field potential has been denoted by 
$V_0=-\mu v^2+{1\over2}gv^4$.
We proceed by dividing the modes in the path integral into slow
and fast modes
separated by an infrared cutoff $k$. This is done by introducing a cutoff
function $R_k(p)$ or regulator, which we keep general for the moment. 
We add the term
\bqa\nonumber
S_{k}[\psi_1,\psi_2]&=&-
\int_{0}^{\beta}d\tau\int d^dx\;
\mbox{$1\over2$}R_k(\sqrt{-\nabla^2})
\\&&
\hspace{-0.7cm}
\times
\left[\psi_1\nabla^2\psi_1+\psi_2\nabla^2\psi_2
\right]
\;,
\label{eterm}
\eqa 
to the action Eq.~(\ref{s1}). The argument 
$p$ of the function $R_k(p)$ has been replaced
by $\sqrt{-\nabla^2}$ in coordinate space.
After performing the shift~(\ref{shift}),
the modified propagator reads
\bqa\nonumber
&&D_k(\omega_{n},p)={1\over\omega_{n}^2+\epsilon^2_k(p)}
\\&&
\hspace{-0.5cm}
\times
\left(\begin{array}{cc}
p^2\left(R_k(p)+1\right)+V_0^{\prime}&\omega_{n}
\vspace{0.2cm}
\\ \nonumber
-\omega_{n}&
\hspace{-0.5cm}
p^2(R_k(p)+1)+V_0^{\prime}+2V_0^{\prime\prime}v^2
\end{array}\right)\;,
\\ &&
\eqa
where a prime on $V_0$ denotes differentiation with respect to $v^2$.
The modified dispersion relation is
\bqa\nonumber
\epsilon_k(p)&=&
\sqrt{\big[p^2(R_k(p)+1)+V_0^{\prime}\big]}
\\ &&\times
\sqrt{\big[p^2(R_k(p)+1)+V_0^{\prime}+2V_0^{\prime\prime}v^2\big]}
\;.
\label{cla2}
\eqa
By a judicious choice of $R_k(p)$, 
we can suppress the low momentum modes in the path integral
and leave the high-momentum modes essentially
unchanged. 
It is useful to introduce a blocking function 
$f_k(p)$ which is defined by 
\bqa
R_k(p)={1-f_k(p)\over f_k(p)}\;.
\eqa\\
The blocking function satisfies
\bqa
\lim_{p\rightarrow 0}f_k(p)=0\;,\hspace{1cm}
\lim_{p\rightarrow \infty}f_k(p)=1\;.
\eqa
These properties ensure that the low--momentum modes are suppressed by making
them very heavy and the high--momentum modes are left essentially unchanged.
Typical blocking functions are shown in Fig.~\ref{cut}.

The sharp cutoff function $R_k(p)$ is defined by the blocking 
function $f_k(p)=\theta (p-k)$, and 
is shown in Fig.~\ref{cut} (solid line). 
It provides a sharp separation between fast and slow modes.
Using the sharp cutoff, the slow modes become 
completely suppressed in the path integral, while the fast
modes are completely unaltered. The advantage of using a sharp cutoff
is that certain integrals can be done analytically and the 
integro-differential RG--equation is reduced to a 
differential RG--equation.

The solutions to approximate renormalization group equations (e.g. the
truncation of the derivative expansion at some finite order) depends
on the regulator function $R_k(p)$. There have been several papers
on the optimal choice of the cutoff function~\cite{jmike,daniel1,daniel2}.
Here we consider the class of smooth blocking functions
\bqa
f^m_k(p)&=&{p^m\over p^2+k^m}\;.
\label{smooth1}
\eqa
In the limit $m\rightarrow\infty$, we recover the sharp blocking function.
In Sec.~\ref{critt}, we shall see that a smooth regulator is 
better than the sharp one, but it comes at the expense of more
complicated numerics.

We return to the one--loop effective potential
in~(\ref{efpot1}). Using the inverse propagator $D^{-1}_k(\omega_n,p)$,
the modified one-loop effective potential becomes
\bqa
V_k&=&V_0+{1\over2}\sumint_P
\log \left[\omega_n^2+\epsilon_k^2(p)\right]\;.
\label{modi}
\eqa
Upon summation over the Matsubara frequencies, we obtain
\bqa\nonumber
V_k&=&
V_0+
\int{d^dp\over(2\pi)^d}\bigg\{\frac{1}{2}\epsilon_k(p)+
T\log\left[1-e^{-\beta\epsilon_k(p)}\right]\bigg\}\;.\\
&&
\label{oneloop}
\eqa
The first term in the brackets is the $T=0$ part
and represents the zero--point fluctuations. The second term includes thermal effects. 
Differentiation with respect to the infrared cutoff $k$ yields:
\bqa\nonumber
k{\partial\over \partial k}V_{k}&=&-
k\int {d^dp\over (2\pi)^d}
\left({\partial R_k(p)\over\partial k}\right)
{1\over2\epsilon_k(p)}\left[1+
2n(\epsilon_k(p))\right]
\\ &&\times
\left[p^2(R_k(p)+1)
+V_0^{\prime}+V_0^{\prime\prime}v^2
\right]\;.
\label{rg1}
\eqa
Eq.~(\ref{rg1}) is an integro-differential equation for the one--loop 
effective potential. 
It is obtained by integrating out each mode independently, 
where the feedback from the fast modes to the slow modes is completely
ignored. Since all modes are integrated out independently, 
this is sometimes called the independent
mode approximation~\cite{mike1}.
The lack of feedback leads to a poor tracking of the effective 
degrees of freedom. The situation is remedied by applying the
renormalization group, which effectively sums up important classes
of diagrams.

\subsubsection{Renormalization group improvement}
\label{rgrgrg}

The easiest way of deriving the RG--improved version of Eq.~(\ref{rg1})
is simply to make it self-consistent by replacing everywhere $V_0$
by $V_k$. This yields:
\bqa\nonumber
k{\partial \over \partial k}V_k
&=&-
k\int {d^dp\over (2\pi)^d}
\left({\partial R_k(p)\over\partial k}\right)
{1\over2\epsilon_k(p)}\left[1+
2n(\epsilon_k(p))\right]
\\ &&\times
\left[p^2(R_k(p)+1)
+V_k^{\prime}+V_k^{\prime\prime}v^2
\right]\;,
\label{rg2}
\eqa
where the RG--improved dispersion relation is
\bqa\nonumber 
\epsilon_k(p)&=&
\sqrt{\left[p^2(R_k(p)+1)+V^{\prime}_k\right]}
\\ &&\times
\sqrt{\left[p^2(R_k(p)+1)+V^{\prime}_k
+2V^{\prime\prime}_kv^2\right]}
\;,
\label{ndisp}
\eqa
and the primes in Eqs.~(\ref{rg2}) and~(\ref{ndisp})
again denote differentiation with respect to $v^2$.
The self--consistent Eq.~(\ref{rg2}) is not a perturbative approximation,
but is exact to leading order in the derivative expansion. 
This equation has been rigorously derived
using the path integral representation and the derivative expansion
of the effective action.~\cite{jmike}. There are other ways of regulating
the one-loop expression(~\ref{efpot1}) and subsequently ``RG-improve'' it.
However, they do not always resum perturbation theory correctly
(with all higher order diagrams included and all combinatorical 
coefficients correct). These issues have been studied in detail by
Litim and Pawlowski~\cite{litp}.

Note that since $V_{k=0}^{\prime}$ vanishes at the minimum of the effective
potential, the dispersion relation in the broken phase reduces to
\bqa
\epsilon_{k=0}(p)&=&p\sqrt{p^2+2V_{k=0}^{\prime\prime}v^2}\;.
\eqa
Thus, the Goldstone theorem is automatically satisfied for temperatures
below $T_c$.

In the calculations that follow, we will restrict ourselves to 
using the sharp cutoff function.
The integral
over $p$ in Eq.~(\ref{rg2}) can be done analytically, resulting
in a differential RG--equation. In this case, Eq.~(\ref{rg2}) 
reduces to
\bqa
k\frac{\partial}{\partial k}V_k=
-{1\over2}S_dk^d\Bigg\{\epsilon(k)+
2T\log\left[1-e^{-\beta\epsilon(k)}\right]\Bigg\}\;,
\label{rg}
\eqa
where
\bqa
S_d&=&{\Omega_d\over(2\pi)^d}\;,
\eqa
and $\Omega_d$ is the area of a $d$--dimensional sphere whose
expression is given in the appendix.

Since the factor ${\partial R_k(p)/\partial k}$
forces $p=k$, we have defined $\epsilon(k)=\epsilon_k(p=k)$,
where the dispersion relation $\epsilon(k)$ is
\bqa
\epsilon(k)&=&\sqrt{\left[k^2+V^{\prime}_k\right]
\left[k^2+V^{\prime}_k+2V^{\prime\prime}_kv^2\right]}\;.
\eqa

In order to solve~(\ref{rg}), one must impose the correct
boundary condition
on the effective potential $V_k$. For $k=\infty$, no modes have been
integrated out and $V_k$ reduces to the classical potential $V_0$.
In practise, one must impose the boundary condition at a large but finite
value $k=\Lambda$.

\begin{figure}[htb]
\epsfysize=7cm
\epsffile{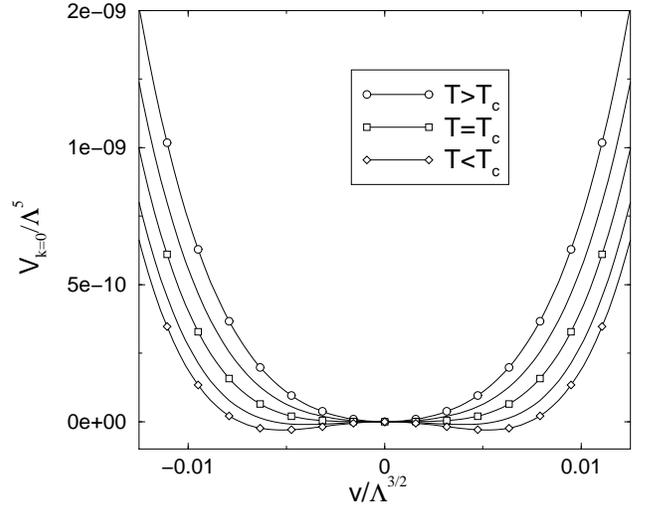}
\caption[a]{
Real part of the RG--improved effective potential $V_{k=0}(v)$ 
for various values of the temperature. The phase transition is clearly
second order.
}
\label{fig2}
\end{figure}

\begin{figure}[htb]
\epsfysize=5cm
\epsffile{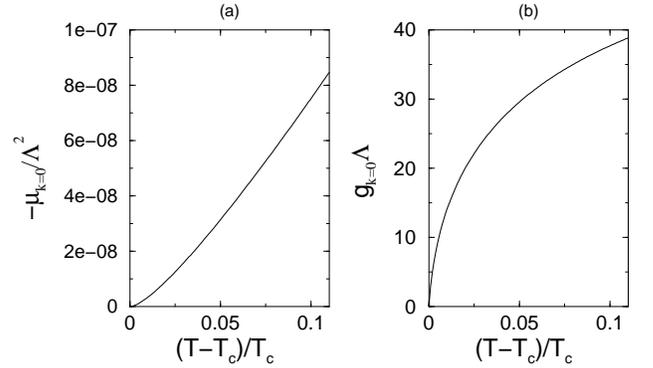}
\caption[a]{
The effective chemical potential $\mu_{k=0}$ and 
the effective quartic coupling $g_{k=0}$ 
near the critical temperature. Both vanish at $T_c$.}
\label{fig3}
\end{figure}

Eq.~(\ref{rg}) has been solved numerically 
for the effective potential $V_{k=0}(v)$ 
with the above boundary condition 
in $d=3$ dimensions
for different values 
of $T$ are shown in Fig.~\ref{fig2}. 
We have normalized the condensate $v$ as well as the effective potential
by the appropriate powers of the ultraviolet cutoff $\Lambda$.
The curves clearly show that the phase
transition is second order. For $T<T_c$, the effective potential has a small
imaginary part, and we have shown only the real part in Fig.~\ref{fig2}.
The imaginary part of the effective
potential does, however, vanish for $T\geq T_c$.
The effective chemical potential $\mu_{k=0}$
as well as the effective quartic coupling constant
$g_{k=0}$ 
are shown in Fig.~\ref{fig3}
and both quantities vanish at the
critical point. The corresponding operators are relevant and must therefore
vanish at $T_c$, and we see that the renormalization group approach correctly
describes the behavior near criticality.
Moreover, it can be shown that the sextic coupling $g_{k=0}^{(6)}$
goes to a nonzero constant at the critical temperature.

\subsubsection{Critical behavior and critical exponents}
\label{critt}

In order to investigate the critical behavior near fixed points,
we write the flow equation in dimensionless form using
the dimensionless quantities

\begin{table}[htb]
\bqa
\begin{tabular}{rrlrrrl}
$\bar{\beta}$&=&$\beta k^2\;,$ &$\;\;\;\;\;\;\;\;\;\;\;\;$&
$\bar{v}$&=&$\beta^{1/2}k^{2-d\over2}v$ \vspace{0.2cm}\\ 
$\bar{V}_k$&=&$\beta k^{-d}V_k\;,$
&&$\bar{\epsilon}(k)$&=&$k^{-2}\epsilon(k)$
\label{coll}
\end{tabular}
\eqa
\end{table}
Eq.~(\ref{rg}) can then be written as
\bqa
0&=&\nonumber
\left[k{\partial\over \partial k}-{1\over 2}(d-2)\bar{v}{\partial\over
\partial\bar{v}}
+d\right]\bar{V}_k+{1\over2}S_d\bar{\beta}\bar{\epsilon}(k)
\\ &&
+S_d\log\left[
1-e^{-\bar{\beta}\bar{\epsilon}(k)}\right]\;.
\label{dr1}
\eqa
The critical potential is obtained by setting to zero
the derivative with respect 
to $k$ in Eq.~(\ref{dr1}).
Expanding in powers of $\bar{\beta}\bar{\epsilon}(k)$, we obtain
\bqa\nonumber
\left[
-{1\over 2}(d-2)\bar{v}{\partial \over\partial\bar{v}}
+d\right]\bar{V}_{k}&=&
-{1\over 2}S_d\bar{\beta}\bar{\epsilon}(k)
\\&&
-S_d\log
\left[\bar{\beta}\bar{\epsilon}(k)\right]\;.
\eqa
Taking the limit $\bar{\beta}\rightarrow 0$ and
ignoring the term which is independent of $v$ leads to
\bqa\nonumber
\left[
-{1\over 2}(d-2)\bar{v}{\partial \over\partial \bar{v}}
+d\right]\bar{V}_k&=&-{1\over2}S_d
\Bigg[\log\left[1+\bar{V}^{\prime}_k\right] 
\\ &&
\hspace{-1cm}
+\log\left[1+\bar{V}^{\prime}_k+2\bar{V}^{\prime\prime}_k\bar{v}^2\right]
\Bigg]\;.
\eqa
This is the same equation as obtained by Morris~\cite{morris}
for a relativistic $O(2)$--symmetric scalar theory in $d$ dimensions
to leading order in the derivative
expansion.
Therefore, the results for the 
critical behavior at leading order in the derivative
expansion will be the same as those obtained in the $d$--dimensional
$O(2)$--model 
at zero temperature.

The above also demonstrates that the system behaves as 
a $d$--dimensional one as the temperature becomes much higher
than any other scale in the problem
Thus the system goes from being $d+1$-dimensional at low temperature
to being effectively $d$-dimensional 
at high temperature and this is often called
dimensional crossover.
It is also referred to as dimensional reduction.
The nonzero Matsubara modes decouple
and the system can be described in terms of a classical 
field theory
for the $n=0$ modes in $d$ dimensions~\cite{lands}.
We return to this subject in Sec.~\ref{tccc}, where we discuss
the calculation of the critical temperature of a dilute Bose gas.

The RG--equation~(\ref{rg2}) satisfied by $V_{k}[v]$ is highly nonlinear
and a direct measurement of the critical exponents from the numerical
solutions is very time-consuming. This becomes even worse as
one goes to higher orders in the derivative expansion and so it is important
to have an additional reliable approximation scheme for calculating critical
exponents. In the following we perform a polynomial expansion~\cite{poly}
of the effective potential, expand around $v=0$, and truncate
at $N$th order:
\beq
\label{pol}
V_k=-\mu_k v^2+
{1\over2}g_{k}v^4+
\sum_{n=3}^{N}
\frac{g^{(2n)}_{k}}{n!}v^{2n}
\eeq
The polynomial expansion turns the partial differential equation~(\ref{rg})
into a set of coupled ordinary differential equations.
In order to demonstrate the procedure we will show how the fixed points
and critical exponents are calculated at the lowest nontrivial order of 
truncation ($N=2$). We write the equations in dimensionless form using
Eqs.~(\ref{coll}) and
\bqa
\bar{\mu}_{k}=k^{-2}\mu_{k}\;, \hspace{0.2cm}
\bar{g}^{}_{k}=\beta^{-1}k^{d-4}g^{}_{k}.
\eqa 
We then obtain the following set of
equations:
\bqa
k\frac{\partial }{\partial k}\bar{\mu}_{k}&=&
-2\bar{\mu}_{k}+S_d\bar{\beta}
\bar{g}_{k}
\left[1+2n(\bar{\epsilon}(k))\right]\\ \nonumber
\label{d}k\frac{\partial }{\partial k}
\bar{g}_{k}
&=&(d-4)\bar{g}_{k}+
S_d\bar{\beta}\bar{g}_{k}^2
\Bigg[\frac{1}{2(1-\bar{\mu}_{k})}
\left[1+2n(\bar{\epsilon}(k))\right]
\\ &&\
+\bar{\beta}n(\bar{\epsilon}(k))\left[1+n(\bar{\epsilon}(k))\right]\Bigg]\;.
\eqa
A similar set of equations was first obtained
in by Bijlsma and 
Stoof~\cite{henk} by considering the one--loop diagrams that contribute
to the running of the effective chemical potential, the effective 
quartic coupling constant etc.

The equations for the fixed points are
\bqa
k\frac{\partial}{\partial k}\bar{\mu}_{k}=0\;,\hspace{1cm}
k\frac{\partial}{\partial k}\bar{g}_{k}
=0\;.
\eqa
If we introduce the variables $r$ and $s$ through the relations
\beq
r={\bar{\mu}_k\over1-\bar{\mu}_k},\hspace{1cm}
s={\bar{g}_k\over(1-\bar{\mu}_k)^2}\;,
\eeq
and expand the equations in powers of 
$\bar{\beta}(1-\bar{\mu}_k)$,
the RG--equations can be written as
\bqa
\label{lin}
\frac{\partial r}{\partial k}&=&-2\left[1+r\right]
\left[r-S_ds\right]\;,
\\ 
\label{lin2}
\frac{\partial s}{\partial k}&=&-s\left[4-d+4r-9S_ds\right]\;.
\eqa
We have the trivial Gaussian fixed point $(r,s)=(0,0)$ as well as the
infinite temperature Gaussian fixed point $(-1,0)$. Finally, for 
$d<4$ there is the
infrared Wilson--Fisher fixed point 
$\left((4-d)/5,(4-d)/\left(5S_d\right)\right)$~\cite{wilson}.

\begin{figure}[htb]
\epsfysize=4.2cm
\epsffile{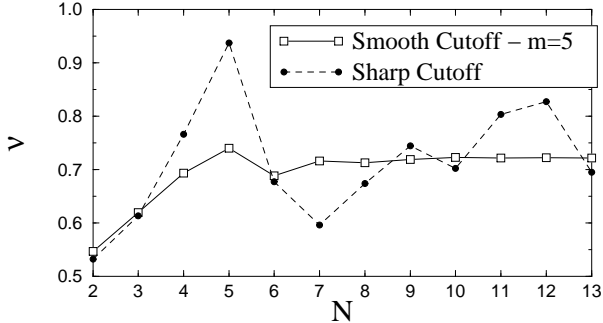}
\caption[a]{
The critical exponent $\nu$ as a function of number of terms $N$
in the polynomial expansion.}
\label{termsn}
\end{figure}

Setting $d=3$ and 
linearizing Eq.~(\ref{lin}) around the fixed point, we find the eigenvalues
$(\lambda_1,\lambda_2)=(-1.278,1.878)$. 
The critical exponent $\nu$ is given by 
the inverse of the largest eigenvalue: $\nu=1/\lambda_2=0.532$.
This procedure can now be repeated including a larger number, $N$, of
terms in the
expansion Eq.~(\ref{pol}).
The result for $\nu$ is plotted in Fig.~\ref{termsn} as a function of the 
number of terms in the polynomial expansion (dashed line).
 Our result agrees with
that of Morris, who considered the relativistic $O(2)$--model
in $d=3$ at zero temperature~\cite{morris}. 
The critical exponent $\nu$ oscillates around the average value $0.73$.
The value of $\nu$ never 
converges as $N\rightarrow\infty$, but continues
to fluctuate. 
For comparison, we have also showed the results using the smooth regulator
function~(\ref{smooth1}) with $m=5$. Clearly, the convergence
properties has improved signficantly.

Our result should be compared to experiment on $^{4}$He and the 
$\epsilon$--expansion which
both give a value of $0.67$~\cite{zinn}. 
The result $\nu=0.73$ is a leading-order result in the
derivative expansion of the effective action~(\ref{ea}).
The next-to-leading order in the derivative expansion
involves a set of coupled equations for the
potential $V_k$ and the wavefunction normalization terms
$Z_k^{(1)}$ and $Z_k^{(2)}$. 
These equations were derived by Morris~\cite{morris}.
A calculation of the critical exponent gives a value of $\nu=0.65$.
One expects the critical exponent $\nu$ to converge towards $0.67$
as one includes more terms in the derivative expansion.

We close this Sec. by 
comparing the two approaches discussed here~\cite{henk,jmike}. 
The renormalization group 
equation~(\ref{rg}) that we obtained in the local potential
approximation is highly nonlinear. We made the further approximation
by expanding the effective potential in a power series, which turns
the partial differential equation~(\ref{rg}) into a set of coupled
ordinary differential equations.
These flow equations were first 
obtained by Bijlsma and Stoof~\cite{henk} by
considering the one-loop diagrams that contribute to
the various $n$--point functions,
while neglecting their
momentum dependence.
However, the two sets of RG--equations differ if
one includes the momentum dependence of the one-loop graphs and 
go beyond the local potential approximation.~\cite{jmike,mike1}.

\subsection{Beliaev-Popov Approximation}
In Sec.~\ref{pf}, we discussed in some detail the Beliaev approximation 
at zero temperature. We recall that this approximation is defined
by all one-loop diagrams contributing to the self-energies. 
It was shown that the resulting dispersion relation is gapless.
Popov~\cite{popov} generalized the Beliaev approximation 
to finite temperature,
and hence it is
often called the Beliaev-Popov (B--P) approximation~\cite{grif}.
He calculated the one-loop diagrams in the limit of zero external 
energy and momentum.
However, one has to be careful because the limits 
$\omega\rightarrow0,\;{\bf p}\rightarrow0$
and ${\bf p}\rightarrow0,\;\omega\rightarrow0$ do not commute at finite
temperature.
Later, 
Shi and Griffin have given a detailed discussion of the B--P approximation
and explicit formal expressions for the normal and anomalous self-energies
at finite temperature and arbitrary external energy $\omega$
and momentum ${\bf p}$. 
An analysis of the B--P approximation similar to the one
presented at zero temperature in Sec.~\ref{t0} would probably be 
significantly simpler. For instance it follows immediately from 
the finite temperature version of Eq.~(\ref{eqs}) 
that the chemical potential 
is free of the infrared 
divergences that plague separately the expressions 
for $\Sigma_{11}(0,0)$
and $\Sigma_{12}(0,0)$. 
The expression for the chemical potential is
\bqa\nonumber
\mu&=&gv^2+{1\over2}g\sumint_{P}{4p^2-4\mu+6gv^2\over\omega_n^2+\epsilon^2(p)}
\\ &=&\nonumber
gv^2+{1\over4}g\left[
3I_{1,1}+I_{-1,-1}\right]
\\ &&
+{1\over2\pi^2}g\int_0^{\infty}
dp\;{p^2\left(2p^2-2\mu+3gv^2\right)\over\epsilon(p)}n(\epsilon(p))\;.
\eqa
In the high-temperature limit, where $\beta gv^2\ll1$, one finds
\bqa
\mu&=&gv^2\left[1-{3\over8\pi^2}{T\over gv^2}\sqrt{2g^3v^2}\right]\;.
\label{htpt}
\eqa
Thus at high temperature, the dimensionless expansion parameter
has an extra factor of $T/gv^2$ compared to the zero-temperature case,
where it is $\sqrt{g^3v^2}$.

So far our description of the dilute Bose gas has been in terms of 
spontaneous breaking of the $U(1)$ symmetry and the use of a chemical
potential to ensure that the mean number of particles is constant.
There exist alternative approaches that are number-conserving.
Such an approach was first introduced by Giradeau and Arnowitt
as early as in 1959~\cite{arno}. This approach is variational in nature
and, like the HBF approximation, 
it exhibits a gap in the dispersion relation.
However, later it was shown by Takano~\cite{taka} that the gap is removed if 
one takes into account cubic terms in the Hamiltonian in a consistent manner.
More recently, Morgan~\cite{sam} has discussed in detail
number-conserving approaches and generalized them to trapped Bose gases. 
The starting point is the Hamiltonian including interaction terms
describing binary collisions written in terms of pair operators
that conserve particle number. These operators are defined by
\bqa
\alpha_{i}&=&\beta_0^{\dagger}a_i\;,
\eqa
where $a_i$ are the standard annihilation operators and 
$\beta_0=(1+a_0^{\dagger}a_0)^{-1/2}a_0$.
The Hamiltonian can then be divided into
terms containing zero, one, two, three, and four creation/annihilation
operators.
The terms that contain up to two operators are then diagonalized, while
the cubic and quartic terms are treated as perturbations using
first and second order perturbation theory. 
The validity of perturbation theory is the usual requirement that 
$\sqrt{na^3}\ll 1$ at $T=0$. At high temperature, it follows
from~(\ref{htpt}) that the requirement be
$T\sqrt{g^3v^2}/gv^2\ll 1$.
In the zero temperature limit, the calculations reproduce several
known results presented in Sec.~\ref{t0}. The Hamiltonian
can be used to calculate the ground state energy density 
${\cal E}$
(rather than the free 
energy density ${\cal F}$) and a calculation involving the quadratic terms
yields the result of Bogoliubov given in Eq.~(\ref{ed}). 
A second-order calculation
reproduces the $na^3\log(na^3)$-term of Wu~\cite{wu}, but one is faced with
a logarithmic ultraviolet divergent term which is proportional
to $na^3$. As we argued in Sec.~\ref{noneff}, 
the correct way to treat this divergence is to absorb
it in the coefficient of the operator $(\psi^*\psi)^3$, which represents
$3\rightarrow3$ scattering. The approach also reproduces 
Beliaev's results for the phonon 
spectrum~(\ref{pre}) and the damping rate~(\ref{imw}).
Finally, we mention that the modified Popov 
approximation~\cite{ext1,ext2} has also been examined within the framework
of number-conserving approaches~\cite{sam}.

\section{Calculations of $T_c$}
\label{tccc}
The critical temperature for an ideal Bose gas is given by  
Eq.~(\ref{tc0}). A natural question to ask is: what is 
the leading-order effect of a weak two-body interaction
on the critical temperature of a homogeneous Bose 
gas? This question has been around for almost fifty years, but only
very recently has the issue been settled. 
It has been discussed in detail by Baym {\it et al.}~\cite{baym2}
and we follow to some extent their paper.
Various approaches to the problem have also been discussed very recently
by Haque~\cite{hakk}.

In the following, we assume that the interaction is repulsive, which
corresponds to a positive scattering length $a$. One might think
that effects of a repulsive interaction is to decrease the 
critical temperature of Bose gas. For instance, the superfluid transition
in liquid $^4$He, takes place at a lower temperature than that of
an ideal gas of the same density. However, liquid $^4$He is not 
weakly interacting and it turns out that the leading effects of
interactions in the dilute Bose gas is to increase $T_c$.

The first paper in which a
quantitative prediction appears, is from 1957 by Lee and Yang~\cite{leeyang}.
In that paper, the authors predict that the critical temperature increases
compared to that of an ideal Bose gas, and that the increase is proportional
to $\sqrt{a}$. Later the same authors predicted that the shift is linear
with $a$~\cite{leeyang2}. 
This prediction was purely qualitative since neither sign nor 
magnitude were given. A couple of years later, 
Glassgold {\it et al}.~\cite{glass} also predicted an increase of $T_c$
which is proportional to $\sqrt{a}$. A couple of decades later, the 
problem was revisited by Toyoda~\cite{toyota}.
He predicted a decrease of the critical temperature which is 
proportional to $\sqrt{a}$. Since the sign agrees with the measurements
on $^4$He, there seemed to be at least qualitative agreement between theory
and experiments.
Long ago, Huang claimed an increase in $T_c$ that is proportional
$a^{3/2}$~\cite{huangbok}, and very recently he claims 
in another paper that 
$T_c$ increases proportional to $\sqrt{a}$~\cite{huang2}.
From this selection of papers, it is clear that there has been
considerable amount of confusion about how $T_c$ depends parametrically
on the scattering length. Common to these results as well 
other attempts to calculate $T_c$~\cite{schakel,illu}, is that they
are based on perturbation theory. However, Bose condensation in
a dilute gas is governed by long-distance physics that is
inherently nonperturbative. These issues will be discussed later.

There have also been other approaches to the calculation of $T_c$.
Stoof and Bijlsma have 
carried out renormalization group calculations
of the critical temperature ~\cite{henk}.
This approach was discussed in the previous section and it predicts
that the leading shift is proportional to $a\log a$. 
Since this approach is purely numerical, it is difficult to take the
limit $a\rightarrow0$ and thus obtain the correct dependence on $a$
in the dilute limit. Very recently, 
a calculation of $T_c$
based on the exact renormalization group by
Ledowski, Hasselmann and 
Kopietz~\cite{ledow}. They calculated the momentum-dependent two-point function
and showed that the leading behavior is proportional to the scattering length
$a$.

The first Monte Carlo simulations for hard-sphere bosons
in the low-density regime were done by Gr\"uter {\it et al.}~\cite{grut}.
In that paper, the authors predict a positive linear shift after extrapolating
to the limit $a\rightarrow 0$. 
This result was somewhat surprising since 
some early Monte Carlo simulations 
as well as the experiments on $^4$He, show a decrease in $T_c$ due to
repulsive interactions.
Later, it has been shown rigorously
using effective field theory methods that the parametric dependence of $T_c$
indeed is linear in $a$~\cite{baym2}. Thus in the 
dilute limit, we can write
\bqa
{\Delta T_c\over T_c^0}&=&cn^{1/3}a\;,
\label{proptcc}
\eqa
where $c$ is a constant that is to be determined.
The problem of determining the constant $c$ has been attacked by
analytical as well as numerical methods in recent years.
These methods include high-precision Monte Carlo simulations,
$1/N$ expansions, self-consistens calculations involving
summation of bubble and ladder diagrams, and variational perturbation theory.
We will discuss these in Secs.~\ref{1n}--\ref{oc}.

\subsection{Hartree-Fock Approximation and Breakdown of Perturbation Theory}
In this subsection, we will briefly discuss the Hartree-Fock approximation and
show that it predicts no shift in the critical temperature. We approach
the phase transition from above, so the condensate density $v$ is zero.

In an ideal gas, the number 
density of excited particles is given by Eq.~(\ref{nf}),
which can be written as 
\bqa
n_{\rm ex}&=&{1\over\lambda_T^3}g_{3/2}(z),
\label{nonn}
\eqa
where the function $g(z)$ is the polylogarithmic function
\bqa
g_l(z)&=&\sum_{n=1}^{\infty}{z^n\over n^l}\;,
\label{non}
\eqa
and $z=e^{\beta\mu}$ is the fugacity. 

\begin{figure}[htb]
\epsfysize=0.95cm
\vspace{0.5cm}
\epsffile{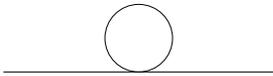}
\caption[a]{Self-energy diagram in the Hartree-Fock approximation.}
\label{hfe}
\end{figure}

The simplest way to include the effects of interactions is to include
the self-energy 
in the Hartree-Fock approximation.
The Feynman graph is the tadpole diagram shown in Fig.~\ref{hfe}.

The tadpole is independent of the external momentum and the
expression for the Hartree-Fock self-energy is
\bqa\nonumber
\Sigma_{\rm HF}&=&2g\sumint_{P}
{p^2-\mu\over\omega_n^2+(p^2-\mu)^2}
\\
&=&2gn\;,
\eqa
where we have used Eqs.~(\ref{nf}) and~(\ref{exs}).
Thus a particle with momentum ${\bf p}$ effectively has the energy
$\epsilon(p)=p^2+2gn$, where the term $2gn$ arises from the mean field
of the other particles. We can now generalize Eq.~(\ref{nonn})
by writing
\bqa
n&=&{1\over\lambda_T^3}g_{3/2}\left(e^{\beta(\mu-2gn)}\right)\;.
\label{sc}
\eqa
Eq.~(\ref{sc}) shows that we must increase the chemical potential
by an amount $\Delta\mu=\Sigma_{\rm HF}$ to 
keep the same number density as that of an ideal gas at the same temperature.
It shows in particular that $\mu$ must approach $\Sigma_{\rm HF}$
from below to obtain the critical number density at a given temperature.
Thus the
critical temperature remains the same. The conclusion is that 
including a constant
mean-field shift in the single-particle energies cannot
change the critical temperature of a Bose gas.
This is an example of the fact that mean-field theories
effectively treat interacting gases as ideal gases with
modified parameters and thus predict the same $T_c$.

Calculations using the Hartree-Fock approximation have been carried out
by Huang. He 
applies a virial expansion to Eq.~(\ref{sc}) and obtains
a change in the critical temperature which is proportional to $a^{1/2}$.
However, this is an artifact of the approximation, as can be seen
by including more terms in the expansion. 
This was discussed in some detail in~\cite{baym2}.
A correct treatment is given
in e.g.~\cite{pet}. 
Similarly, the summation of the ring diagrams~\cite{haug} 
in the effective potential does not change $T_c$.
The reason is that
these diagrams are evaluated at zero external momentum
and therefore merely corresponds to a 
redefinition of the chemical potential.

\begin{figure}[htb]
\epsfysize=2.2cm
\vspace{0.5cm}
\epsffile{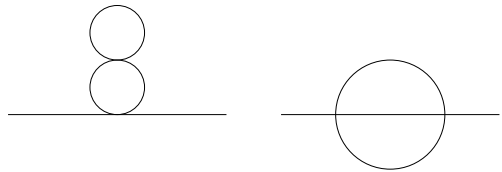}
\caption[a]{Two-loop self-energy diagrams.}
\label{ptself}
\end{figure}

We have seen that a leading perturbative calculation in the scattering
length $a$
gives no corrections
to the critical temperature of a dilute Bose gas. One might try to improve
on this result by going to higher orders in perturbation theory.
The Feynman diagrams contributing to the self-energy at second order in 
perturbation theory are shown in Fig.~\ref{ptself}. 

If we focus on the
contribution from the $n=0$ Matsubara mode, 
the diagrams read
\bqa\nonumber
\Sigma^{2a}(0,{\bf p})
&=&-4g^2T^2
\int{d^dk\over(2\pi)^d}\int{d^dq\over(2\pi)^d}
\\ &&\times
{1\over(k^2-\mu)
(q^2-\mu)^2}\;,\\ \nonumber
\Sigma^{2b}(0,{\bf p})
&=&-2g^2T^2
\int{d^dk\over(2\pi)^d}\int{d^dq\over(2\pi)^d}
\\ &&\hspace{-0.4cm}\times
{1\over(k^2-\mu)(q^2-\mu)(|{\bf p}+{\bf q}+{\bf k}|^2-\mu)}\;,
\eqa
where the chemical potential acts
as an infrared cutoff in the integral. 
The left diagram is independent of the external momentum.
If we use the mean-field criterion that  
$\mu\rightarrow0$ at the transition, the integral is linearly
divergent in the infrared. 
The right diagram depends on the external momentum
${\bf k}$. For $\mu=0$, it 
is logarithmically divergent as the external momentum 
goes to zero. As one goes to higher orders in the perturbation expansion,
the diagrams become increasingly infrared divergent for $\mu=0$.
If we denote a generic self-energy
diagram with $n$ loops by $\Sigma_n$, we have~\cite{baym2}
\bqa
\Sigma_n&\sim&T\left({a\over\lambda_T}\right)^2
\left({a^2\over\mu\lambda_T^4}\right)^{n-2\over2}\;.
\eqa
This shows that perturbation theory breaks 
down in the critical region
due to the infrared
divergences. 
Physically, these infrared divergences are screened and this can be
taken into account by summing certain classes of diagrams from all orders
of perturbation theory. Examples of this is summation of bubble or ladder
diagrams. We return to this issue at the end of this section.

\subsection{Dimensional Reduction}
In Sec.~\ref{rgapp}, 
we saw that the renormalization-group equations at high temperature
reduce to those of a three-dimensional $O(2)$-symmetric theory.
This is an example of dimensional reduction and for the dilute Bose
gas, it can be understood as follows.
In the imaginary time formalism, the fields are decomposed into 
modes which are characterized by their Matsubara frequency $\omega_n=2\pi nT$.
At distances much larger than the thermal wavelength
and for temperatures sufficiently close to 
the critical temperature, the time derivative term for $n\neq0$
is much larger than both the kinetic energy term and the effective chemical 
potential term. 
This implies that the nonstatic Matsubara modes
decouple and the long-distance physics can be described in terms of an
effective
three-dimensional field theory for the $n=0$ mode. 
The fact that the long-distance physics associated with the
phase transition is well separated from the typical momentum scale $T$
associated with the nonzero Matsubara modes, makes effective field theory
methods ideal to study the phase transition.


The effective three-dimensional theory 
that describes the long-distance physics can be constructed
using the methods of effective field theory~\cite{eft}. 
Once the symmetries of the theory have been identified, one writes down the 
most general Lagrangian ${\cal L}_{\rm eff}$
that is consistent with these symmetries. In the present
case, we simply have a 
complex scalar field with an $O(2)$-symmetry.
In addition, there a is three-dimensional rotational symmetry.
The effective three-dimensional theory is then described by the action
\bqa\nonumber
S_{\rm eff}&=&\int d^3x\bigg[
-{1\over2}\phi^*\nabla^2\phi+{1\over2}\mu_3\phi^*\phi
\\ &&
\hspace{3.5cm}
+{1\over24}u(\phi^*\phi)^2+...\bigg]\;,
\label{3eff}
\eqa
where we have used the conventional normalization of an $O(2)$-invariant
theory.
The dots indicate operators with more derivatives and more fields. Examples
are $\left[\nabla(\phi^*\phi)\right]^2$ and $(\phi^*\phi)^3$.
The relation between the parameters in the effective theory and in the 
full theory can be determined by perturbative
matching; one requires that the effective
theory~(\ref{3eff}) reproduces static correlators at long distances 
$R\gg1/T$ to a specified accuracy. 
The reason why the parameters of the effective three-dimensional theory
can be determined in perturbation theory, stems from the fact 
that the coefficients
of the effective theory encode the short-distance physics 
at the scale $T$ which is perturbative and that the 
matching procedure does not involve the nonperturbative long-distance
physics.

At the tree level, the matching can be done simply by inspection.
By comparing the action~(\ref{s1})
that describes the full four-dimensional theory with the 
action~(\ref{3eff}) that describes the effective three-dimensional theory,
we can read off the relation between the fields in the two theories.
This yields
\bqa
\psi&=&\sqrt{T\over2}\phi
\label{fmat}
\;,
\eqa
Similarly, by matching the other terms in the action in the two theories
at the tree level and using~(\ref{fmat}), one easily finds
\bqa
\mu_3&=&-\mu 
\;,\\
u&=&3gT\;,
\eqa
Higher order operators as well as corrections to the coefficients of the
operators in Eq.~(\ref{3eff}) can be 
ignored at the order of interest
in the diluteness expansion.
For instance, the coefficient of the operator
$\left[\nabla(\phi^*\phi)\right]^2$ is proportional to $a^2\lambda_T$.
From dimensional analysis, it follows that 
the contribution to physical quantities
from this operator is then suppressed by a factor
of $n^{1/3}a$ compared to the operator $(\phi^*\phi)^2$.
The contribution to physical quantities from other operators are analyzed
in a similar manner.

For an ideal Bose gas, the critical 
temperature is given by Eq.~(\ref{tc0}). Equivalently, the 
critical number density $n_c^0$ at fixed temperature 
satisfies
$n_c^0\lambda_T^3=\zeta\left({3\over2}\right)$. 
Due to the repulsive interactions in the
dilute Bose gas, the critical number densitiy changes.
The first-order change in the critical temperature
$\Delta T_c=T_c-T_c^0$
is related to the first-order change $\Delta n_c=n_c-n_c^0$
in the critical number density
at fixed $T_c$
by~\cite{baym2}:
\bqa\nonumber
{\Delta T_c\over T_c^0}&=&-{2\over3}{\left[n_c(T_c)-n_c^0(T_c)\right]\over n}\\
&=&
-{1\over3}{T^0\Delta\langle\phi^*\phi\rangle\over n_c^0}\;.
\label{delten}
\eqa
The factor $2/3$ in the first line
comes from the relation $T^0\propto (n^0)^{2/3}$.
The last equality follows from the fact that $n=\langle\psi^*\psi\rangle$
and that the contribution from the the zeroth Matsubara
mode is ${1\over2}T\langle\phi^*\phi\rangle$, which
follows from Eq.~(\ref{fmat}). 

In order to calculate the critical temperature, we must evaluate the 
quantity $\Delta\langle\phi^*\phi\rangle$. We discuss this next.

\subsubsection{$1/N$ expansion}
\label{1n}
The $1/N$ expansion is a nonperturbative method thas has been widely used
in high-energy and condensed matter physics~\cite{largen,moshe}.
In condensed matter physics, it has been used to 
study the critical behavior of $O(N)$ spin models and calculate 
their
critical exponents~\cite{zinn}.
The idea is to generalize a Lagrangian with a fixed number of fields to
$N$ fields and then let $N$ be a variable. The expansion is defined
as an expansion in powers of $1/N$ while $gN$ is held fixed ($g$ is 
the coupling constant). The method is nonperturbative in the sense that
calculations at every order in $1/N$, there are Feynman diagrams
contributing from all orders of perturbation theory.
In this way, one sums up graphs from all orders of 
perturbation theory. One hopes that this expansion captures some of the 
essential physics that cannot be captured by e.g. perturbative methods.

The critical temperature was recently calculated by Baym, Blaizot, and 
Zinn-Justin~\cite{baym3} in the large-$N$ limit. The next-to-leading-order 
result was obtained by 
Arnold and Tom\'a\^sik~\cite{at}.

In the present case, the 
Lagrangian~(\ref{3eff}) is generalized to a 
scalar field theory with $N$ real 
components. The Lagranigan is now $O(N)$ invariant 
and reads
\bqa
{\cal L }_{\rm eff}&=&-{1\over2}\phi_i\nabla^2\phi_i
+{1\over2}\mu_3\phi_i^2
+{1\over24}u(\phi_i\phi_i)^2
\;,
\eqa
where $i$ runs from 1 to $N$. Summation over $i$ is implicitly 
understood.
The large-$N$ limit is obtained by taking $N\rightarrow\infty$, while
keeping $uN$ constant.

\begin{figure}[htb]
\epsfysize=2.2cm
\vspace{0.2cm}
\epsffile{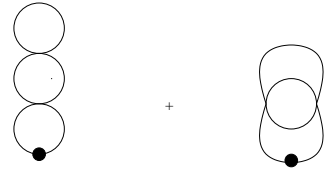}
\caption[a]{Three-loop
Feynman diagrams contributing to $\Delta\langle\phi^2\rangle$.}
\label{ex}
\end{figure}

Generally, the diagrams that contribute to $\Delta\langle\phi^2\rangle$
can be obtained from vacuum diagrams by 
inserting an operator $\phi^2$.
For example, at the three-loop level, there are two
diagram that contribute and they are shown in Fig
diagrams in Fig.~\ref{ex}. 
It can be shown that the first diagram is suppressed by a factor of $1/N$
relative to the second.

\begin{figure}[htb]
\epsfysize=1.4cm
\vspace{0.2cm}
\epsffile{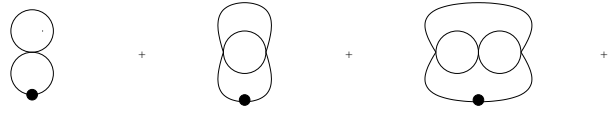}
\caption[a]{Feynman diagrams contributing to $\Delta\langle\phi^2\rangle$
to leading order in $1/N$.}
\label{nden}
\end{figure}

\begin{figure}[htb]
\epsfysize=0.95cm
\vspace{0.5cm}
\epsffile{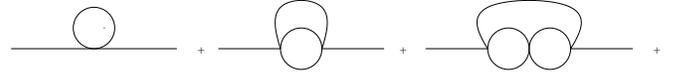}
\caption[a]{Feynman graphs contributing to the self-energy to leading order
in $1/N$.}
\label{nself}
\end{figure}

The Feynman diagrams that contribute to $\Delta\langle\phi^2\rangle$
at leading order in $1/N$ are shown in Fig.~\ref{nden}.
The dot denotes an insertion of the operator $\phi^2$. 
These diagrams are called bubble diagrams and the bubble summation
is thus exact in the large-$N$ limit.
The expression for the expectation value $\Delta\langle\phi_i^2\rangle$
then becomes
\bqa
\Delta\langle\phi_i^2
\rangle&=&
-N\int{d^dp\over(2\pi)^d}{1\over p^4}\Sigma(p)\;,
\label{sigmaapp}
\eqa
where $\Sigma(p)$ is the self-energy. 
The Feynman diagrams for the self-energy
are obtained from those in Fig.~\ref{nden} by 
cutting the propagator line that goes through the dots.
The self-energy diagrams are shown in Fig.~\ref{nself}.

After mass renormalization, so that $\Sigma(0)=0$, 
the expression for the self-energy is~\cite{zinn}
\bqa\nonumber
\Sigma(p)&=&{2\over N}
\int{d^dk\over(2\pi)^d}
{1\over6/Nu+{\cal T}_1(k)}
\left[
{1\over|{\bf p}+{\bf k}|^2}-{1\over k^2}
\right]\;,
\\ &&
\label{lnself}
\eqa
where ${\cal T}_1(k)$ is the one-loop contribution to the 
four-point function:
\bqa
{\cal T}_1(k)&=&
\int{d^dq\over(2\pi)^d}{1\over q^2|{\bf q}+{\bf k}|^2}\;.
\label{tdeff}
\eqa
In the appendix, we show how to calculate 
the function ${\cal T}_1(k)$ in dimensional regularization. One finds
\bqa
{\cal T}_1(k)&=&
M^{2\epsilon}
{\Gamma\left(2-{d\over2}\right)\Gamma\left({d\over2}-1\right)
\over2^{2d-3}\pi^{{d-1\over2}}\Gamma\left({d-1\over2}\right)}
k^{d-4}
\;.
\label{tdr}
\eqa
The next step is to evaluate the integral over $p$ in Eq.~(\ref{sigmaapp}).
Using Eq.~(\ref{pint}), we obtain
\bqa \nonumber
\int{d^dp\over(2\pi)^d}{1\over p^4}
\left[
{1\over|{\bf p}+{\bf k}|^2}-{1\over k^2}
\right]&=& 
\\ &&\hspace{-2cm}
M^{2\epsilon}
{\Gamma\left(3-{d\over2}\right)\Gamma\left({d\over2}-2\right)
\over2^{2d-4}\pi^{d-1\over2}\Gamma\left({d-3\over2}\right)}k^{d-6}
\;.
\label{pdr}
\eqa
Inserting Eqs.~(\ref{tdr}) and~(\ref{pdr}) into Eq.~(\ref{lnself}),
we obtain
\bqa\nonumber
\Delta\langle\phi_i^2\rangle&=&-M^{2\epsilon}
{\Gamma\left(3-{d\over2}\right)\Gamma\left({d\over2}-2\right)
\over2^{2d-5}\pi^{d-1\over2}\Gamma\left({d-3\over2}\right)}
\\ &&\times
\int{d^dk\over(2\pi)^d}{k^{d-6}\over a+bk^{d-4}}
\;,
\label{above2}
\eqa
where
\bqa
a&=&6/Nu\;,\\
b&=&
M^{2\epsilon}
{\Gamma\left(2-{d\over2}\right)\Gamma\left({d\over2}-1\right)
\over2^{2d-3}\pi^{{d-1\over2}}\Gamma\left({d-1\over2}\right)}
\;.
\eqa
The final step consists of integrating over $k$.
Using Eq.~(\ref{neeed}) in the Appendix
we obtain
\bqa\nonumber
\Delta\langle\phi^2_i\rangle&=&M^{4\epsilon}
a^{d-2\over d-4}b^{3-d\over d-4}
\\ &&\hspace{-1cm}\times
{\Gamma\left(3-{d\over2}\right)\Gamma\left({d\over2}-2\right)
\Gamma\left({2-d\over4-d}\right)
\Gamma\left({6-d\over4-d}\right)
\over2^{3d-6}\pi^{d-{1\over2}}\Gamma\left({d-3\over2}\right)}\;.
\eqa
The limit $d\rightarrow3$ is regular and one obtains
\bqa
\Delta\langle\phi^2_i\rangle
&=&-{Nu\over96\pi^2}
\label{1/n}
\;.
\eqa
Inserting the result~(\ref{1/n}) into~(\ref{delten}),
we obtain the critical temperature to leading order in $1/N$: 
\bqa\nonumber
{\Delta T_c\over T_c^0}&=&{8\pi\over3
\left[\zeta\left({3\over2}\right)\right]^{4/3}}an^{1/3}
\\ &\approx&2.33n^{1/3}a
\;,
\label{lotc}
\eqa
where we have used that $u=3gT$ and $g=8\pi a$ and set $N=2$.
Thus $T_c$ {\it increases} linearly with $a$. As noted
before~\cite{baym3}, the result~(\ref{lotc}) is independent
of $N$. However, it is only valid in the limit $N\rightarrow\infty$.

The $1/N$ correction to the above result has recently been calculated by 
Arnold and 
Tom\'a\^sik~\cite{at}.
This is a very lengthy and technically complicated calculation.
For instance the integrals that one encounters 
at order $1/N$ are difficult to evaluate in $3-2\epsilon$ dimensions.
Instead, Arnold and Tom\'a\^sik always reduced their diagrams
to unambigious integrals in three dimensions that are simpler to evaluate.
We shall not review the calculation, but merely state the result.
Through next-to-leading order in $1/N$, the shift in $T_c$ is 
\bqa
{\Delta T_c\over T_c^0}&=&
{8\pi\over3\left[\zeta\left({3\over2}\right)\right]^{4/3}}\left[
1-{0.5272\over N}\right]n^{1/3}a
\;.
\eqa
For $N=2$, it gives a correction of only $26\%$:
\bqa
{\Delta T_c\over T_c^0}&=&1.71n^{1/3}a
\;.
\eqa

\subsubsection{Monte Carlo simulations}
The action~(\ref{3eff}) can also be used as the starting point for numerical
calculations of the critical temperature and other
nonuniversal effects of a dilute Bose gas.
This is done by putting the theory on a lattice and using Monte Carlo methods
to solve the theory
nonperturbatively. Such numerical simulations have been carried out by
several groups~\cite{grut,kraut,arnold1,arnlat,svis}. 
This approach has been discussed in considerable detail by Arnold and 
Moore~\cite{arnold1}.
The value of the coefficient in~(\ref{proptcc}) reported by Gru\"ter
{\it et al.}~\cite{grut} is $c\simeq0.34$, while Holzmann and 
Krauth~\cite{kraut} obtained $c\simeq2.3$. 
The most recent values reported by 
Arnold and Moore~\cite{arnlat} and by Kashurnikov {\it et al.}~\cite{svis}
are $c\simeq1.32$ and $c\simeq1.29$, respectively.
One source of discrepancy lies in the nonlinear corrections
to $T_c$ as a function of $a$ at the densities where the 
simulations of~\cite{grut} were carried performed.
In~\cite{kraut}, the authors expand the integrand in the path
integral in powers of the interaction and keep only the first
term in that expansion. This perturbative treatment of the
interaction is incorrect since the physics close the phase transition
is inherently nonperturbative. This is in contrast to the
Monte Carlo simulations based on the action~(\ref{3eff})
carried out~\cite{arnlat,svis}. These calculations agree
within error bars. We will regard these lattice results as the
correct result for the coefficient $c$ in Eq.~(\ref{proptcc}).

\subsubsection{Other calculations}
\label{oc}
In this subsection, we will briefly discuss other calculations
of the shift in $T_c$ in the dilute Bose gas.
Generally, the expectation value $\langle\phi^2\rangle$ can be written as 
\bqa
\langle\phi^2\rangle&=&
\int{d^dk\over(2\pi)^d}
{1\over k^2+\mu_3+\Sigma(k)}
\;,
\eqa
where $\Sigma(k)$ is the self-energy function. The critical point
is determined by the condition that 
the correlation length becomes infinite, or equivalently that 
the effective chemical potential $\mu_3+\Sigma(0)$ vanishes:
\bqa
\mu_3+\Sigma(0)&=&0\;.
\label{epcon}
\eqa
In absence of interactions, the condition~(\ref{epcon}) 
reduces to the well known $\mu_3=0$. Using Eq.~(\ref{epcon}) to eliminate
the chemical potential $\mu_3$, we can write Eq.~(\ref{delten}) as
\bqa
\Delta\langle\phi^2\rangle
&=&
\int{d^dk\over(2\pi)^d}
\left[{1\over k^2+\Sigma(k)-\Sigma(0)}-{1\over k^2}\right]\;.
\label{selfeq}
\eqa
The next step is to make an approximation for the
self-energy function appearing in Eq.~(\ref{selfeq}). In the previous
section, we used the large-$N$ expression for the self-energy~(\ref{lnself}).
Baym {\it et al.}~\cite{baym2} considered three different
equations for the self-energy. 
\begin{itemize}
\item{One-bubble approximation:}
\end{itemize}
The self-energy is approximated by the second diagram in Fig.~\ref{ptself}:
\bqa\nonumber
\Sigma(p)-\Sigma(0)&=&-2g^2T
\int{d^dk\over(2\pi)^d}{\cal T}_1(k)
\\ && \nonumber
\hspace{-2.4cm}
\times
\left[
{1\over|{\bf p}+{\bf k}|^2+\Sigma(|{\bf p}+{\bf k}|)
-\Sigma(0)}-{1\over k^2+\Sigma(k)-\Sigma(0)}
\right]\;,
\\ &&
\label{selfcon1}
\eqa

\begin{itemize}
\item{Bubble-summation approximation:}
\end{itemize}
The self-energy is approximated by the bubble sum in Fig.~\ref{nself}:
\bqa\nonumber
\Sigma(p)-\Sigma(0)&=&-2g^2T
\int{d^dk\over(2\pi)^d}{{\cal T}_1(k)\over1+2g{\cal T}_1(k)}
\\ && \nonumber
\hspace{-2.4cm}
\times
\left[
{1\over|{\bf p}+{\bf k}|^2
+\Sigma(|{\bf p}+{\bf k}|)-\Sigma(0)}-{1\over k^2+\Sigma(k)-\Sigma(0)}
\right]\;,
\\ &&
\label{selfcon2}
\eqa

\begin{itemize}
\item{Ladder-summation approximation:}
\end{itemize}
The self-energy is approximated by the ladder sum similar to the bubble sum.
\bqa\nonumber
\Sigma(p)-\Sigma(0)&=&
\int{d^dk\over(2\pi)^d}{{\cal T}_1(k)\over1+g{\cal T}_1(k)}
\\ && \nonumber
\hspace{-2.4cm}
\times
\left[
{1\over|{\bf p}+{\bf k}|^2
+\Sigma(|{\bf p}+{\bf k}|)-\Sigma(0)}-{1\over k^2+\Sigma(k)-\Sigma(0)}
\right]\;,
\\ &&
\label{selfcon3}
\eqa
Note that Eqs.~(\ref{selfcon1})--(\ref{selfcon3}) have been made
self-consistent
by replacing the free propagators on the right hand side by the
interacting propagators. Note also that the only difference between the
bubble sum and the ladder sum is a factor of two.

Eqs.~(\ref{selfcon1})--(\ref{selfcon3}) have been solved numerically and the
results were used to evaluate Eq.~(\ref{selfeq}) to obtain the
corresponding shifts in $T_c$. The results for the coefficient $c$ are
$3.8$, $2.5$, and $1.6$, respectively and thus
within a factor three. Note also that the prediction for the
shift in $T_c$ from the non self-consistent bubble summation
(the leading $1/N$ result with $N=2$) is very close to the result from 
self-consistent bubble summation.

We can gain more insight into the mechanism behind the increase in $T_c$
by looking at the modification of the spectrum for small momenta.
Let us consider the non self-consistent bubble sum:
\bqa
\epsilon(k)&=&k^2+\Sigma(k)-\Sigma(0)\;,
\eqa
where 
\bqa
\Sigma(k)&=&
\int{d^dk\over(2\pi)^d}{{\cal T}_1(k)\over1+g{\cal T}_1(k)}
{1\over|{\bf p}+{\bf k}|^2}\;.
\eqa
For small momenta $k$, 
the difference $\Sigma(k)-\Sigma(0)$ can be approximated by~\cite{baym2} 
\bqa
\Sigma(k)-\Sigma(0)&=&-{2\over3\pi^2}
k^2\left(\log{k\over k_c}-{1\over3}\right)\;,\hspace{0.2cm}k\ll k_c\;,
\label{sigmahard}
\eqa
where $k_c=8\pi^2a/\lambda_T^2$ is the screening wave number.
The logarithmic term in~(\ref{sigmahard}) indicates a modified
spectrum for small wave numbers $\sim k^{2-\eta}$
and thus a hardening (The spectrum is of the form $k^{\alpha}$, 
where $\alpha<2$) 
of the spectrum 
compared to the noninteracting case. 
The hardening results from correlations among particles with low momentum,
which leads to a
decrease in the critical density and thus an increase in the
critical temperature. Other approximations show a different functional
dependence of the difference $\Sigma(k)-\Sigma(0)$ at small $k$, but
the basic mechanism remains the same, namely a hardening of the spectrum
at the critical temperature. It has been pointed out~\cite{gordon}
that the hardening of the spectrum for the $n=0$ Matsubara mode
only takes place exactly at $T_c$. The Bogoliubov operator 
inequality
guarantees that the spectrum remains quadratic away from $T_c$.

Variational methods have also been used 
recently~\cite{ram1,ram2,tceric}
to calculate $T_c$. The basic idea is to compute~(\ref{selfeq}) using
an effective three-dimensional Lagrangian that has been reorganized
according to the discussion in Sec.~\ref{subopt}.
The Lagrangian is written as ${\cal L}={\cal L}_0+{\cal L}_{\rm int}$, where
\bqa
{\cal L}_0&=&-{1\over2}\phi_i\nabla^2\phi_i+{1\over2}m^2\phi^2_i\;,
\label{vpt1def}\\
{\cal L}_{\rm int}&=&
{1\over2}\delta(\mu_3-m^2)\phi_i^2+{1\over24}\delta u(\phi_i\phi_i)^2\;,
\label{vpt2def}
\eqa
and $i$ runs from 1 to $N$.
Calculations are carried out by using $\delta$ as a formal expansion parameter,
expanding to a given order in $\delta$ and setting $\delta=1$ at the end
of the calculation. Finally, we need to give a prescription for the mass
parameter $m$. In the calculations below, we will use the PMS criterion.
In this context, it reads
\bqa
{\partial \Delta\langle\phi^2\rangle\over\partial m}&=&0\;.
\label{pmsm}
\eqa
If we wish to apply variational methods,
to calculating the shift in $T_c$, we need to generalize 
the quantity $\Delta\langle\phi^2\rangle$ appearing in
Eq.~(\ref{selfeq})
to the field theory defined by Eqs.~(\ref{vpt1def}) and~(\ref{vpt2def}).
One must be able to expand this quantity in powers of $\delta$ and it must
reduce to $\Delta\langle\phi^2\rangle$ when $\delta=1$.
Several prescriptions for generalizing $\Delta\langle\phi^2\rangle$ 
have been proposed in the literature\cite{ram1,ram2,tceric}.
Some of these prescriptions are well behaved in the limit $N\rightarrow\infty$
and some of them are not. Since the result for the shift in $T_c$ is
known analytically in this limit, this a desireable property of a
prescription. Two generalizations that have this property have been
considered~\cite{tceric}:
\bqa
\Delta_a\langle\phi^2\rangle&=& 
N\int{d^dk\over(2\pi)^d}\left[
{1\over k^2+\Sigma(k)-\Sigma(0)}-{1\over k^2}
\right]\;,
\label{propa1}
\\\nonumber
\Delta_b\langle\phi^2\rangle&=&
N\int{d^dk\over(2\pi)^d}\left[
{1\over k^2+m^2(1-\delta)+\Sigma(k)-\Sigma(0)}\right.
\\ &&
\left.
-{1\over k^2+m^2(1-\delta)}
\right]\;.
\eqa
In the following, we will consider $\Delta_a\langle\phi^2\rangle$.
The strategy is to calculate the difference 
$\Sigma(k)-\Sigma(0)$ in a powers series in $\delta$, substitute
the result into Eq.~(\ref{propa1}), and finally expand the resulting
integral in powers of $\delta$. The Feynman diagram that contributes
to the self-energy to first order in $\delta$ is the leftmost diagram
in Fig.~\ref{ptself}. It is independent of the external momentum and 
so the difference $\Sigma(k)-\Sigma(0)$ vanishes.
The first nonzero contribution to the
quantity $\Sigma(k)-\Sigma(0)$ is then given by the two-loop diagram
in Fig.~\ref{ptself}. The expression is
\bqa\nonumber
\Sigma_2(k)-\Sigma_2(0)&=&\delta^2{N(N+2)\over6}u^2
\int{d^3p\over(2\pi)^3}\int{d^3q\over(2\pi)^3}
\\ && \nonumber
\hspace{-1cm}
\times
\left[
{1\over p^2+m^2}{1\over q^2+m^2}{1\over(p+q+k)^2+m^2}
\right.
\\ &&
\hspace{-1cm}
\left.
-{1\over p^2+m^2}{1\over q^2+m^2}{1\over(p+q)^2+m^2}
\right]\;,
\eqa
where the subscript $n$ indicates the order in the loop expansion.
We have set $d=3$ since the integral finite in three dimensions.
The integral is calculated in the appendix. Using Eq.~(\ref{selfdiff}),
we obtain
\bqa\nonumber
\Sigma_2(k)-\Sigma_2(0)&=&\delta^2{N(N+2)\over6}u^2
{1\over(4\pi)^2}\Bigg[
1
\\ &&
\hspace{-1cm}
-{3m\over k}\arctan{k\over3m}
-{1\over2}\log{k^2+9m^2\over9m^2}
\Bigg]\;.
\label{selfdel}
\eqa
The self-energy~(\ref{selfdel}) is itself second order in $\delta$.
The second-order result for $\Delta_a^{(2)}\langle\phi^2\rangle$
is then obtained by expanding~(\ref{propa1}) in powers of the subtracted
self-energy and keeping only the first term:
\bqa
\Delta_a^{(2)}\langle\phi^2\rangle&=&
\int{d^3k\over(2\pi)^3}{1\over k^4}\left[
\Sigma_2(k)-\Sigma_2(0)
\right]\;,
\eqa
where the superscript indicates the order of $\delta$.
Using Eq.~(\ref{delta2c})
\bqa
\Delta_a^{(2)}\langle\phi^2\rangle&=&
-{1\over108(4\pi)^3}{1\over m}
\delta^2u^2N(N+3)\;.
\label{pms2}
\eqa
The PMS criterion~(\ref{pmsm}) has no solution at second order in $\delta$ 
because~(\ref{pms2}) is a monotonic function of $m$.
Thus one has
to go the third order
in order to obtain a value for $m$. The result 
for 
$\Delta_a^{(3)}\langle\phi^2\rangle$
reads~\cite{tceric}
\bqa\nonumber
\Delta_a^{(3)}\langle\phi_i^2\rangle&=&
-{1\over108(4\pi)^3}{1\over m}
\delta^2u^2N(N+3)\left(1+{1\over2}\delta\right)
\\ &&
-\delta^3{N(N+2)(N+8)}u^3I_3\;,
\eqa
where 
\bqa\nonumber
I_3&=&-{1\over24(4\pi)^4}\left[
\pi^2+16\log{3\over4}+12{\rm Li}_2(-1/3)
\right]{1\over m^2}\;.
\\ &&
\eqa
Here 
\bqa
{\rm Li}_n(x)&=&\sum_{i=1}^{\infty}{x^i\over i^n} 
\eqa
is the polylogarithmic function.
At this order, the PMS criterion has a single real solution:
\bqa
m&=&1.04u{N+8\over24\pi}\;.
\eqa
The resulting value for 
$\Delta_a^{(2)}\langle\phi_i^2\rangle$ is
\bqa
\Delta_a^{(2)}\langle\phi_i^2\rangle&=&
0.4813{N+2\over N+8}\left(-{Nu\over96\pi^2}\right)\;.
\eqa
Setting $N=2$, the prefactor becomes $0.19$.
The lattice results~\cite{arnlat} is
\bqa
\Delta\langle\phi_i^2\rangle&=&
\left(0.284\pm0.004\right)
\left(-{Nu\over96\pi^2}\right)\;.
\eqa
Thus the third-order approximation differs from the lattice Monte Carlo 
result by $66$\%. Similarly, it was found~\cite{tceric} that the fourth
order approximation differs from the numerical simulations by 61\%
In the paper by Braaten and 
Radescu~\cite{tceric}, the convergence of the linear $\delta$
expansion to the exact result in the large--$N$ limit was studied as well.
It was shown that it converges to the lattice Monte-Carlo result~(\ref{lotc}),
but that the convergence is rather slow.

A straightforward 
application of the linear delta-expansion to field theoretic
problems have recently been criticized by Kleinert~\cite{kleinert,kleinertb}.
One must take into account the correct Wegner 
exponent~\cite{we} that governs
the approach to the 
strong-coupling limit. The method is then called variational perturbation
theory (VPT)~\cite{vptdef1}. The failure to take into account the
correct Wegner exponent is the reason for why one finds complex-valued
solutions in the linear delta-expansion.
These matters are highly technical and
beyond the scope of this paper. Interested
readers are referred to the textbook by Kleinert and
Schulte-Frolinde~\cite{kleinert2}.
VPT has very 
reently been applied in a seven-loop calculation by Kastening~\cite{boris2}
(See also~\cite{kleinert,boris} for five and and six-loop calculations.).
The result for the coefficient is $c=1.28\pm 0.1$ in excellent agreement
with lattice field theory results. We also note that seven-loop 
calculations have been carried out for the case $N=1$ and $N=4$ as well.
The values for the coefficient $c$ are $1.07\pm0.10$ and
$1.54\pm0.11$, respectively. These results are also in good agreement
with the lattice simulation of Sun~\cite{sun} who obtained the
values $1.09$ and $1.59$, respectively.

We close this section by listing in Table.~\ref{tctable}
the predictions for the shift
in the critical temperature that has been obtained by the various methods
discussed in this section.

\begin{table}[htb]
\begin{center}
\begin{tabular}{llllll}\hline
\\
${\Delta T_c\over T_0}$&=&$2.33n^{1/3}a$,
& Leading order $1/N$
\\
&&&~\cite{baym3}.
\\
&&&\\
${\Delta T_c\over T_0}$&=&$1.71n^{1/3}a$,\hspace{0.3cm}&
Next-to-leading order $1/N$
\\
&&&~\cite{at}.\\
&&&\\
${\Delta T_c\over T_0}$&=&$\left(1.32\pm0.02\right)n^{1/3}a$,
&
Lattice~\cite{arnlat}.
\\
&&&\\
${\Delta T_c\over T_0}$&=&$\left(1.29\pm0.05\right)n^{1/3}a$,
&
Lattice~\cite{svis}. 
\\
&&&\\
${\Delta T_c\over T_0}$&=&$0.7n^{1/3}a$,
&
One-bubble approximation
\\
&&&~\cite{laloe}.\\
&&&\\
${\Delta T_c\over T_0}$&=&$3.8n^{1/3}a$,
&
One-bubble self-consistent 
\\
&&&approximation
\\
&&&~\cite{baym2}.\\
&&&\\
${\Delta T_c\over T_0}$&=&$2.5n^{1/3}a$,
&
Ladder-summation approximation
\\
&&&~\cite{baym2}.\\
&&&\\
${\Delta T_c\over T_0}$&=&$1.6n^{1/3}a$,
&
Bubble summation approximation
\\
&&&~\cite{baym2}.\\
&&&\\${\Delta T_c\over T_0}$&=&$(1.27\pm0.11)n^{1/3}a$,
&
7-loop VPT~\cite{boris2}.
\\
&&&\\${\Delta T_c\over T_0}$&=&$(1.23\pm)n^{1/3}a$,
&
RG in three dimensions\\
&&&~\cite{ledow}.
\\
&&&\\${\Delta T_c\over T_0}$&=&$(1.3\pm0.4)n^{1/3}a$,
&
Simulations of classical field theory
\\
&&&~\cite{tcdavis}.
\\
&&&\\ \hline
\end{tabular}
\end{center}
\caption{The critical temperature for a dilute Bose 
gas obtained by various analytic and numerical methods.}
\label{tctable}
\end{table}

\section{Conclusions}
In the present paper, we have extensively discussed the dilute Bose gas
at zero and finite temperature. Using effective field theory methods,
a systematic perturbative framework that can be used to 
calculate any quantity of the dilute Bose gas at zero temperature
was set up.
For instance, many of the classic results for the weakly interacting
Bose gas are derived in an efficient and economical manner. Nonuniversal
effects are another application where effective field theory methods 
are ideal. 
For instance, it was used to solve the long-standing problem of 
calculating the full order--$na^3$ correction to the ground state energy
density of a weakly interacting Bose gas.
Similarly, it would be of interest to calculate higher-order corrections to
the condensate density and compare it with numerical results~\cite{georg}.
The strength of the effective field theory approach
lies in the fact
that it is 
a systematic approach. To any given order in the low energy expansion, only
a finite number of terms 
in the effective Lagrangian contribute to a physical quantity
and this allows us to make definite predictions.
Although effective field theories are nonrenormalizable in the traditional
sense of the word, one can carry out renormalization systematically
order by order in the low-energy expansion.

We have also discussed different approaches to the dilute Bose gas
at finite temperature. Although both the Bogoliubov and Popov approximations
are gapless, they break down at sufficiently high temperature. Since the 
interactions between excited bosons are ignored in the
Bogoliubov approximation, it is only valid at very low temperatures, where one 
can neglect the thermal depletion of the condensate.
The Popov approximation can be used at much higher temperatures.
However, being a mean-field theory, it breaks down in the critical region.
It incorrectly predicts a first-order phase transition with the
same $T_c$ as the ideal Bose gas.
The full HFB approximation violates the Hugenholz-Pines theorem,
which any reasonable approximation should incorporate.
The modified Popov approximation based on the many-body $T$-matrix
is an improvement over the original Popov approximation because
it correctly predicts a second-order phase transition.
The renormalization group equations that have been derived~\cite{jmike,henk}
show that the critical properties of the dilute Bose gas
are give by a three-dimensional spin model with a continuous $O(2)$
symmetry. Explicit numerical calculations demonstrate
that the system undergoes a second order phase transition
where the effective coupling constant vanishes at the critical
point. 
Of the other approaches to the thermodynamics of the dilute Bose, 
optimized perturbation
theory is probably the most promising. It is a systematically improvable
expansion with significant flexibility with respect to the choice
of parameters. This approximation also respects the Goldstone theorem
order by order in the perturbative expansion.
To go beyond one loop for the free energy
in optimized perturbation theory is a difficult, but
not impossible task. The complicated setting sun diagrams that appear
at the two-loop level are most easily calculated using the method
where one separates the diagrams into contributions from
zero, one, and two Bose-Einstein distribution functions~\cite{abs}.

The problem of calculating the shift in the critical 
temperature of a weakly interacting Bose gas has been a long-standing
problem.
Using effective field theory methods to obtain an effective three-dimensional
field theory combined with high precision lattice calculations has settled 
the issue in a very elegant way.
The next-to-leading order result in the $1/N$ expansion, shows that
this expansion works surprisingly well for $T_c$; for $N=2$, the result
is 29\% higher than the predictions from the lattice simulations.
Among other notable calculations, we mention the application of 
variational perturbation theory through seven loops. The agreement with
lattice data for $N=1,2$ and $4$  is convincing.

This summarizes our current understanding of some aspects of the 
homogeneous Bose gas at zero and finite temperature. Significant progress
has been made in the last ten years, but there are still open problems.
We hope that this review has stimulated the reader to further research in the
field.

 \section{Acknowledgments}
The author would like to thank P, Arnold, K. Burnett, E. Braaten, \
D. Litim, B. Kastening, 
H. Kleinert, F. Lal\"oe, C.J. Pethick, H.T.C. Stoof and V.I. Yukalov
for useful discussions and suggestions.  
It is also a pleasure to thank Michael Strickland for the collaboration
on which Sec.~\ref{rgapp} is based.

\appendix
\section{Calculational Details, Notation and Conventions}
\renewcommand{\theequation}{\thesection.\arabic{equation}}
In this appendix, we give some calculational details that may be useful for
the reader who is interested in going through the calculations
in detail.
We also define our notation and conventions used throughout the paper.
\subsection{Zero temperature}
All the zero-temperature calculations are carried out in real time. Loop
integrals are integrals over real energies $\omega$ and over 
three-dimensional momenta ${\bf k}$. The integrals over $\omega$
are evaluated using contour integration. 

The specific integrals needed are
\bqa
\int{d\omega\over2\pi}\log\left[\omega^2-\epsilon^2(p)\right]&=&
i\epsilon(p)\;, 
\label{wi1}\\
\int{d\omega\over2\pi}{1\over[\omega^2-\epsilon^2(p)]}&=&
-{i\over2\epsilon(p)}\;, \\
\int{d\omega\over2\pi}{1\over[\omega^2-\epsilon^2(p)]^2}&=&
{i\over4\epsilon^3(p)}\;.
\eqa

Some momentum integrals are divergent in the infrared or in the ultraviolet,
or both.
Dimensional regularization can be used to regularize both the ultraviolet
and infrared divergences in three-dimensional integrals over momenta.
The spatial dimension is generalized to $d=3-2\epsilon$ dimensions.
The continuum limit is taken by replacing sums over wave vectors by integrals 
in $d=3-2\epsilon$
dimensions:
\bqa
{1\over V}\sum_{\bf p}
\rightarrow M^{2\epsilon}
\int{d^dp\over(2\pi)^d}\;. 
\eqa 
Here, $M$ is a renormalization scale that ensures that the integral has
the canonical dimension also for $d\neq3$. 
In the following, we absorb the factor $M^{2\epsilon}$ 
in the measure and so it will not appear explicitly.
Integrals are evaluated at 
a value of $d$, for which they converge and then analytically continued to
$d=3$.

The integral $I_{m,n}$ is defined by
\bqa
\label{idef}
I_{m,n}(\Lambda)=
M^{2\epsilon}
\int{d^dp\over(2\pi)^d}{p^{2m}\over p^n(p^2+\Lambda^2)^{n/2}}\;.
\eqa
With dimensional regularization, $I_{m,n}$ is given by the formula
\bqa\nonumber
I_{m,n}(\Lambda)&=&{\Omega_d\over(2\pi)^d}
M^{2\epsilon}
\Lambda^{d+2m-2n}
\\ &&\times
{\Gamma
\left({{d-n\over2}+m})\Gamma(n-m-{d\over2}\right)
\over2\Gamma\left({n\over2}\right)}\;,
\eqa
where $\Omega_d=2\pi^{d/2}/\Gamma\left({d\over2}\right)$ is the 
area of a $d$-dimensional sphere.

The integrals $I_{m,n}$ satisfy the relations
\bqa
{\!\!\!d\over d\Lambda^2}I_{m,n}&=&-{n\over2}I_{m+1,n+2}\;,
\label{alge1}
\\
\left(d+2m-n\right)I_{m,n}&=&nI_{m+2,n+2}\;,\\
\label{alge2}
\Lambda^2I_{m,n}&=&I_{m-1,n-2}-I_{m+1,n}\;.
\label{alge3}
\eqa
The first relation follows directly from the definition of $I_{m,n}$.
The second relation follows
from integration by parts, while the last
is simply an algebraic relation.

The specific integrals we need in the calculations are listed  below. 
In the limit $d\rightarrow3$, they become
\bqa
I_{0,-1}&=&{\Lambda^5\over15\pi^2}\;,
\label{freedr}
\\
I_{1,1}&=&{\Lambda^3\over3\pi^2}\;,
\label{mu11}
\\
I_{-1,-1}&=&-{\Lambda^3\over6\pi^2}\;,
\label{mu22}
\\
I_{4,5}&=&-{4\Lambda\over3\pi^2}\;,\\
I_{2,3}&=&-{\Lambda\over\pi^2}\;,\\
I_{0,1}&=&-{\Lambda\over2\pi^2}\;,\\
I_{-2,-1}&=&-{\Lambda\over4\pi^2}\left[{1\over\epsilon}+4-L
-\gamma+\log(\pi)
\right]\;,
\label{ir1}
\\
I_{-1,1}&=&-{1\over4\Lambda\pi^2}\left[{1\over\epsilon}+2-L
-\gamma+\log(\pi)\right]\;,
\label{ir2}
\eqa
where $L=\log(\Lambda^2/M^2)$. The integrals $I_{-1,1}$ and
$I_{-2,-1}$ are both logarithmically divergent in the infrared and this 
shows up as a pole in $\epsilon$.

We also need to calculate some integrals in $d$ dimensions that 
depend on the external momentum $k$. The integral ${\cal T}_1(k)$ defined in 
Eq.~(\ref{tdeff}) is
\bqa
{\cal T}_1(k)&=&
\int{d^dq\over(2\pi)^d}{1\over q^2|{\bf q}+{\bf k}|^2}\;.
\eqa
By introducing a Feynman parameter $y$, we can write the integral as
\bqa
T(k)&=&
\int_0^1dy\int{d^dq\over(2\pi)^d}{1\over\left(q^2+m^2\right)^2}\;,
\eqa
where $m^2=y(1-y)k^2$. First integrating over $q$, and then over $y$ gives
\bqa\nonumber
T(k)&=&M^{2\epsilon}
{\Gamma(2-{d\over2})\over(4\pi)^{d/2}}
\int_0^1dy\;
m^{d-4} \\
&=&M^{2\epsilon}
{\Gamma\left(2-{d\over2}\right)\Gamma\left({d\over2}-1\right)
\over2^{2d-3}\pi^{{d-1\over2}}\Gamma\left({d-1\over2}\right)}
k^{d-4}
\;.
\eqa
Other integrals needed are calculated in the same manner. Specifically, we need
the integrals
\bqa\nonumber
&&\int{d^dp\over(2\pi)^d}{1\over p^4}
\left[{1\over|{\bf p}+{\bf k}|^2}-{1\over k^2}\right]
=
\\ &&\hspace{2cm}
M^{2\epsilon}
{\Gamma\left(3-{d\over2}\right)\Gamma\left({d\over2}-2\right)
\over2^{2d-4}\pi^{d-1\over2}\Gamma\left({d-3\over2}\right)}k^{d-6}
\label{pint}
\;,\\\nonumber
&&\int{d^dp\over(2\pi)^d}{p^{d-6}\over a+bp^{d-4}}
=
\\ &&\hspace{1.5cm}
-M^{2\epsilon}
{\Gamma\left({2-d\over4-d}\right)
\Gamma\left({6-d\over4-d}\right)
\over2^{d-1}\pi^{d\over2}\Gamma\left({d\over2}\right)}
a^{d-2\over d-4}b^{3-d\over d-4}
\;.
\label{neeed}
\eqa
Note that the integral in~(\ref{pint}) vanishes
in $d=3$ dimensions due to the
factor $\Gamma(-\epsilon)$ in the denominator. The integral~(\ref{neeed})
has a pole in $\epsilon$. When these integrals are combined, the
limit $d\rightarrow3$ is regular.

We need to evaluate the subtracted two-loop self-energy
in the linear delta expansion. The expression is 
\bqa\nonumber
\Sigma_2(k)-\Sigma_2(0)&=&\int{d^dp\over(2\pi)^d}\int{d^dq\over(2\pi)^d}
\\ && \nonumber
\hspace{-1cm}
\times
\left[
{1\over p^2+m^2}{1\over q^2+m^2}{1\over(p+q+k)^2+m^2}
\right.
\\ &&
\hspace{-1cm}
\left.
-{1\over p^2+m^2}{1\over q^2+m^2}{1\over(p+q)^2+m^2}
\right]\;.
\eqa
This integral is ultraviolet finite and can be calculated 
directly in three dimensions by going to 
coordinate space:
\bqa
\Sigma_2(k)-\Sigma_2(0)&=&\int{d^3r}\left[e^{i{\bf k}\!\cdot\!{\bf r}}-1\right]
V^3(r)\;,
\eqa
where $V(r)$ is the Fourier transform of the propagator:
\bqa\nonumber
V(r)&=&\int{d^dp\over(2\pi)^d}
{e^{i{\bf p}\!\cdot\!{\bf r}}\over p^2+m^2}
\\
&=&
{1\over4\pi r}e^{-mr}\;.
\eqa
This is the usual Yukawa potential.
Integrating over $r$ yields
\bqa\nonumber
\Sigma_2(k)-\Sigma_2(0)&=&
{1\over(4\pi)^2}\left[
1-{3m\over k}\arctan{k\over3m}
\right.\\ &&
\left.
-{1\over2}\log{k^2+9m^2\over9m^2}
\right]\;.
\label{selfdiff}
\eqa
Finally, we need to evaluate the integral
\bqa\nonumber
\Delta_a^{(2)}\langle\phi^2\rangle&=&
\int{d^dk\over(2\pi)^d}{1\over k^4}\left[
\Sigma_2(k)-\Sigma_2(0)
\right]\;.
\eqa
The integral is finite in the ultraviolet and can thus be evaluated directly
in three dimensions. One obtains
\bqa
\Delta_a^{(2)}\langle\phi^2\rangle
&=&-{1\over6(4\pi)^3}{1\over m}\;.
\label{delta2c}
\eqa

\subsection{Finite temperature}
In the imaginary-time formalism for thermal field theory, 
the 4-momentum $P=(\omega_n,{\bf p})$ is Euclidean with 
$P^2=\omega_n^2+{\bf p}^2$. 
The Euclidean energy $p_0$ has discrete values:
$\omega_n=2n\pi T$ for bosons,
where $n$ is an integer. 
Loop diagrams involve sums over $\omega_n$ and integrals over ${\bf p}$. 
With dimensional regularization, the integral is generalized
to $d = 3-2 \epsilon$ spatial dimensions.
We define the dimensionally regularized sum-integral by
\bqa
  \hbox{$\sum$}\!\!\!\!\!\!\int_{P}& \;\equiv\; &
M^{2\epsilon}
  T\sum_{\omega_n=2n\pi T}\:\int {d^dp \over (2 \pi)^{d}}\;,
\label{sumint-def}
\eqa
Again, the factor $M^{2\epsilon}$ 
is absorbed in the measure for convenience.

We also need to evaluate the various sums over 
Matsubara frequencies. They can be calculated by a standard contour
trick, where one rewrites the sum as a contour integral in the complex 
energy plane~\cite{kapusta}.
Specifically, we need the following sums 
\bqa
\sum_{n}\log\left[\omega_n^2+\omega^2\right]&=&\beta\omega
+2\log\left[1-e^{-\beta\omega}\right]\;,\\
\sum_{n}{1\over\omega_n^2+\omega^2}&=&
{\beta\over2\omega}\left[1+2n(\omega)\right]
\;,
\label{exs}
\eqa
where $n(\omega)=1/(e^{\beta\omega}-1)$ 
is the Bose-Einstein distribution function.

We also need to expand some sum-integrals about zero temperature.
The phonon part of the spectrum then dominates the
temperature-dependent part of the sum-integral. We can therefore approximate 
the Bogoliubov dispersion relation $\epsilon(p)$ by $p\sqrt{2\mu}$
and this gives the leading temperature correction. The
specific sum-integrals needed are
\bqa\nonumber
\sumint_P\log\left[\omega_n^2+\epsilon^2(p)\right]
&=&I_{0,-1} 
\\ \nonumber&&
+{T\over\pi^2}\int_0^{\infty}dp\;p^2\log\left[1-e^{-\beta\epsilon(p)}\right]\\
&=&{(2\mu)^{5/2}\over15\pi^2}-{\pi^2T^4\over45(2\mu)^{3/2}}+...\;. 
\label{ftt1}
\\ \nonumber
\sumint_P{\epsilon^2(p)/p^2\over\omega_n^2+\epsilon^2(p)}
&=&{1\over2}I_{-1,-1}+{1\over2\pi^2}\int_0^{\infty}dp\;\epsilon(p)n(\epsilon(p))\\ 
&=&-{(2\mu)^{3/2}\over12\pi^2}+{T^2\over12\sqrt{2\mu}}+...\;,
\label{ftt2}
\\
\nonumber
\sumint_P{p^2\over\omega_n^2+\epsilon^2(p)}
&=&{1\over2}I_{1,1}+{1\over2\pi^2}\int_0^{\infty}dp\;{p^4\over\epsilon(p)}n(\epsilon(p))\\ 
&=&{(2\mu)^{3/2}\over6\pi^2}+{\pi^2T^4\over30(2\mu)^{5/2}}+...\;,
\label{ftt3} \\
\nonumber
\sumint_P{1\over\omega_n^2+\epsilon^2(p)}
&=&{1\over2}I_{0,1}+{1\over2\pi^2}\int_0^{\infty}dp\;{p^2\over\epsilon(p)}n(\epsilon(p))\\ 
&=&-{\sqrt{2\mu}\over4\pi^2}+{\pi^2T^2\over12(2\mu)^{3/2}}+...\;.
\label{ftt4}
\eqa

\bibliographystyle{apsrmp}

\begin{thebibliography}{99}


\expandafter\ifx\csname natexlab\endcsname\relax\def\natexlab#1{#1}\fi
\expandafter\ifx\csname bibnamefont\endcsname\relax
  \def\bibnamefont#1{#1}\fi
\expandafter\ifx\csname bibfnamefont\endcsname\relax
  \def\bibfnamefont#1{#1}\fi
\expandafter\ifx\csname citenamefont\endcsname\relax
  \def\citenamefont#1{#1}\fi
\expandafter\ifx\csname url\endcsname\relax
  \def\url#1{\texttt{#1}}\fi
\expandafter\ifx\csname urlprefix\endcsname\relax\def\urlprefix{URL }\fi
\providecommand{\bibinfo}[2]{#2}
\providecommand{\eprint}[2][]{\url{#2}}


\bibitem[{\citenamefont{Al Khawaja \emph{et~al.}}(2002)}]{all}
\bibinfo{author}{\bibnamefont{Al Khawaja}, \bibfnamefont{U.}},
\bibinfo{author}{\bibnamefont{Andersen}, \bibfnamefont{J.O.}},
\bibinfo{author}{\bibnamefont{Proukakis}, \bibfnamefont{N.P.}}, and
\bibinfo{author}{\bibnamefont{Stoof}, \bibfnamefont{H.T.C.}},
  \bibinfo{year}{2002}, \bibinfo{journal}{Phys Rev.}
  \textbf{\bibinfo{volume}{A66}}, \bibinfo{pages}{013615}.


\bibitem[{\citenamefont{Amelino-Camelia and Pi}(1993)}]{camel}
\bibinfo{author}{\bibnamefont{Amelino-Camelia}, \bibfnamefont{G.}},
\bibinfo{author}{\bibnamefont{Pi}, \bibfnamefont{S.-Y}},
  \bibinfo{year}{1993}, \bibinfo{journal}{Phys. Rev.}
  \textbf{\bibinfo{volume}{D47}}, \bibinfo{pages}{2356}.

\bibitem[{\citenamefont{Andersen and Strickland}(1999)}]{jmike}
\bibinfo{author}{\bibnamefont{Andersen}, \bibfnamefont{J.O.}}, and
\bibinfo{author}{\bibnamefont{Strickland}, \bibfnamefont{M.}},
  \bibinfo{year}{1999}, \bibinfo{journal}{Phys. Rev.}
  \textbf{\bibinfo{volume}{A60}}, \bibinfo{pages}{1442}.


\bibitem[{\citenamefont{Andersen, Braaten and Strickland}(2000)}]{abs}
\bibinfo{author}{\bibnamefont{Andersen}, \bibfnamefont{J.O.}},
\bibinfo{author}{\bibnamefont{Braaten}, \bibfnamefont{E.}}, and
\bibinfo{author}{\bibnamefont{Strickland}, \bibfnamefont{M.}},
  \bibinfo{year}{2000}, \bibinfo{journal}{Phys. Rev.}
  \textbf{\bibinfo{volume}{D62}}, \bibinfo{pages}{045004}.

\bibitem[{\citenamefont{Andersen \emph{et~al.}}(2002)}]{jhenk}
\bibinfo{author}{\bibnamefont{Andersen}, \bibfnamefont{J.O.}},
\bibinfo{author}{\bibnamefont{Al Khawaja}, \bibfnamefont{U.}}, and
\bibinfo{author}{\bibnamefont{Stoof}, \bibfnamefont{H.T.C.}},
  \bibinfo{year}{202}, \bibinfo{journal}{Phys. Rev. Lett.}
  \textbf{\bibinfo{volume}{88}}, \bibinfo{pages}{070407}.




\bibitem[{\citenamefont{Anderson \emph{et~al.}}(1995)}]{bec1}
\bibinfo{author}{\bibnamefont{Anderson}, \bibfnamefont{M.H}},
\bibinfo{author}{\bibnamefont{Ensher}, \bibfnamefont{J.R}},
\bibinfo{author}{\bibnamefont{Matthews}, \bibfnamefont{M.R}},
\bibinfo{author}{\bibnamefont{Wieman}, \bibfnamefont{C}}, and
\bibinfo{author}{\bibnamefont{Cornell}, \bibfnamefont{E.A}},
  \bibinfo{year}{1995}, \bibinfo{journal}{Science}
  \textbf{\bibinfo{volume}{269}}, \bibinfo{pages}{198}.



\bibitem[{\citenamefont{Arnold and Tomasik}(2000)}]{at}
\bibinfo{author}{\bibnamefont{Arnold}, \bibfnamefont{P.}}, and
\bibinfo{author}{\bibnamefont{Tomazik}, \bibfnamefont{B.}},
  \bibinfo{year}{2000}, \bibinfo{journal}{Phys. Rev.}
  \textbf{\bibinfo{volume}{A62}}, \bibinfo{pages}{63604}.

\bibitem[{\citenamefont{Arnold and Moore}(2001)}]{arnlat}
\bibinfo{author}{\bibnamefont{Arnold}, \bibfnamefont{P.}}, and
\bibinfo{author}{\bibnamefont{Moore}, \bibfnamefont{G.D.}},
  \bibinfo{year}{2001}, \bibinfo{journal}{Phys. Rev. Lett.}
  \textbf{\bibinfo{volume}{87}}, \bibinfo{pages}{120401};
  \bibinfo{year}{2001}, \bibinfo{journal}{Phys. Rev.}
  \textbf{\bibinfo{volume}{A64}}, \bibinfo{pages}{066113}.



\bibitem[{\citenamefont{Arnold \emph{et~al.}}(2002)}]{arnold1}
\bibinfo{author}{\bibnamefont{Arnold}, \bibfnamefont{P.}},
\bibinfo{author}{\bibnamefont{Moore}, \bibfnamefont{G.D.}}, and
\bibinfo{author}{\bibnamefont{Tomazik}, \bibfnamefont{B.}},
  \bibinfo{year}{2002}, \bibinfo{journal}{Phys. Rev.}
  \textbf{\bibinfo{volume}{65}}, \bibinfo{pages}{013606}.









\bibitem[{\citenamefont{Baym}(1962)}]{baym}
\bibinfo{author}{\bibnamefont{Baym}, \bibfnamefont{G.}},
  \bibinfo{year}{1962}, \bibinfo{journal}{Phys. Rev.}
  \textbf{\bibinfo{volume}{127}}, \bibinfo{pages}{1391}.





\bibitem[{\citenamefont{Baym \emph{et~al.}}(1999,2001)}]{baym2}
\bibinfo{author}{\bibnamefont{Baym}, \bibfnamefont{G.}},
\bibinfo{author}{\bibnamefont{Blaizot}, \bibfnamefont{J.-P.}},
\bibinfo{author}{\bibnamefont{Holzmann}, \bibfnamefont{M.}},
\bibinfo{author}{\bibnamefont{Laloe}, \bibfnamefont{F.}}, and
\bibinfo{author}{\bibnamefont{Vautherin}, \bibfnamefont{D.}},
  \bibinfo{year}{1999}, \bibinfo{journal}{Phys. Rev. Lett.}
  \textbf{\bibinfo{volume}{83}}, \bibinfo{pages}{1703},
  \bibinfo{year}{2001}, \bibinfo{journal}{Eur Phys. J.}
  \textbf{\bibinfo{volume}{B24}}, \bibinfo{pages}{107}.


\bibitem[{\citenamefont{Baym \emph{et~al.}}(2000)}]{baym3}
\bibinfo{author}{\bibnamefont{Baym}, \bibfnamefont{G.}},
\bibinfo{author}{\bibnamefont{Blaizot}, \bibfnamefont{J.-P}}, and
\bibinfo{author}{\bibnamefont{Zinn-Justin}, \bibfnamefont{J.}},
  \bibinfo{year}{2000}, \bibinfo{journal}{Europhys. Lett. }
  \textbf{\bibinfo{volume}{49}}, \bibinfo{pages}{150}.

\bibitem[{\citenamefont{Baym and Holzmann}(2003)}]{gordon}
\bibinfo{author}{\bibnamefont{Baym}, \bibfnamefont{G.}}, and
\bibinfo{author}{\bibnamefont{Holzmann}, \bibfnamefont{M.}},
  \bibinfo{year}{2003}, \bibinfo{journal}{Phys. Rev. Lett.}
  \textbf{\bibinfo{volume}{90}}, \bibinfo{pages}{040402}.

\bibitem[{\citenamefont{Beliaev}(1958)}]{beli}
\bibinfo{author}{\bibnamefont{Beliaev}, \bibfnamefont{S.T}},
  \bibinfo{year}{1958}, \bibinfo{journal}{Sov. J. Phys.}
  \textbf{\bibinfo{volume}{7}}, \bibinfo{pages}{289}.


\bibitem[{\citenamefont{Bijlsma and Stoof}(1996a)}]{henk}
\bibinfo{author}{\bibnamefont{Bijlsma}, \bibfnamefont{M.}}, and
\bibinfo{author}{\bibnamefont{Stoof}, \bibfnamefont{H.T.C.}},
  \bibinfo{year}{1996}, \bibinfo{journal}{Phys. Rev.}
  \textbf{\bibinfo{volume}{A54}}, \bibinfo{pages}{5085}.


\bibitem[{\citenamefont{Bijlsma and Stoof}(1996a)}]{sb}
\bibinfo{author}{\bibnamefont{Bijlsma}, \bibfnamefont{M.}}, and
\bibinfo{author}{\bibnamefont{Stoof}, \bibfnamefont{H.T.C.}},
  \bibinfo{year}{1996}, \bibinfo{journal}{Phys. Rev.}
  \textbf{\bibinfo{volume}{A55}}, \bibinfo{pages}{498}.




\bibitem[{\citenamefont{Blaizot and Ripka}(1986)}]{rip}
\bibinfo{author}{\bibnamefont{Blaizot}, \bibfnamefont{J.-P}}, and
\bibinfo{author}{\bibnamefont{Ripka}, \bibfnamefont{G.}},
  \bibinfo{year}{1986}, \bibinfo{journal}{{\it Quantum Theory of Finite
Systems}, MIT, Cambridge (1986).}

\bibitem[{\citenamefont{Bogoliubov}(1947)}]{bogo}
\bibinfo{author}{\bibnamefont{Bogoliubov}, \bibfnamefont{N.N}},
  \bibinfo{year}{1947}, \bibinfo{journal}{J. Phys.(USSR)}
  \textbf{\bibinfo{volume}{11}}, \bibinfo{pages}{23}.

\bibitem[{\citenamefont{Bose}(1924)}]{b}
\bibinfo{author}{\bibnamefont{Bose}, \bibfnamefont{S.N}},
  \bibinfo{year}{1924}, \bibinfo{journal}{Z. Phys.}
  \textbf{\bibinfo{volume}{26}}, \bibinfo{pages}{178}.



\bibitem[{\citenamefont{Braaten and Nieto}(1997)}]{eric}
\bibinfo{author}{\bibnamefont{Braaten}, \bibfnamefont{E.}}, and
\bibinfo{author}{\bibnamefont{Nieto}, \bibfnamefont{A.}},
  \bibinfo{year}{1997}, \bibinfo{journal}{Phys. Rev.}
  \textbf{\bibinfo{volume}{B55}}, \bibinfo{pages}{8090}.



\bibitem[{\citenamefont{Braaten and Nieto}(1999)}]{ericag}
\bibinfo{author}{\bibnamefont{Braaten}, \bibfnamefont{E.}}, and
\bibinfo{author}{\bibnamefont{Nieto}, \bibfnamefont{A.}},
  \bibinfo{year}{1999}, \bibinfo{journal}{Eur. Phys J}
  \textbf{\bibinfo{volume}{B11}}, \bibinfo{pages}{143}.


\bibitem[{\citenamefont{Braaten \emph{et~al.}}(2001)}]{e2}
\bibinfo{author}{\bibnamefont{Braaten}, \bibfnamefont{E.}},
\bibinfo{author}{\bibnamefont{Hammer}, \bibfnamefont{H.-W.}}, and
\bibinfo{author}{\bibnamefont{Hermans}, \bibfnamefont{S.}},
  \bibinfo{year}{2001}, \bibinfo{journal}{Phys. Rev.}
  \textbf{\bibinfo{volume}{A63}}, \bibinfo{pages}{063609}.

\bibitem[{\citenamefont{Braaten and Radescu}(2002)}]{tceric}
\bibinfo{author}{\bibnamefont{Braaten}, \bibfnamefont{E.}}, and
\bibinfo{author}{\bibnamefont{Radescu}, \bibfnamefont{E.}},
  \bibinfo{year}{2002}, \bibinfo{journal}{Phys. Rev. }
  \textbf{\bibinfo{volume}{A66}}, \bibinfo{pages}{063601}.

\bibitem[{\citenamefont{Braaten and Hammer}(2003)}]{e+h}
\bibinfo{author}{\bibnamefont{Braaten}, \bibfnamefont{E.}}, and
\bibinfo{author}{\bibnamefont{Hammer}, \bibfnamefont{H.-W.}},
  \bibinfo{year}{2003}, \bibinfo{journal}{Phys. Rev.}
  \textbf{\bibinfo{volume}{A67}}, \bibinfo{pages}{042706}.



\bibitem[{\citenamefont{Bradley \emph{et~al.}}(1995)}]{bec3}
\bibinfo{author}{\bibnamefont{Bradley}, \bibfnamefont{C.C.}},
\bibinfo{author}{\bibnamefont{Sackett}, \bibfnamefont{C.A}},
\bibinfo{author}{\bibnamefont{Tollett}, \bibfnamefont{J.J}}, and
\bibinfo{author}{\bibnamefont{Hulet}, \bibfnamefont{R.G.}},
  \bibinfo{year}{1995}, \bibinfo{journal}{Phys. Rev. Lett.}
  \textbf{\bibinfo{volume}{75}}, \bibinfo{pages}{1687}.

\bibitem[{\citenamefont{Bre\'zin and Wadia}(1993)}]{largen}
\bibinfo{journal}{{\it The Large--N Expansion in Quantum Field Theory
and Statistical Physics}, E. Bre\'zin and S.R. Wadia, eds., World Scientific
Publ. Co. (1993).}


\bibitem[{\citenamefont{Caswell and Lepage}(1986)}]{nrqed}
\bibinfo{author}{\bibnamefont{Caswell}, \bibfnamefont{W.E.}}, and
\bibinfo{author}{\bibnamefont{Lepage}, \bibfnamefont{G.P}},
  \bibinfo{year}{1986}, \bibinfo{journal}{Phys. Lett.}
  \textbf{\bibinfo{volume}{B167}}, \bibinfo{pages}{437}.

\bibitem[{\citenamefont{Brueckner and Sawada}(1957)}]{bs}
\bibinfo{author}{\bibnamefont{Bruecker}, \bibfnamefont{K.A.}}, and
\bibinfo{author}{\bibnamefont{Sawada}, \bibfnamefont{K.}},
  \bibinfo{year}{1957}, \bibinfo{journal}{Phys. Rev.}
  \textbf{\bibinfo{volume}{106}}, \bibinfo{pages}{1117}.



\bibitem[{\citenamefont{Chiku and Hatsuda}(1998)}]{CK-98}
\bibinfo{author}{\bibnamefont{Chiku}, \bibfnamefont{S.}}, and
\bibinfo{author}{\bibnamefont{Hatsuda}, \bibfnamefont{T.}},
  \bibinfo{year}{1998}, \bibinfo{journal}{Phys. Rev.}
  \textbf{\bibinfo{volume}{D58}}, \bibinfo{pages}{076001}.



\bibitem[{\citenamefont{Chiku}(2000)}]{chick2}
\bibinfo{author}{\bibnamefont{Chiku}, \bibfnamefont{S.}},
  \bibinfo{year}{2000}, \bibinfo{journal}{Prog. Theor. Phys.}
  \textbf{\bibinfo{volume}{104}}, \bibinfo{pages}{1129}.


\bibitem[{\citenamefont{Cornish \emph{et~al.}}(2000)}]{corny}
\bibinfo{author}{\bibnamefont{Cornish}, \bibfnamefont{L.}},
\bibinfo{author}{\bibnamefont{Claussen}, \bibfnamefont{N.R.}},
\bibinfo{author}{\bibnamefont{Roberts}, \bibfnamefont{J.L}},
\bibinfo{author}{\bibnamefont{Cornell}, \bibfnamefont{E.A}}, and
\bibinfo{author}{\bibnamefont{Wieman}, \bibfnamefont{C.}},
  \bibinfo{year}{2000}, \bibinfo{journal}{Phys. Rev. Lett.}
  \textbf{\bibinfo{volume}{85}}, \bibinfo{pages}{1795}.

\bibitem[{\citenamefont{Cornwall \emph{et~al.}}(1974)}]{CJT-74}
\bibinfo{author}{\bibnamefont{Cornwall}, \bibfnamefont{J.M.}},
\bibinfo{author}{\bibnamefont{Jackiw}, \bibfnamefont{R.}}, and
\bibinfo{author}{\bibnamefont{Tomboulis}, \bibfnamefont{E.}},
  \bibinfo{year}{1974}, \bibinfo{journal}{Phys. Rev.}
  \textbf{\bibinfo{volume}{D10}}, \bibinfo{pages}{2428}.

\bibitem[{\citenamefont{Dalfovo \emph{et~al.}}(1999)}]{string}
\bibinfo{author}{\bibnamefont{Dalfovo}, \bibfnamefont{F.}},
\bibinfo{author}{\bibnamefont{Giorgini}, \bibfnamefont{S.}},
\bibinfo{author}{\bibnamefont{Pitaevskii}, \bibfnamefont{L.P.}}, and
\bibinfo{author}{\bibnamefont{Stringari}, \bibfnamefont{S.}},
  \bibinfo{year}{1999}, \bibinfo{journal}{Rev. Mod. Phys.}
  \textbf{\bibinfo{volume}{71}}, \bibinfo{pages}{463}.

\bibitem[{\citenamefont{Davis \emph{et~al.}}(1995)}]{bec2}
\bibinfo{author}{\bibnamefont{Davis}, \bibfnamefont{K.B}},
\bibinfo{author}{\bibnamefont{Mewes}, \bibfnamefont{M.O}},
\bibinfo{author}{\bibnamefont{Andrews}, \bibfnamefont{M.R}},
\bibinfo{author}{\bibnamefont{van Druten}, \bibfnamefont{N.J}},
\bibinfo{author}{\bibnamefont{Durfee}, \bibfnamefont{D.S}},
\bibinfo{author}{\bibnamefont{Kurn}, \bibfnamefont{D.M}}, and
\bibinfo{author}{\bibnamefont{Ketterle}, \bibfnamefont{W}},
 \bibinfo{year}{1995}, \bibinfo{journal}{Phys. Rev. Lett.}
 \textbf{\bibinfo{volume}{75}}, \bibinfo{pages}{3969}.

\bibitem[{\citenamefont{Davis and Morgan}(2003)}]{tcdavis}
\bibinfo{author}{\bibnamefont{Davis}, \bibfnamefont{M.J}}, and
\bibinfo{author}{\bibnamefont{Morgan}, \bibfnamefont{S.A}},
  \bibinfo{year}{2003}, \bibinfo{journal}{Phys. Rev.}
  \textbf{\bibinfo{volume}{A??}}, \bibinfo{pages}{}.





\bibitem[{\citenamefont{Duncan and Moshe}(1988)}]{lindel}
\bibinfo{author}{\bibnamefont{Duncan }, \bibfnamefont{A.}}, and
\bibinfo{author}{\bibnamefont{Moshe}, \bibfnamefont{M.}},
  \bibinfo{year}{1988}, \bibinfo{journal}{Phys. Lett.}
  \textbf{\bibinfo{volume}{B215}}, \bibinfo{pages}{352}.

\bibitem[{\citenamefont{Einstein}(1925)}]{e}
\bibinfo{author}{\bibnamefont{Einstein}, \bibfnamefont{A.}},
  \bibinfo{year}{1924}, \bibinfo{journal}{{\it Sitzungsberichte der 
Preussischen Akademie der Wissenschaften, 
Physikalisch-matematische klasse}} \bibinfo{pages}{261},
\bibinfo{year}{1925}, \bibinfo{pages}{3}.

\bibitem[{\citenamefont{Feshbach}(1962)}]{fesh}
\bibinfo{author}{\bibnamefont{Feshbach}, \bibfnamefont{H.}}, and
  \bibinfo{year}{1962}, \bibinfo{journal}{ Ann. Phys.}
  \textbf{\bibinfo{volume}{19}}, \bibinfo{pages}{287}.


\bibitem[{\citenamefont{Fetter and Walecka}(1971)}]{fetter}
\bibinfo{author}{\bibnamefont{Fetter}, \bibfnamefont{A.}}, and
\bibinfo{author}{\bibnamefont{Walecka}, \bibfnamefont{J.D}},
  \bibinfo{year}{}, \bibinfo{journal}{{\it Quantum Theory of 
Many-particle Systems}, McGraw-Hill 1971. }

\bibitem[{\citenamefont{Gasser and Leutwyler}(1984)}]{chiral}
\bibinfo{author}{\bibnamefont{Gasser}, \bibfnamefont{J.}}, and
 \bibinfo{author}{\bibnamefont{Leutwyler}, \bibfnamefont{H.}},
 \bibinfo{year}{1984}, \bibinfo{journal}{Annals Phys.}
  \textbf{\bibinfo{volume}{158}}, \bibinfo{pages}{142};
  \bibinfo{year}{1985}, \bibinfo{journal}{ Nucl. Phys.}
  \textbf{\bibinfo{volume}{B250}}, \bibinfo{pages}{465}.


\bibitem[{\citenamefont{Gavoret and Nozieres}(1964)}]{gav}
\bibinfo{author}{\bibnamefont{Gavoret}, \bibfnamefont{J.}}, and
\bibinfo{author}{\bibnamefont{Nozieres}, \bibfnamefont{P.}},
  \bibinfo{year}{1964}, \bibinfo{journal}{ Ann. Phys. (N.Y.)}
  \textbf{\bibinfo{volume}{28}}, \bibinfo{pages}{349}.

\bibitem[{\citenamefont{Gell-Mann and Brueckner}(1957)}]{gm}
\bibinfo{author}{\bibnamefont{Gell-Mann}, \bibfnamefont{M.}}, and
\bibinfo{author}{\bibnamefont{Brueckner}, \bibfnamefont{K.A.}},
  \bibinfo{year}{1957}, \bibinfo{journal}{Phys. Rev.}
  \textbf{\bibinfo{volume}{106}}, \bibinfo{pages}{364}.


\bibitem[{\citenamefont{Georgi}(1993)}]{eft}
\bibinfo{author}{\bibnamefont{Georgi}, \bibfnamefont{H.}},
  \bibinfo{year}{1993}, \bibinfo{journal}{ Ann. Rev. Nucl. Part. Sci.}
  \textbf{\bibinfo{volume}{43}}, \bibinfo{pages}{209}.

\bibitem[{\citenamefont{Giorgini \emph{et~al.}}(1999)}]{georg}
\bibinfo{author}{\bibnamefont{Giorgini}, \bibfnamefont{S.}},
\bibinfo{author}{\bibnamefont{Boronat}, \bibfnamefont{J.}}, and
\bibinfo{author}{\bibnamefont{Casulleras}, \bibfnamefont{J.}},
  \bibinfo{year}{1999}, \bibinfo{journal}{Phys. Rev.}
  \textbf{\bibinfo{volume}{A60}}, \bibinfo{pages}{5129}


\bibitem[{\citenamefont{Giradeau and Arnowitt}(1959)}]{arno}
\bibinfo{author}{\bibnamefont{Giradeau}, \bibfnamefont{M.}}, and
\bibinfo{author}{\bibnamefont{Arnowitt}, \bibfnamefont{R.}},
  \bibinfo{year}{}, \bibinfo{journal}{Phys. Rev.}
  \textbf{\bibinfo{volume}{113}}, \bibinfo{pages}{755}.

\bibitem[{\citenamefont{Glassgold \emph{et~al.}}(1960)}]{glass}
\bibinfo{author}{\bibnamefont{Glassgold}, \bibfnamefont{A.E.}},
\bibinfo{author}{\bibnamefont{Kaufmann}, \bibfnamefont{A.N}}, and
\bibinfo{author}{\bibnamefont{Watson}, \bibfnamefont{K.M}},
  \bibinfo{year}{1960}, \bibinfo{journal}{Phys. Rev.}
  \textbf{\bibinfo{volume}{120}}, \bibinfo{pages}{660}


\bibitem[{\citenamefont{Goldstone}(1961)}]{gold}
\bibinfo{author}{\bibnamefont{Goldstone}, \bibfnamefont{J.}},
  \bibinfo{year}{1961}, \bibinfo{journal}{Nuovo Cim.}
  \textbf{\bibinfo{volume}{19}}, \bibinfo{pages}{154}.


\bibitem[{\citenamefont{Griffin}(1996)}]{hbf1}
\bibinfo{author}{\bibnamefont{Griffin}, \bibfnamefont{A.}},
  \bibinfo{year}{1996}, \bibinfo{journal}{Phys. Rev.}
  \textbf{\bibinfo{volume}{B53}}, \bibinfo{pages}{9341}.

\bibitem[{\citenamefont{Gr\"uter \emph{et~al.}}(1997)}]{grut}
\bibinfo{author}{\bibnamefont{Gr\"uter}, \bibfnamefont{P.}},
\bibinfo{author}{\bibnamefont{Ceperly}, \bibfnamefont{D.}}, and
\bibinfo{author}{\bibnamefont{Laloe}, \bibfnamefont{F.}},
  \bibinfo{year}{1997}, \bibinfo{journal}{Phys. Rev. Lett.}
  \textbf{\bibinfo{volume}{79}}, \bibinfo{pages}{3549}.



\bibitem[{\citenamefont{Haque}(2003)}]{hakk}
\bibinfo{author}{\bibnamefont{Haque}, \bibfnamefont{M.}},
  \bibinfo{year}{2003}, \bibinfo{journal}{cond-mat/0302076.}

\bibitem[{\citenamefont{Haugset \emph{et~al.}}(1998)}]{haug}
\bibinfo{author}{\bibnamefont{Haugset}, \bibfnamefont{T.}},
\bibinfo{author}{\bibnamefont{Haugerud}, \bibfnamefont{H.}}, and
\bibinfo{author}{\bibnamefont{Ravndal}, \bibfnamefont{F.}},
  \bibinfo{year}{1998}, \bibinfo{journal}{Ann. Phys.}
  \textbf{\bibinfo{volume}{266}}, \bibinfo{pages}{27}.




\bibitem[{\citenamefont{Hohenberg and Martin}(1965)}]{hohen}
\bibinfo{author}{\bibnamefont{Hohenberg }, \bibfnamefont{P.H}}, and
\bibinfo{author}{\bibnamefont{Martin}, \bibfnamefont{P.H}},
  \bibinfo{year}{1965}, \bibinfo{journal}{Ann. Phys. (N.Y.)}
  \textbf{\bibinfo{volume}{34}}, \bibinfo{pages}{291}.


\bibitem[{\citenamefont{Holzmann \emph{et al.}}(1999)}]{laloe}
\bibinfo{author}{\bibnamefont{Holzmann}, \bibfnamefont{M.}}, and
\bibinfo{author}{\bibnamefont{Gr\"uter}, \bibfnamefont{P.}},
\bibinfo{author}{\bibnamefont{Laloe}, \bibfnamefont{F.}},
  \bibinfo{year}{1999}, \bibinfo{journal}{Eur. J. Phys.}
  \textbf{\bibinfo{volume}{10}}, \bibinfo{pages}{739}.

\bibitem[{\citenamefont{Holzmann and Krauth}(1999)}]{kraut}
\bibinfo{author}{\bibnamefont{Holzmann}, \bibfnamefont{M.}}, and
\bibinfo{author}{\bibnamefont{Krauth}, \bibfnamefont{W.}},
  \bibinfo{year}{1999}, \bibinfo{journal}{Phys. Rev. Lett.}
  \textbf{\bibinfo{volume}{83}}, \bibinfo{pages}{2687}.



\bibitem[{\citenamefont{'t Hooft and Veltman}(1972)}]{jerry}
\bibinfo{author}{\bibnamefont{'t Hooft}, \bibfnamefont{G.}}, and
\bibinfo{author}{\bibnamefont{Veltman}, \bibfnamefont{J.M.G.}},
  \bibinfo{year}{1972}, \bibinfo{journal}{Nucl. Phys.}
  \textbf{\bibinfo{volume}{B44}}, \bibinfo{pages}{189}.

\bibitem[{\citenamefont{Huang and Yang}(1957)}]{huangyang}
\bibinfo{author}{\bibnamefont{Huang}, \bibfnamefont{K.}}, and
\bibinfo{author}{\bibnamefont{Yang}, \bibfnamefont{C.N.}},
  \bibinfo{year}{1957}, \bibinfo{journal}{Phys. Rev.}
  \textbf{\bibinfo{volume}{105}}, \bibinfo{pages}{767}.


\bibitem[{\citenamefont{Huang}(1964)}]{huangbok}
\bibinfo{author}{\bibnamefont{Huang}, \bibfnamefont{K.}},
  \bibinfo{year}{1964}, 
\bibinfo{journal}{{\it Imperfect Bose Gas} in {\it Studies
in Statistical Mechanics}, Vol. 2, edited by J. de Boer and G.E.
Uhlenbeck, North-Holland.}

\bibitem[{\citenamefont{Huang}(1999)}]{huang2}
\bibinfo{author}{\bibnamefont{Huang}, \bibfnamefont{K.}},
  \bibinfo{year}{1999}, \bibinfo{journal}{ Phys. Rev. Lett.}
  \textbf{\bibinfo{volume}{83}}, \bibinfo{pages}{3770}.



\bibitem[{\citenamefont{Hugenholz and Pines}(1958)}]{hug}
\bibinfo{author}{\bibnamefont{Hugenholz}, \bibfnamefont{N.M}}, and
\bibinfo{author}{\bibnamefont{Pines}, \bibfnamefont{D.}},
  \bibinfo{year}{1958}, \bibinfo{journal}{Phys. Rev.}
  \textbf{\bibinfo{volume}{116}}, \bibinfo{pages}{489}.


\bibitem[{\citenamefont{Hugenholz and Pines}(1959)}]{hug2}
\bibinfo{author}{\bibnamefont{Hugenholz}, \bibfnamefont{N.M}}, and
\bibinfo{author}{\bibnamefont{Pines}, \bibfnamefont{D.}},
  \bibinfo{year}{1959}, \bibinfo{journal}{Phys. Rev.}
  \textbf{\bibinfo{volume}{116}}, \bibinfo{pages}{489}.


\bibitem[{\citenamefont{Hutchinson \emph{et al.}}(1998)}]{ext2}
\bibinfo{author}{\bibnamefont{Hutchinson}, \bibfnamefont{D.A.W.}},
\bibinfo{author}{\bibnamefont{Dodd}, \bibfnamefont{R.J.}}, and
\bibinfo{author}{\bibnamefont{Burnett}, \bibfnamefont{K.}},
  \bibinfo{year}{1998}, \bibinfo{journal}{Phys. Rev. Lett.}
  \textbf{\bibinfo{volume}{81}}, \bibinfo{pages}{2198}.

\bibitem[{\citenamefont{Hutchinson \emph{et al.}}(2000)}]{hottot}
\bibinfo{author}{\bibnamefont{Hutchinson}, \bibfnamefont{D.A.W.}},
\bibinfo{author}{\bibnamefont{Burnett}, \bibfnamefont{K.}},
\bibinfo{author}{\bibnamefont{Dodd}, \bibfnamefont{R.J.}}, 
\bibinfo{author}{\bibnamefont{Morgan}, \bibfnamefont{S.A.}},
\bibinfo{author}{\bibnamefont{Rusch}, \bibfnamefont{M.}},
\bibinfo{author}{\bibnamefont{Zaremba}, \bibfnamefont{E.}},
\bibinfo{author}{\bibnamefont{Proukakis}, \bibfnamefont{N.P}},
\bibinfo{author}{\bibnamefont{Edwards}, \bibfnamefont{M.}}, and
\bibinfo{author}{\bibnamefont{Clark}, \bibfnamefont{C.W.}},
  \bibinfo{year}{2000}, \bibinfo{journal}{J. Phys. B: At. Mol. Opt. Phys.}
  \textbf{\bibinfo{volume}{33}}, \bibinfo{pages}{3825}.

\bibitem[{\citenamefont{Inouye \emph{et al.}}(1998)}]{ino}
\bibinfo{author}{\bibnamefont{Inouye}, \bibfnamefont{S.}},
\bibinfo{author}{\bibnamefont{Andrews}, \bibfnamefont{M.R}},
\bibinfo{author}{\bibnamefont{Stenger}, \bibfnamefont{J.}}, 
\bibinfo{author}{\bibnamefont{Miesner}, \bibfnamefont{H.-J.}},
\bibinfo{author}{\bibnamefont{Stamper-Kurn}, \bibfnamefont{D.M}}, and
\bibinfo{author}{\bibnamefont{Ketterle}, \bibfnamefont{W.}},
  \bibinfo{year}{1998}, \bibinfo{journal}{Nature}
  \textbf{\bibinfo{volume}{392}}, \bibinfo{pages}{151}.








\bibitem[{\citenamefont{Kaplan}(1995)}]{kap}
\bibinfo{author}{\bibnamefont{Kaplan}, \bibfnamefont{D.}},
  \bibinfo{year}{1995}, \bibinfo{journal}{Lectures given at 
7th Summer School in 
Nuclear Physics Symmetries, Seattle, WA, 18-30 Jun 1995.
e-Print Archive: nucl-th/9506035.}



\bibitem[{\citenamefont{Kapusta}(1989)}]{kapusta}
\bibinfo{author}{\bibnamefont{Kapusta}, \bibfnamefont{J.I.}},
  \bibinfo{year}{1989}, 
\bibinfo{journal}{{\it Finite Temperature Field Theory},
Cambridge University Press.}


\bibitem[{\citenamefont{Kashurnikov \emph{et~al.}}(2001)}]{svis}
\bibinfo{author}{\bibnamefont{Kashurnikov}, \bibfnamefont{V.A.}},
\bibinfo{author}{\bibnamefont{Prokof'ev}, \bibfnamefont{N.V}}, and
\bibinfo{author}{\bibnamefont{Svistunov}, \bibfnamefont{B.V}},
  \bibinfo{year}{2001}, \bibinfo{journal}{Phys. Rev. Lett.}
  \textbf{\bibinfo{volume}{87}}, \bibinfo{pages}{120302}.




\bibitem[{\citenamefont{Kastening}(2003a)}]{boris}
\bibinfo{author}{\bibnamefont{Kastening}, \bibfnamefont{B.}},
  \bibinfo{year}{2003}, \bibinfo{journal}{Phys. Rev.}
  \textbf{\bibinfo{volume}{A68}}, \bibinfo{pages}{061601R}.

\bibitem[{\citenamefont{Kastening}(2003b)}]{boris2}
\bibinfo{author}{\bibnamefont{Kastening}, \bibfnamefont{B.}},
  \bibinfo{year}{2003}, \bibinfo{journal}{cond-mat/0309060.}


\bibitem[{\citenamefont{Kleinert}(1998)}]{vptdef1}
\bibinfo{author}{\bibnamefont{Kleinert}, \bibfnamefont{H.}},
  \bibinfo{year}{1998}, \bibinfo{journal}{Phys. Lett.}
  \textbf{\bibinfo{volume}{B434}}, \bibinfo{pages}{74}
  \bibinfo{year}{1998}; \bibinfo{journal}{Phys. Rev.}
  \textbf{\bibinfo{volume}{D57}}, \bibinfo{pages}{2264}
(addendum \textbf{\bibinfo{volume}{D58}}, \bibinfo{pages}{107702})
  \bibinfo{year}{1999}; \bibinfo{journal}{Phys. Rev.}
  \textbf{\bibinfo{volume}{D60}}, \bibinfo{pages}{085001}.


\bibitem[{\citenamefont{Kleinert and Schulte-Frolinde}(2001)}]{kleinert2}
\bibinfo{author}{\bibnamefont{Kleinert}, \bibfnamefont{H.}}, and
\bibinfo{author}{\bibnamefont{Schulte-Frolinde}, \bibfnamefont{V.}},
  \bibinfo{year}{2001}, 
\bibinfo{journal}{{\it Critical properties of $\phi^4$--Theories}, 1st ed.
World Scientific, Singapore (2001). 
}


\bibitem[{\citenamefont{Kleinert}(2003a)}]{kleinert}
\bibinfo{author}{\bibnamefont{Kleinert}, \bibfnamefont{H.}},
  \bibinfo{year}{2003}, \bibinfo{journal}{Mod. Phys. Lett.}
\textbf{\bibinfo{volume}{B17}}, \bibinfo{pages}{1011}.

\bibitem[{\citenamefont{Kleinert}(2003b)}]{kleinertb}
\bibinfo{author}{\bibnamefont{Kleinert}, \bibfnamefont{H.}},
\bibinfo{author}{\bibnamefont{Hamprecht}, \bibfnamefont{B.}},
  \bibinfo{year}{2003}, \bibinfo{journal}{Phys. Rev.}
\textbf{\bibinfo{volume}{D68}}, \bibinfo{pages}{065001}.





\bibitem[{\citenamefont{Landau}(1947)}]{landau22}
\bibinfo{author}{\bibnamefont{Landau}, \bibfnamefont{L.D.}},
  \bibinfo{year}{1947}, \bibinfo{journal}{ J. Phys. (USSR)}
  \textbf{\bibinfo{volume}{11}}, \bibinfo{pages}{91}.


\bibitem[{\citenamefont{Landau and Lifshitz}(1980)}]{landau}
\bibinfo{author}{\bibnamefont{Landau}, \bibfnamefont{L.D.}}, and
\bibinfo{author}{\bibnamefont{Lifshitz}, \bibfnamefont{E.M}},
  \bibinfo{year}{1980}, \bibinfo{journal}{{\it Quantum Mechanics},
Volume 3, Pergamon Press N.Y. (1980).}




\bibitem[{\citenamefont{Landsman}(1989)}]{lands}
\bibinfo{author}{\bibnamefont{Landsman}, \bibfnamefont{N.P.}},
  \bibinfo{year}{1989}, \bibinfo{journal}{Nucl. Phys.}
  \textbf{\bibinfo{volume}{B322}}, \bibinfo{pages}{498}.




\bibitem[{\citenamefont{Ledowski \emph{et~al.}}(2003)}]{ledow}
\bibinfo{author}{\bibnamefont{Ledowski}, \bibfnamefont{S.}},
\bibinfo{author}{\bibnamefont{Hasselmann}, \bibfnamefont{N.}}, and
\bibinfo{author}{\bibnamefont{Kopietz}, \bibfnamefont{P.}}
  \bibinfo{year}{2003}, \bibinfo{journal}{Cond-mat/0311043.}



\bibitem[{\citenamefont{Leggett}(2001)}]{leggett}
\bibinfo{author}{\bibnamefont{Leggett}, \bibfnamefont{A.J}},
  \bibinfo{year}{2001}, \bibinfo{journal}{Rev. Mod. Phys.}
  \textbf{\bibinfo{volume}{73}}, \bibinfo{pages}{307}.



\bibitem[{\citenamefont{Lee and Yang}(1957)}]{leeyang}
\bibinfo{author}{\bibnamefont{Lee}, \bibfnamefont{T.D.}}, and
\bibinfo{author}{\bibnamefont{Yang}, \bibfnamefont{C.N.}},
  \bibinfo{year}{1957}, \bibinfo{journal}{Phys. Rev.}
  \textbf{\bibinfo{volume}{105}}, \bibinfo{pages}{1119}.

\bibitem[{\citenamefont{Lee Huang and Yang}(1957)}]{leehu}
\bibinfo{author}{\bibnamefont{Lee}, \bibfnamefont{T.D.}}, 
\bibinfo{author}{\bibnamefont{Huang}, \bibfnamefont{K.}}, and
\bibinfo{author}{\bibnamefont{Yang}, \bibfnamefont{C.N.}},
  \bibinfo{year}{1957}, \bibinfo{journal}{Phys. Rev.}
  \textbf{\bibinfo{volume}{106}}, \bibinfo{pages}{1135}.


\bibitem[{\citenamefont{Lee and Yang}(1958)}]{leeyang2}
\bibinfo{author}{\bibnamefont{Lee}, \bibfnamefont{T.D}}, and
\bibinfo{author}{\bibnamefont{Yang}, \bibfnamefont{C.N}},
  \bibinfo{year}{1958}, \bibinfo{journal}{Phys. Rev.}
  \textbf{\bibinfo{volume}{112}}, \bibinfo{pages}{1419}.

\bibitem[{\citenamefont{Lepage}(1989)}]{lep2}
\bibinfo{author}{\bibnamefont{Lepage}, \bibfnamefont{G.P.}},
  \bibinfo{year}{1989}, \bibinfo{journal}{Invited lectures given at 
TASI'89 Summer School, Boulder, CO, Jun 4-30, 1989, Boulder ASI 483}
  \textbf{\bibinfo{volume}{A16}}, \bibinfo{pages}{483}.

\bibitem[{\citenamefont{Liao and Strickland}(1995)}]{mike1}
\bibinfo{author}{\bibnamefont{Liao}, \bibfnamefont{S.-B.}}, and
\bibinfo{author}{\bibnamefont{Strickland}, \bibfnamefont{M.}},
  \bibinfo{year}{1995}, \bibinfo{journal}{ Phys. Rev.}
  \textbf{\bibinfo{volume}{D52}}, \bibinfo{pages}{3653}.

\bibitem[{\citenamefont{Lieb}(1963)}]{lieb}
\bibinfo{author}{\bibnamefont{Lieb}, \bibfnamefont{E.H.}},
  \bibinfo{year}{1963}, \bibinfo{journal}{Phys. Rev.}
  \textbf{\bibinfo{volume}{130}}, \bibinfo{pages}{2518}.



\bibitem[{\citenamefont{Litim and Pawlowski}(2002)}]{litp}
\bibinfo{author}{\bibnamefont{Litim}, \bibfnamefont{D.F.}},
\bibinfo{author}{\bibnamefont{Pawlowski}, \bibfnamefont{J.M.}},
  \bibinfo{year}{2002}, \bibinfo{journal}{Phys. Rev.}
  \textbf{\bibinfo{volume}{D65}}, \bibinfo{pages}{081701}.

\bibitem[{\citenamefont{Litim}(2001)}]{daniel2}
\bibinfo{author}{\bibnamefont{Litim}, \bibfnamefont{D.F.}},
  \bibinfo{year}{2001}, \bibinfo{journal}{Phys. Rev.}
  \textbf{\bibinfo{volume}{D64}}, \bibinfo{pages}{105007}.

\bibitem[{\citenamefont{Litim}(2000)}]{daniel1}
\bibinfo{author}{\bibnamefont{Litim}, \bibfnamefont{D.F.}},
  \bibinfo{year}{2000}, \bibinfo{journal}{Phys. Lett.}
  \textbf{\bibinfo{volume}{B486}}, \bibinfo{pages}{92}.



\bibitem[{\citenamefont{Liu}(1997)}]{w1}
\bibinfo{author}{\bibnamefont{Liu}, \bibfnamefont{W.V}},
  \bibinfo{year}{1997}, \bibinfo{journal}{Phys. Rev. Lett.}
  \textbf{\bibinfo{volume}{79}}, \bibinfo{pages}{4056}.

\bibitem[{\citenamefont{Liu}(1998)}]{w2}
\bibinfo{author}{\bibnamefont{Liu}, \bibfnamefont{W.V}},
  \bibinfo{year}{1998}, \bibinfo{journal}{Int. J. Mod. Phys.}
  \textbf{\bibinfo{volume}{B12}}, \bibinfo{pages}{2103}.

\bibitem[{\citenamefont{Lundh and Rammer}(2002)}]{lundh}
\bibinfo{author}{\bibnamefont{Lundh}, \bibfnamefont{E.}}, and
\bibinfo{author}{\bibnamefont{Rammer}, \bibfnamefont{J.}},
  \bibinfo{year}{2002}, \bibinfo{journal}{Phys. Rev.}
  \textbf{\bibinfo{volume}{A66}}, \bibinfo{pages}{033607}.


\bibitem[{\citenamefont{Luttinger and Ward}(1960)}]{lut}
\bibinfo{author}{\bibnamefont{Luttinger}, \bibfnamefont{J.M.}}, and
\bibinfo{author}{\bibnamefont{Ward}, \bibfnamefont{J.D.}},
  \bibinfo{year}{1960}, \bibinfo{journal}{Phys. Rev.}
  \textbf{\bibinfo{volume}{118}}, \bibinfo{pages}{1417}.











\bibitem[{\citenamefont{Metikas and Alber}(2002)}]{alber}
\bibinfo{author}{\bibnamefont{Metikas}, \bibfnamefont{G.}}, and
\bibinfo{author}{\bibnamefont{Alber}, \bibfnamefont{G.}},
  \bibinfo{year}{2002}, \bibinfo{journal}{J. Phys}
  \textbf{\bibinfo{volume}{B35}}, \bibinfo{pages}{4223}.









\bibitem[{\citenamefont{Morgan}(1999)}]{sam}
\bibinfo{author}{\bibnamefont{Morgan}, \bibfnamefont{S.}},
  \bibinfo{year}{1999}, \bibinfo{journal}{J. Phys}
  \textbf{\bibinfo{volume}{B33}}, \bibinfo{pages}{3847}.



\bibitem[{\citenamefont{Morris}(1994)}]{morris22}
\bibinfo{author}{\bibnamefont{Morris}, \bibfnamefont{T.R.}},
  \bibinfo{year}{1994}, \bibinfo{journal}{Phys. Lett.}
  \textbf{\bibinfo{volume}{B329}}, \bibinfo{pages}{241}.
  \bibinfo{year}{1994}, \bibinfo{journal}{Phys. Lett.}
  \textbf{\bibinfo{volume}{B334}}, \bibinfo{pages}{355}.


\bibitem[{\citenamefont{Morris}(1998)}]{morris}
\bibinfo{author}{\bibnamefont{Morris}, \bibfnamefont{T.R.}},
  \bibinfo{year}{1998}, \bibinfo{journal}{Int. J. Mod. Phys.}
  \textbf{\bibinfo{volume}{B12}}, \bibinfo{pages}{1343}.

\bibitem[{\citenamefont{Moshe and Zinn-Justin}(2003)}]{moshe}
\bibinfo{author}{\bibnamefont{Moshe}, \bibfnamefont{M.}},
\bibinfo{author}{\bibnamefont{Zinn-Justin}, \bibfnamefont{J.}},
  \bibinfo{year}{1998}, \bibinfo{journal}{Phys. Rept.}
  \textbf{\bibinfo{volume}{385}}, \bibinfo{pages}{69}.







\bibitem[{\citenamefont{Negele and Orland}(1988)}]{nege}
\bibinfo{author}{\bibnamefont{Negele}, \bibfnamefont{J.W.}}, and
\bibinfo{author}{\bibnamefont{Orland}, \bibfnamefont{H.}},
  \bibinfo{year}{1988}, \bibinfo{journal}{{\it Quantum Many-Particle Systems}, Addison-Wesley, New York, (1988).
}


\bibitem[{\citenamefont{Nicoll \emph{et~al.}}(1974)}]{poly}
\bibinfo{author}{\bibnamefont{Nicoll}, \bibfnamefont{J.F.}},
\bibinfo{author}{\bibnamefont{Chang}, \bibfnamefont{T.S.}}, and
\bibinfo{author}{\bibnamefont{Stanley}, \bibfnamefont{H.E.}},
  \bibinfo{year}{1974}, \bibinfo{journal}{Phys. Rev. Lett.}
  \textbf{\bibinfo{volume}{33}}, \bibinfo{pages}{540}.


\bibitem[{\citenamefont{Okopinska}(1987)}]{oko}
\bibinfo{author}{\bibnamefont{Okopinska}, \bibfnamefont{A.}},
  \bibinfo{year}{1987}, \bibinfo{journal}{Phys. Rev.}
  \textbf{\bibinfo{volume}{D35}}, \bibinfo{pages}{1835}.


\bibitem[{\citenamefont{Pethick and Smith}(2002)}]{pet}
\bibinfo{author}{\bibnamefont{Pethick}, \bibfnamefont{C.J}}, and
\bibinfo{author}{\bibnamefont{Smith}, \bibfnamefont{H.}},
  \bibinfo{year}{2002}, \bibinfo{journal}{
{\it Bose-Einstein Condensation
in Dilute Gases}, Cambridge University Press, UK.}



\bibitem[{\citenamefont{Petrov \emph{et~al.}}(2000)}]{2d}
\bibinfo{author}{\bibnamefont{Petrov}, \bibfnamefont{D.S.}},
\bibinfo{author}{\bibnamefont{Holzmann}, \bibfnamefont{M.}}, and
\bibinfo{author}{\bibnamefont{Shlyapnikov}, \bibfnamefont{G.V}},
  \bibinfo{year}{2000}, \bibinfo{journal}{Phys. Rev. Lett.}
  \textbf{\bibinfo{volume}{84}}, \bibinfo{pages}{2551};
%
\bibitem[{\citenamefont{Petrov \emph{et~al.}}(2000)}]{1d}
\bibinfo{author}{\bibnamefont{Petrov}, \bibfnamefont{D.S.}},
\bibinfo{author}{\bibnamefont{Shlyapnikov}, \bibfnamefont{G.V}}, and
 \bibinfo{author}{\bibnamefont{Walraven}, \bibfnamefont{J.T.M.}},
\bibinfo{year}{2000}, \bibinfo{journal}{Phys. Rev. Lett.}
  \textbf{\bibinfo{volume}{85}}, \bibinfo{pages}{3745}.


\bibitem[{\citenamefont{Pitaevskii and Stringari}(2003)}]{pitab}
\bibinfo{author}{\bibnamefont{Pitaevskii}, \bibfnamefont{L.P.}},
\bibinfo{author}{\bibnamefont{Stringarii}, \bibfnamefont{S.}},
  \bibinfo{year}{2003}, \bibinfo{journal}{{\it Bose-Einstein
Condensation}, Oxford University Press, UK}

\bibitem[{\citenamefont{Polchinski}(1984)}]{joe}
\bibinfo{author}{\bibnamefont{Polchinski}, \bibfnamefont{J.}},
  \bibinfo{year}{1984}, \bibinfo{journal}{ Nucl. Phys.}
  \textbf{\bibinfo{volume}{231}}, \bibinfo{pages}{269}.
.

\bibitem[{\citenamefont{Popov}(1983)}]{popov}
\bibinfo{author}{\bibnamefont{Popov}, \bibfnamefont{V.N.}},
  \bibinfo{year}{1983}, \bibinfo{journal}{{\it Functional 
Integrals in Quantum Field Theory
and Statistical Physics}, (Reidel, Doordrecht 1983).}

\bibitem[{\citenamefont{Popov}(1987)}]{popov2}
\bibinfo{author}{\bibnamefont{Popov}, \bibfnamefont{V.N.}},
  \bibinfo{year}{}, \bibinfo{journal}{{\it Functional Integrals and Collective 
excitations}, Cambridge University Press, Cambridge (1987)}


\bibitem[{\citenamefont{Proukakis \emph{et~al.}}(1998)}]{ext1}
\bibinfo{author}{\bibnamefont{Proukakis}, \bibfnamefont{N.P.}},
\bibinfo{author}{\bibnamefont{Morgan}, \bibfnamefont{S.A.}},
\bibinfo{author}{\bibnamefont{Choi}, \bibfnamefont{S.}}, and
\bibinfo{author}{\bibnamefont{Burnett}, \bibfnamefont{K.}},
  \bibinfo{year}{1998}, \bibinfo{journal}{Phys. Rev.}
  \textbf{\bibinfo{volume}{58}}, \bibinfo{pages}{2435}.

\bibitem[{\citenamefont{Sawada}(1959)}]{saw}
\bibinfo{author}{\bibnamefont{Sawada}, \bibfnamefont{K.}},
  \bibinfo{year}{2001}, \bibinfo{journal}{Phys. Rev.}
  \textbf{\bibinfo{volume}{116}}, \bibinfo{pages}{1344}.


\bibitem[{\citenamefont{Schakel}(1994)}]{schakel}
\bibinfo{author}{\bibnamefont{Schakel}, \bibfnamefont{A.M.J.}},
  \bibinfo{year}{1994}, \bibinfo{journal}{Int. J. Mod. Phys.}
  \textbf{\bibinfo{volume}{B8}}, \bibinfo{pages}{2021}.



\bibitem[{\citenamefont{Shi and Griffin}(1998)}]{grif}
\bibinfo{author}{\bibnamefont{Shi}, \bibfnamefont{H.}}, and
\bibinfo{author}{\bibnamefont{Griffin}, \bibfnamefont{A.}},
  \bibinfo{year}{1998}, \bibinfo{journal}{Phys. Rept.}
  \textbf{\bibinfo{volume}{304}}, \bibinfo{pages}{1}.




\bibitem[{\citenamefont{de Souza Cruz \emph{et~al.}}(2001)}]{ram1}
\bibinfo{author}{\bibnamefont{de Souza Cruz}, \bibfnamefont{F.F.}},
\bibinfo{author}{\bibnamefont{Pinto}, \bibfnamefont{M.B.}}, and
\bibinfo{author}{\bibnamefont{Ramos}, \bibfnamefont{R.O}}
  \bibinfo{year}{2001}, \bibinfo{journal}{Phys. Rev.}
  \textbf{\bibinfo{volume}{B64}}, \bibinfo{pages}{014515}.

\bibitem[{\citenamefont{de Souza Cruz \emph{et~al.}}(2002)}]{ram2}
\bibinfo{author}{\bibnamefont{de Souza Cruz}, \bibfnamefont{F.F.}},
\bibinfo{author}{\bibnamefont{Pinto}, \bibfnamefont{M.B.}},
\bibinfo{author}{\bibnamefont{Ramos}, \bibfnamefont{R.O}} and
\bibinfo{author}{\bibnamefont{Sena}, \bibfnamefont{P.}},
  \bibinfo{year}{2002}, \bibinfo{journal}{Phys. Rev.}
  \textbf{\bibinfo{volume}{A65}}, \bibinfo{pages}{053613}.


\bibitem[{\citenamefont{Stancu and Stevenson}(1990)}]{steve}
\bibinfo{author}{\bibnamefont{Stancu}, \bibfnamefont{I.}}, and
\bibinfo{author}{\bibnamefont{Stevenson}, \bibfnamefont{P.M.}},
  \bibinfo{year}{1990}, \bibinfo{journal}{Phys. Rev.}
  \textbf{\bibinfo{volume}{D42}}, \bibinfo{pages}{2710}.

\bibitem[{\citenamefont{Stevenson}(1981)}]{s-81}
\bibinfo{author}{\bibnamefont{Stevenson}, \bibfnamefont{P.M.}},
  \bibinfo{year}{1981}, \bibinfo{journal}{Phys. Rev.}
  \textbf{\bibinfo{volume}{D23}}, \bibinfo{pages}{2916}.

\bibitem[{\citenamefont{Stwalley}(1976)}]{stw}
\bibinfo{author}{\bibnamefont{Stwalley}, \bibfnamefont{W.C.}},
  \bibinfo{year}{1976}, \bibinfo{journal}{Phys. Rev. Lett.}
  \textbf{\bibinfo{volume}{37}}, \bibinfo{pages}{1628}.







\bibitem[{\citenamefont{Sun}(2003)}]{sun}
\bibinfo{author}{\bibnamefont{Sun}, \bibfnamefont{X.}},
  \bibinfo{year}{2003}, \bibinfo{journal}{Phys. Rev.}
  \textbf{\bibinfo{volume}{E67}}, \bibinfo{pages}{066702}.

\bibitem[{\citenamefont{Takano}(1961)}]{taka}
\bibinfo{author}{\bibnamefont{Takan}, \bibfnamefont{F.}},
  \bibinfo{year}{1961}, \bibinfo{journal}{Phys. Rev.}
  \textbf{\bibinfo{volume}{123}}, \bibinfo{pages}{699}.



\bibitem[{\citenamefont{Tiesinga, Verhaar and Stoof}(1993)}]{ties}
\bibinfo{author}{\bibnamefont{Tiesinga}, \bibfnamefont{E.}}, 
\bibinfo{author}{\bibnamefont{Verhaar}, \bibfnamefont{B.J.}}, and
\bibinfo{author}{\bibnamefont{Stoof}, \bibfnamefont{H.T.C.}},
  \bibinfo{year}{1993}, \bibinfo{journal}{Phys. Rev.}
  \textbf{\bibinfo{volume}{A47}}, \bibinfo{pages}{4114}.




\bibitem[{\citenamefont{Toyoda}(1982)}]{toyota}
\bibinfo{author}{\bibnamefont{Toyoda}, \bibfnamefont{T.}},
  \bibinfo{year}{1982}, \bibinfo{journal}{Ann. Phys. (N.Y)}
  \textbf{\bibinfo{volume}{141}}, \bibinfo{pages}{154}.

\bibitem[{\citenamefont{Wegner}(1972)}]{we}
\bibinfo{author}{\bibnamefont{Wegner}, \bibfnamefont{F.J.}},
  \bibinfo{year}{1972}, \bibinfo{journal}{Phys. Rev.}
  \textbf{\bibinfo{volume}{B5}}, \bibinfo{pages}{1891}.


\bibitem[{\citenamefont{Wilkens \emph{et~al.}}(2000)}]{illu}
\bibinfo{author}{\bibnamefont{Wilkens}, \bibfnamefont{M.}},
\bibinfo{author}{\bibnamefont{Illuminati}, \bibfnamefont{F.}}, and
\bibinfo{author}{\bibnamefont{Kra\"mer}, \bibfnamefont{M.}},
  \bibinfo{year}{2000}, \bibinfo{journal}{J. Phys.}
  \textbf{\bibinfo{volume}{B33}}, \bibinfo{pages}{L779}.


\bibitem[{\citenamefont{Wilson and Kogut}(1974)}]{wilson}
\bibinfo{author}{\bibnamefont{Wilson}, \bibfnamefont{K.}}, and
\bibinfo{author}{\bibnamefont{Kogut}, \bibfnamefont{J.}},
  \bibinfo{year}{1974}, \bibinfo{journal}{Phys. Rep.}
  \textbf{\bibinfo{volume}{12}}, \bibinfo{pages}{75}.


\bibitem[{\citenamefont{Wu}(1959)}]{wu}
\bibinfo{author}{\bibnamefont{Wu}, \bibfnamefont{T.T}},
  \bibinfo{year}{1959}, \bibinfo{journal}{Phys. Rev.}
  \textbf{\bibinfo{volume}{115}}, \bibinfo{pages}{1390}.

\bibitem[{\citenamefont{Yan \emph{et al.}}(1996)}]{atom2}
\bibinfo{author}{\bibnamefont{Yan}, \bibfnamefont{Z.-C.}},
\bibinfo{author}{\bibnamefont{Babb}, \bibfnamefont{J.F.}},
\bibinfo{author}{\bibnamefont{Dalgarno}, \bibfnamefont{A.}}, and
\bibinfo{author}{\bibnamefont{Drake}, \bibfnamefont{G.W.F.}},
  \bibinfo{year}{1996}, \bibinfo{journal}{Phys. Rev. }
  \textbf{\bibinfo{volume}{A54}}, \bibinfo{pages}{2824}.

\bibitem[{\citenamefont{Yukalov}(1976)}]{yukyuk}
\bibinfo{author}{\bibnamefont{Yukalov}, \bibfnamefont{V.I}},
  \bibinfo{year}{1976}, \bibinfo{journal}{Moscow Univ. Phys. Bull.}
  \textbf{\bibinfo{volume}{31}}, \bibinfo{pages}{10}.


\bibitem[{\citenamefont{Zinn-Justin}(1989)}]{zinn}
\bibinfo{author}{\bibnamefont{Zinn-Justin}, \bibfnamefont{J.}},
  \bibinfo{year}{}, \bibinfo{journal}{{\it Quantum field Theory and Critical
Phenomena}, Oxford University Press Inc., New York.}



 

\end{thebibliography}

\end{document}